\documentclass[12pt]{article}

\usepackage{amsmath}
\usepackage{amssymb}
\usepackage{amsfonts}
\usepackage{graphicx}
\usepackage{subcaption}
\usepackage{slashed} 

\numberwithin{equation}{section}
\begin{document}

\setcounter{page}{0}
\thispagestyle{empty}

\begin{center}
{\large \bf{ The QCD vacuum as a disordered chromomagnetic condensate }}
\end{center}

\vspace*{2cm}

\begin{center}
{
Paolo Cea\protect\footnote{Electronic address:
{\tt paolo.cea@ba.infn.it}}  \\[0.5cm]
{\em INFN - Sezione di Bari, via G. Amendola 173, I-70126 Bari,
Italy} }
\end{center}

\vspace*{1.5cm}

\renewcommand{\abstractname}{\normalsize ABSTRACT}
\begin{abstract}
\noindent
An attempt is made to describe from first principles the large-scale structure of the confining vacuum in quantum
chromodynamics. Starting from our previous variational studies of the SU(2) pure gauge theory in an external
Abelian chromomagnetic field and extending the Feynman's qualitative analysis in (2+1)-dimensional SU(2)
gauge theory, we show that the SU(3) vacuum in three-space and one-time dimensions behaves like a disordered
chromomagnetic condensate. Color confinement is assured by the presence of a mass gap together with the absence
of color long-range correlations. We offer a clear physical picture for the formation of the flux tube between static quark charges that
allowed to determine the color structure and the transverse profile of the flux-tube chromoelectric field.
The transverse profile of the flux-tube chromoelectric field turns out to be in reasonable agreement with   lattice data. 
We, also, show that our quantum vacuum allows for  both the color and ordinary Meissner effect. We find that for massless quarks 
 the quantum vacuum could accommodate a finite non-zero density of fermion zero modes leading to the dynamical
 breaking of the chiral  symmetry.
\end{abstract}

\vspace*{0.5cm}
\begin{flushleft}
\end{flushleft}
\newpage

\tableofcontents
\newpage
\section{Introduction}
\label{S1}
The most challenging aspect of high energy physics is understanding color and quark confinement
(for an overview see Refs.~\cite{Joos:1979,Mandelstam:1980,Bander:1981,Zachariasen:1986,Haymaker:1999,Greensite:2003,Kogut:2004,Ripka:2004,Shifman:2009,Greensite:2011}).
 Despite the fact that the theory of the strong interactions, Quantum Chromodynamics (QCD), is known since decades,
and  not withstanding large efforts with numerical simulations aimed to unravelling  the nature of the QCD
vacuum,  we still lack a fundamental understanding of the corresponding physics.
As a matter of fact, the mechanism that leads to color confinement remains an open question in spite of intense
non-perturbative lattice studies for more than three decades. \\
According to a model conjectured long time  ago by G.~'t~Hooft~\cite{tHooft:1976,tHooft:1980,tHooft:1982} and 
S.~Mandelstam~\cite{Mandelstam:1976} the confining vacuum behaves as a coherent state of color magnetic monopoles, 
or, equivalently, the vacuum resembles a magnetic (dual) superconductor  (for a more complete discussion, 
see Refs.~\cite{Haymaker:1999,Ripka:2004,Greensite:2011,Baker:1991,Kondo:2015} and references therein). 
Up to now there  have been extensive numerical studies of monopole condensation. However, even if magnetic monopoles do condense in the confinement mode, the actual mechanism of confinement could  depend on additional dynamical forces. \\
An alternative model for color confinement is based on the special role of center vortices. By means of lattice
simulations evidence has been accumulated that center vortices are responsible for confinement (for instance, 
see Refs.~\cite{Greensite:2003,Greensite:2011} and references therein). 
A different confinement picture was advanced by V. N. Gribov~\cite{Gribov:1978,Gribov:2001} and further
elaborated by D. Zwanzinger~\cite{Zwanziger:1989} where it is argued that the gauge field configurations which are relevant
for confinement are concentrated on the so-called Gribov horizon (for a pedagogic overview, see Ref.~\cite{Vandersickel:2012}
and references therein).
As a matter of fact, it has been
suggested~\cite{Greensite:2004a,Gattnar:2004,Greensite:2005} that both Gribov-Zwanzinger and center vortex picture of
confinement are compatible. On the other hand, interestingly enough in Refs.~\cite{Feuchter:2004,Reinhardt:2005,Reinhardt:2012} 
pure Yang-Mills theories 
have been investigated within the Schr\"odinger representation in the Coulomb gauge. These authors, by using vacuum functionals
which are strongly peaked at the Gribov horizon, showed that the vacuum functional becomes field independent in the
infrared consistent with  a stochastic vacuum at large distances. \\
One could conclude that there is no totally convincing explanation of the confinement phenomenon  and that a full
understanding of the QCD vacuum dynamics is still lacking. However, previous different pictures of confinement can be
reconciled if the confining vacuum behaves like a disordered chromomagnetic condensate. In this case the condensation of
vortices and chromomagnetic monopoles are  only a symptom rather than the origin of confinement. \\
In a seminal paper, R. P. Feynman~\cite{Feynman:1981} argued that in (2+1)-dimensions the SU(2) confining vacuum at large
distances is a chromomagnetic condensate disordered by the gauge symmetry. The confinement of colours comes from the
existence of a mass gap and the absence of color long range order. In this paper we will show that, indeed, the
Yang-Mills vacuum in (3+1)-dimensions does display at large distance a mass gap and no color long range order. On
the other hand, we will argue that the color dynamics at very short distances is governed by the perturbative regime. \\
In Refs.~\cite{Cea:1997a,Cea:1997b,Cea:1997c}, by means of the lattice Schr\"odinger functional, it was introduced a gauge invariant effective action 
for external static background fields that allowed to probe non-perturbatively the dynamics of gauge theories.
In particular~\cite{Cea:2003,Cea:2005,Cea:2006,Cea:2007} the Yang-Mills vacuum was probed by means of an external constant
Abelian chromomagnetic field on the lattice. Actually, up to now, the lattice studies have been limited to the SU(3) pure gauge theory or
to QCD with two degenerate massive quarks.
It turned out  that, by increasing the strength of the applied external field, the
deconfinement temperature decreases towards zero. In other words, there is a critical field $gH_c$ such that, for
$gH>gH_c$, the gauge system is in the deconfined phase. More precisely, it resulted that:
\begin{equation}
\label{1.1}
\frac{T_c(gH)}{T_c(0)} \; = \; 1 \; - \; \frac{\sqrt{gH}}{\sqrt{gH_c}} \; .
\end{equation}
As a consequence, one is lead to suspect  that there must be  an intimate connection between Abelian chromomagnetic fields and color
confinement. The existence of a critical chromomagnetic field is not easily understandable within the coherent magnetic
monopole or vortex condensate picture of the confining vacuum, but it could be compatible with the disordered chromomagnetic
condensate picture, for strong enough chromomagnetic field strengths enforce long-range color order thereby
destroying confinement. On the other hand, the peculiar  color Meissner effect, Eq.~(\ref{1.1}), could be easily explained
 if the vacuum behaves as an ordinary relativistic color superconductor. 
 Thus, we have to reconcile two apparently different aspects. From one hand, the
confining vacuum does display condensation of both Abelian magnetic monopoles and vortices, on the other hand the relation
between the deconfinement temperature and the applied Abelian chromomagnetic field, Eq.~(\ref{1.1}), would imply that the
magnetic length $a_H \sim 1/\sqrt{gH}$ is the only relevant scale of the problem. 
Indeed, Eq.~(\ref{1.1}) suggests that the vacuum behaves as a condensate of a color charged scalar field whose mass is
proportional to the inverse of the magnetic length. Interestingly enough, long time ago it has been pointed out that
in a  Yang-Mills theory a ferro-chromomagnetic state could have a lower energy with respect to the perturbative 
vacuum~\cite{Savvidy:1977,Matinyan:1978,Pagels:1978,Savvidy:2020}.  However, it was soon shown by N. K. Nielsen and 
P. Olesen~\cite{Nielsen:1978a,Nielsen:1978b} (also, see Ref.~\cite{Leutwyler:1981}) that such a state is affected by unstable modes.
Thus, we see that a natural candidate for the tachyonic color charged scalar fields are
the  Nielsen-Olesen unstable modes, for the eventual condensation of these modes should make the
vacuum a dynamical color superconductor. This kind of arguments led long time ago to the proposal of the so-called
Copenhagen 
vacuum~\cite{Ambjorn:1979,Nielsen:1979a,Nielsen:1979b,Ambjorn:1980a,Ambjorn:1980b,Nielsen:1980,Olesen:1981}, 
where vacuum field configurations which differ from the classical external field only in the unstable mode sector would result in 
the formation of a quantum liquid. 
However, in Refs.~\cite{Cea:1987,Cea:1988}
we showed that, by using variational techniques on a class of gauge-invariant Gaussian wave functionals,
the stabilization of the Nielsen-Olesen  modes contributes to the energy density with a negative classical term which
cancels the classical magnetic energy. Moreover the stabilization of the Nielsen-Olesen modes induces a further background
field which behaves non analytically in the coupling constant and screens almost completely the external chromomagnetic
field. As a consequence, even in the strong-field regime, the naive perturbative regime, one deals with the
non-perturbative regime. This means that the calculation of the energy density even in the one-loop approximation is
non-perturbative. Nevertheless, we believe that the gauge-invariant variational-perturbative approach presented in 
Refs.~\cite{Cea:1987,Cea:1988}
convincingly showed that the Nielsen-Olesen instability is an artefact of the one-loop approximation and that the correct
treatment of this one-loop instability leads to a drastic reduction of the energy density of the trial vacuum functional.
Indeed, this kind of effect has been already checked  by means of non-perturbative numerical simulations in lattice gauge 
theories~\cite{Cea:1991,Cea:1993,Cea:1997d,Cea:1998,Cea:1999b}. 
On the other hand, the inevitable presence of the induced
background field needed to overcome the one-loop instability was not appreciated until the remarkable evidence from the
lattice of the color Meissner effect that leads us to reconsider the dynamics of the stabilizing
chromomagnetic background field and finally to unravelling the nature of the confining vacuum. \\
\indent
The main aims of the present paper are twofold. Firstly, we shall extend  to the SU(3) gauge theory in presence of an external constant Abelian
chromomagnetic field directed along the third direction in color space the variational-perturbative  calculations following
our previous papers in SU(2)~\cite{Cea:1987,Cea:1988}.  In Section~\ref{S2}, following Ref.~\cite{Cea:1988}
(henceforth  referred to as I), we consider the SU(3) pure gauge theory in the Schr\"odinger representation and we show explicitly that
in the one-loop approximation there are three different kinds of unstable modes.
 Section~\ref{S3} is devoted to the variational minimization of the ground state energy density.  We will find that the tachyonic condensation
 of the unstable modes lead to stabilizing chromomagnetic background fields with  peculiar kink structures. We will, also, show that
 the resulting ground state wavefunctional is not energetically favoured with respect to perturbative vacuum. 
In Section~\ref{S4} we present a careful analysis of the induced background fields and we show that they allow the presence
of vacuum chromomagnetic charges. \\
In the second part of the present paper we will set up the QCD vacuum wavefunctional and try a punctual comparison with 
hadron phenomenology and lattice data.  In Section~\ref{S5} we present our picture on the structure of the vacuum as a disordered
chromomagnetic condensate.  This is achieved after  taking into account the kink structure of  the background fields dynamically generated
by the unstable mode condensation together with  Feynman's analysis adapted to the SU(3) gauge theory in three spatial dimensions.
In Section~\ref{S6} we show that our  QCD vacuum is characterised by  non-zero gluon condensate,  mass gap and
absence of color long range correlations. We  reach a clear physical picture for the generation of a squeezed flux tube between
static quark pair that allows to determine both the color structure and the transverse profile of the flux-tube chromoelectric fields.
In addition, we show that our proposal for the confining QCD vacuum allows for the color Meissner effect as well as the Meissner effect
recently suggested by numerical simulations of full QCD on the lattice in presence of extreme magnetic fields.
In Section~\ref{S7}  we evaluate the contributions of dynamical quarks to the vacuum energy in the one-loop approximation.  Moreover,
we  suggest that  for massless quarks the dynamically generated background fields could  account for a finite non-zero density of fermion zero
 modes fulfilling, thereby,  the breaking of the chiral  symmetry. Finally, the summary of the main results and some concluding remarks are 
 relegated to   Section~\ref{S8}. \\
It is worthwhile to stress that we are not claiming that we have solved completely the confinement problem in QCD. However, we feel
that the results of the present paper could offer a promising path  towards the complete understanding of confinement in 
quantum chromodynamics.
\section{\normalsize{Constant chromomagnetic background field in SU(3)}}
\label{S2}
In this Section we consider the (3+1)~dimensional pure SU(3) gauge theory. We shall work in the temporal gauge and
follow closely the approach developed  in I.  In the temporal gauge $A^a_0 =0$  the Hamiltonian reads:
\begin{equation}
\label{2.1}
{\cal{H}} = \frac{1}{2} \int d \, \vec{x}  \,\, \left\{ \left( E^a_i(\vec{x}) \right)^2 + \left( B^a_i(\vec{x}) \right)^2 \right\}
\; ,
\end{equation}
where
\begin{equation}
\label{2.2}
B_i^a(\vec{x}) = \frac{1}{2} \epsilon_{ijk} F_{ij}^a (\vec{x}) \; ,
\end{equation}
and
\begin{equation}
\label{2.3}
F_{ij}^a (\vec{x}) = \partial_i A^a_j(\vec{x}) -
\partial_j A^a_i(\vec{x}) +
g f^{abc} A^b_i(\vec{x}) A^c_j(\vec{x}) \;  .
\end{equation}
We shall use the fixed-time Schr\"odinger representation for quantized fields. In fact, the  Schr\"odinger approach to quantum field theories
permits  a direct study of the vacuum structure through the analysis of the vacuum wavefunctional of the theory. The structure of the
vacuum wavefunctional should reflect the nature of color confinement and other non-perturbative features of QCD, more notably the
quark confinement and the chiral symmetry breaking. Actually, the Schr\"odinger approach may allow the study of the vacuum structure through
variational methods implemented by means of trial wavefunctionals. \\
 In the fixed-time  Schr\"odinger representation the chromoelectric field $E^a_i(\vec{x})$ acts as functional derivative:
\begin{equation}
\label{eq:2.4}
E^a_i(\vec{x}) = + i \frac{\delta}{\delta A^a_i(\vec{x}) }
\end{equation}
on the physical states which are functionals obeying the Gauss  law (see, eg, Ref.~\cite{Jackiw:1984}):
\begin{equation}
\label{2.5}
 \left[ \partial_i \delta^{ab} + g f^{acb} A^c_i(\vec{x}) \right] \frac{\delta}{\delta A^b_i(\vec{x}) }
{\cal G}[A] \; = \; 0 \; .
\end{equation}
The effects of an external background field are incorporated by writing
\begin{equation}
\label{2.6}
A^a_i(\vec{x}) =  \bar{A}^a_i(\vec{x}) + \eta^a_i(\vec{x})
\end{equation}
where $\bar{A}^a_i(\vec{x})$ is the background field, and $\eta^a_i(\vec{x})$ the fluctuating field. We are interested in
a constant Abelian chromomagnetic field:
\begin{equation}
\label{2.7}
\bar{A}^a_k(\vec{x}) = \bar{A}_k(\vec{x}) \,  \delta^{a,3} \; , \quad \bar{A}_k(\vec{x}) = \delta_{k,2} \, x_1 H \; \; , \; \; \vec{x} \; = \; (x_1, x_2, x_3) \; .
\end{equation}
It is straightforward to rewrite the Hamiltonian in terms of the fluctuating fields. We get:
\begin{equation}
\label{2.8}
{\cal{H}} = V \; \frac{H^2}{2} \;  + \; {\cal{H}}^{(2)} \; + \; {\cal{H}}^{(3)} \; + \; {\cal{H}}^{(4)} \; ,
\end{equation}
where
\begin{equation}
\label{2.9}
 {\cal{H}}^{(2)} = \frac{1}{2} \int d \, \vec{x}  \,\, \left\{ - \frac{\delta^2}{\delta \eta^a_i(\vec{x}) \delta
\eta^a_i(\vec{x})} + \eta^a_i(\vec{x}) {\cal O}^{ab}_{ij}(\vec{x}) \eta^b_j(\vec{x}) - \left[ D^{ab}_{i}(\vec{x})
\eta^b_i(\vec{x}) \right]^2 \right\}   \; ,
\end{equation}
\begin{equation}
\label{2.10}
{\cal{H}}^{(3)} = \frac{g}{2} \epsilon_{ijk} \epsilon_{ij^{\prime} k^{\prime}} f^{abc} \int d \, \vec{x} \,\,
\eta^b_j(\vec{x}) \eta^c_k(\vec{x}) D^{ad}_{j^{\prime}} (\vec{x}) \eta^d_{k^{\prime}}(\vec{x}) \; ,
\end{equation}
\begin{equation}
\label{2.11}
{\cal{H}}^{(4)} = \frac{g^2}{8} \epsilon_{ijk} \epsilon_{ij^{\prime} k^{\prime}} f^{abc}
f^{ab^{\prime}c^{\prime}} \int d \, \vec{x}  \,\,  \eta^b_j(\vec{x}) \eta^c_k(\vec{x})
\eta^{b^{\prime}}_{j^{\prime}}(\vec{x}) \eta^{c^{\prime}}_{k^{\prime}}(\vec{x}) \; .
\end{equation}
In Equations~(\ref{2.9}) and  (\ref{2.10})  $D^{ab}_i(\vec{x})$ is the covariant derivative with respect to the background field:
\begin{equation}
\label{2.12}
D^{ab}_i(\vec{x}) = \partial_i \delta^{ab} + g  f^{acb} \bar{A}^c_i (\vec{x}) \; ,
\end{equation}
 while ${\cal O}^{ab}_{ij}(\bar{A},\vec{x})$ is the operator
\begin{equation}
\label{2.13}
{\cal O}^{ab}_{ij}(\vec{x}) = - \delta_{ij} D^{ac}_k(\vec{x}) D^{cb}_k(\vec{x}) + 2 g H \epsilon_{3ij} f^{3ab} \; .
\end{equation}
The Gauss constraints, Eq.~(\ref{2.5}), can be rewritten as:
\begin{equation}
\label{2.14}
 \left[ D^{ab}_i(\vec{x}) \;  + \; g f^{acb} \,  \eta^c_i(\vec{x}) \right ] \,  \frac{\delta}{\delta \eta^b_i(\vec{x}) }
{\cal G}[\eta] \; = \; 0 \; .
\end{equation}
As is well known (see, for instance, Ref.~\cite{Jackiw:1984}), the Gauss constraints ensure that the physical states are invariant against
time-independent gauge transformations. It follows, then, that the physical states are not normalizable. 
To overcome this problem one must fix the residual gauge invariance. Following I, we impose the covariant Coulomb constraints:
\begin{equation}
\label{2.15}
D^{ab}_{i}(\vec{x})  \; \eta^b_i(\vec{x}) \; = \; 0 \; .
\end{equation}
Accordingly, the functional measure in the scalar product between two physical states gets modified  by the Faddeev-Popov
determinant associated to the gauge-fixing Eq.~(\ref{2.15}). So that we are left with the following scalar product  between physical states:
\begin{equation}
\label{2.16}
<{\cal G}_1 | {\cal G}_2 >  \; = \;   \int {\cal D} \eta \;  \;  {\cal G}_1^*[ \eta ] \,  {\cal G}_2[ \eta] \; \Delta_{F-P}[\eta] \; 
\delta [ D^{ab}_i(\vec{x}) \eta^b_i(\vec{x}) ] \; \; .
\end{equation}
Long time ago it was suggested~\cite{Savvidy:1977,Matinyan:1978,Pagels:1978} that states with a constant chromomagnetic field
could lie below the perturbative ground state. This was the main motivation that led us to investigate non-perturbatively the 
structure of the vacuum  functional in presence of Abelian chromomagnetic background field, Eq.~(\ref{2.7}).
To this end we need to set up trial vacuum wavefunctionals and, then, to evaluate the expectation value of the Hamiltonian
on these states. This strategy, however, is not easily implementable due to the Gauss constraints on physical states.
Nevertheless, if one assumes that the quantum fluctuations over the background field can be dealt with perturbatively,
then there is a natural strategy to follow. In fact, the lowest order approximation (the one-loop approximation)
amounts to consider the quadratic piece of the Hamiltonian,  Eq.~(\ref{2.8}):  
\begin{equation}
\label{2.17}
{\cal{H}}_0 \;  = \; V \, \frac{ H^2}{2} \;  + \; {\cal{H}}^{(2)}    \; .
\end{equation}
In the same approximation the Gauss constraint reduces to:
\begin{equation}
\label{2.18}
D^{ab}_i(\vec{x}) \, \frac{ \delta {\cal{ G}} [\eta] }{ \delta \eta^b_i(\vec{x} ) } \; = \; 0 \; .
\end{equation}
It is quite easy to show that Eq.~(\ref{2.18}) is satisfied by wavefunctionals that  depend only on transverse fields, i.e. satisfying Eq.~(\ref{2.15}).
To diagonalize ${\cal{H}}_0$, it suffices to solve the eigenvalue equations:
\begin{equation}
\label{2.19}
 {\cal O}^{ab}_{ij}(\vec{x}) \, \phi^b_j(\vec{x}) \; = \; \lambda  \, \phi^a_i(\vec{x})
\end{equation}
with the conditions
\begin{equation}
\label{2.20}
D^{ab}_{i}(\vec{x}) \, \phi^b_i(\vec{x}) \; = \; 0   \; .
\end{equation}
Fortunately, the solutions of Eqs.~(\ref{2.19}) and (\ref{2.20}) have been discussed in details in I. This will greatly simplify the analyses in the
present case. To see this, let us consider Eq.~(\ref{2.19}) assuming that $a, b = 1, 2$. From Eq.~(\ref{2.13})  we get:
\begin{equation}
\label{2.21}
{\cal O}^{ab}_{ij}(\vec{x}) =  \left [ -  \partial^2 \delta^{a b} + 2 gH f^{3ab} x_1 \partial_2 + g^2 H^2 x_1^2 \, f^{3ca} \, f^{3cb}  \right ] \delta_{ij} 
+ 2 g H \epsilon_{3ij} f^{3ab} \; .
\end{equation}
Since for  $a, b = 1, 2$ we have $f^{3ab} = \epsilon^{3ab}$, Eq.~(\ref{2.21}) reduces to the SU(2) operator discussed in I, Appendix B.
As a consequence, we get the following spectrum:
\begin{eqnarray}
\label{2.22}
\lambda(n, p_2, p_3, \alpha = 1) \;  =  \;  p_3^2 \; + \; gH (2 n \, + \, 1)   \; \; , \; \;  n  \; \ge  \;  0 
\nonumber \\
\lambda(n, p_2, p_3, \alpha = 2) \;  =  \;    p_3^2  \; + \; gH (2 n \, + \, 1)  \; \; , \; \; n \; \ge \; 1  
\end{eqnarray}
with transverse eigenvector $\phi^a_j(N_1; \vec{x})$,  $N_1 = (n, p_2, p_3, \alpha )$. We have also the tachyonic modes (u-modes):
\begin{equation}
\label{2.23}
\lambda(N_u) \;  =  \;   p_3^{ 2}  \; -  \;  gH  \; \; , \; \; N_u = ( p_2, p_3 ) \; \;
\end{equation}
with transverse eigenvectors  $\phi^a_j(N_u; \vec{x})$. The explicit construction of the transverse eigenfunctions is discussed  in
details in I, Appendix B.
\\
For  $a, b =  3$ we get: 
\begin{equation}
\label{2.24}
{\cal O}^{ab}_{ij}(\vec{x}) =  -  \partial^2 \delta^{a 3} \delta^{b 3}  \delta_{ij} 
\end{equation}
with eigenvalues:
\begin{equation}
\label{2.25}
\lambda(N_2) \;  =  \;  \vec{p}^{\, 2}  \;  \; , \; \;  \; N_2 = (p_1, p_2, p_3, \alpha ) \; \; , \; \; \alpha \, = \pm \, 1 
\end{equation}
and the eigenstates $\phi^a_j(N_2; \vec{x})$ are transverse plane waves. \\ 
For  $ a, b =  4, 5$ we have that $f^{3ab} = \frac{1}{2} \epsilon^{3ab}$, so that:
\begin{equation}
\label{2.26}
{\cal O}^{ab}_{ij}(\vec{x}) =  \left [ -  \partial^2 \delta^{a b} + 2 gH \frac{ \epsilon^{3ab} }{2} x_1 \partial_2 + \frac{g^2 H^2}{4}  x_1^2 \, \delta^{ab}  \right ] 
\delta_{ij}  + 2 g H \epsilon_{3ij}  \frac{ \epsilon^{3ab} }{2}  \; 
\end{equation}
that coincides with the operator ${\cal O}^{ab}_{ij}(\vec{x})$ in  Eq.~(\ref{2.21}) with $gH$ replaced by $\frac{gH}{2}$. As a consequence
we easily find the spectrum:
\begin{eqnarray}
\label{2.27}
\lambda(n, p_2, p_3, \alpha = 1) \;  =  \; \sqrt{ p_3^2 \; + \; \frac{gH}{2} (2 n \, + \, 1)} \; \; , \; \; n \; \ge \; 0 
\nonumber \\
\lambda(n, p_2, p_3, \alpha = 2) \;  =  \; \sqrt{ p_3^2 \; + \; \frac{gH}{2} (2 n \, + \, 1)} \; \; , \; \; n \; \ge \; 1  
\end{eqnarray}
with transverse eigenvector $\phi^a_j(N_3; \vec{x})$,  $N_3 = (n, p_2, p_3, \alpha )$. Obviously, we have also the  tachyonic v-modes:
\begin{equation}
\label{2.28}
\lambda(N_v) \;  =  \;    p_3^{ 2}  \; -  \;  \frac{gH}{2}   \; \; , \; \; N_v = ( p_2, p_3 ) \; \;
\end{equation}
with transverse eigenvectors  $\phi^a_j(N_v; \vec{x})$. \\
For   $ a, b =  6, 7$,  using $f^{3ab} =  - \frac{1}{2} \epsilon^{3ab}$,  we get:
\begin{equation}
\label{2.29}
{\cal O}^{ab}_{ij}(\vec{x}) =  \left [ -  \partial^2 \delta^{a b} - 2 gH \frac{ \epsilon^{3ab} }{2} x_1 \partial_2 + \frac{g^2 H^2}{4}  x_1^2 \, \delta^{ab}  \right ] 
\delta_{ij}  - 2 g H \epsilon_{3ij}  \frac{ \epsilon^{3ab} }{2}  \; ,
\end{equation}
with spectrum:
\begin{eqnarray}
\label{2.30}
\lambda(n, p_2, p_3, \alpha = 1) \;  =  \; \sqrt{ p_3^2 \; + \; \frac{gH}{2} (2 n \, + \, 1)} \; \; , \; \; n \; \ge \; 0 
\nonumber \\
\lambda(n, p_2, p_3, \alpha = 2) \;  =  \; \sqrt{ p_3^2 \; + \; \frac{gH}{2} (2 n \, + \, 1)} \; \; , \; \; n \; \ge \; 1  
\end{eqnarray}
with transverse eigenvector $\phi^a_j(N_4; \vec{x})$,  $N_4 = (n, p_2, p_3, \alpha )$.   We have also the  tachyonic w-modes:
\begin{equation}
\label{2.31}
\lambda(N_w) \;  =  \;    p_3^{ 2}  \; -  \;  \frac{gH}{2}   \; \; , \; \; N_w = ( p_2, p_3 ) \; \;
\end{equation}
with transverse eigenvectors  $\phi^a_j(N_w; \vec{x})$. \\
Finally, for    $a, b =  8$ we have: 
\begin{equation}
\label{2.32}
{\cal O}^{ab}_{ij}(\vec{x}) =  -  \partial^2 \delta^{a 8}  \delta^{b 8} \delta_{ij} 
\end{equation}
with eigenvalues:
\begin{equation}
\label{2.33}
\lambda(N_5) \;  =  \;  \vec{p}^{\, 2}  \;  \; , \; \;  \; N_5 = (p_1, p_2, p_3, \alpha ) \; \; , \; \; \alpha \, = \pm \, 1 
\end{equation}
and the transverse eigenvectors  $\phi^a_j(N_5; \vec{x})$ correspond to the familiar  plane waves. \\ 
The eigenvectors   $\phi^a_j(N_3; \vec{x})$,   $\phi^a_j(N_4; \vec{x})$ as well as the unstable mode eigenvectors 
 $\phi^a_j(N_v; \vec{x})$,   $\phi^a_j(N_w; \vec{x})$ can be inferred from the results in I, Appendix B after
 replacing $gH$ with $\pm \frac{gH}{2}$. \\
It is, now, straightforward to find the ground state wavefunctional and energy of the quadratic Hamiltonian, Eq.~(\ref{2.17}).
Evidently we can write:
\begin{equation}
\label{2.34} 
{\cal{W}}_0[\eta] = {\cal{V}}_0[\eta_s] \,\, {\cal{Z}}_0[\eta_u]   \; ,
\end{equation}
where the quantum fluctuations have been separated into stable and unstable modes according to:
\begin{equation}
\label{2.35} 
\eta^a_i(\vec{x}) = \eta^a_{si}(\vec{x}) + \eta^a_{ui}(\vec{x})  \; 
\end{equation}
and
\begin{equation}
\label{2.36}
 \eta^a_{si}(\vec{x}) =  \sum_{k=1}^{5} \;  \sum_{N_k} \; c(N_k)  \, \phi^a_i(N_k;\vec{x})   \; ,
\end{equation}
\begin{equation}
\label{2.37}
 \eta^a_{ui}(\vec{x}) = \sum_{\beta= u, v, w} \; \sum_{N_{\beta}} \; c(N_{\beta}) \, \phi^a_i(N_{\beta};\vec{x})   \; .
\end{equation}
Evidently, we have:
\begin{equation}
\label{2.38}
 {{\cal{V}}}_0[\eta_s] = \exp \left \{ -\frac{1}{4}  \int d \vec{x} \, d \vec{y}  \,\, \eta^a_{si}(\vec{x}) \,  \left( G_s
\right)^{ab}_{ij}(\vec{x},\vec{y}) \,  \eta^b_{sj}(\vec{y})  \right\}     \; ,
\end{equation}
with
\begin{equation}
\label{2.39} 
\left( G_s \right)^{ab}_{ij}(\vec{x},\vec{y}) =  \sum_{k=1}^{5} \;  \sum_{N_k} \;  2 \lambda^{\frac{1}{2}}(N_k) \;
\phi^{a*}_i(N_k;\vec{x}) \, \phi^{b}_j(N_k;\vec{y})   \; .
\end{equation}
The ground state energy of the stable modes is:
\begin{eqnarray}
\label{2.40} 
E_S(gH) \;  = \; \frac{1}{2} \; \sum_{k=1}^{5} \;  \sum_{N_k} \;  \lambda^{\frac{1}{2}}(N_k)  \; = \; 
 V \,\left \{  2 \, \int \frac{ d \vec{p}}{(2 \pi)^3}  \sqrt{\vec{p}^{\, 2}} \;      \right . \hspace{4.8 cm}
 \nonumber \\
 \; + \; \frac{gH}{4 \pi^2}  \int_{- \infty}^{+ \infty} d p_3  \; \left [ \;  \sum_{n=0}^{\infty} \sqrt{p_3^2 + gH (2 n + 1)} \; +  \;
  \sum_{n=1}^{\infty} \sqrt{p_3^2 + gH (2 n + 1)} \;  \right ] \; 
  \nonumber \\
 \left . 
+ \,  2 \, \times \,  \frac{gH}{8 \pi^2}  \int_{- \infty}^{+ \infty} d p_3  \; \left [ \;  \sum_{n=0}^{\infty} \sqrt{p_3^2 + \frac{gH}{2} (2 n + 1)} \; +  \;
  \sum_{n=1}^{\infty} \sqrt{p_3^2 + \frac{gH}{2} (2 n + 1)} \; \right ] \;    \right \} . \;
\end{eqnarray}
As concerns the unstable mode sector, since the eigenvalues $\lambda(N_u)$,   $\lambda(N_v)$ and  $\lambda(N_w)$ are not
positive definite, the analogous of Eqs.~(\ref{2.38}) and (\ref{2.39}) for the wavefunctional ${\cal{Z}}_0(\eta_u)$ would led to an
unphysical ground-state wavefunctional. At this point it is necessary to  mention that, at variance with the SU(2) gauge theory where
one deals with only the Nielsen-Olesen unstable modes, in the SU(3) gauge theory we exposed for the first time the presence
of three different kinds of instabilities. Curiously enough, such a circumstance was never mentioned in the literature. On the contrary,
it is widely believed that the unique instabilities in both SU(2) and SU(3) gauge theories are due to the Nielsen-Olesen
unstable modes. In any cases, the origin of the instability is due to the fact that our u, v and w-modes behave like charged
scalar fields with negative squared masses.  In I this led us to assume that the unstable modes were naturally driven to a dynamical
Bose-Einstein condensation very similar to the Higgs mechanism where the condensation gets stabilized by the short-range
repulsive interaction due to the positive quartic self-coupling. However, as discussed at length in I, one must take care of the
gauge invariance assured by the Gauss constraints to obtain physically meaningful results. For completeness, we briefly recap the
strategy we followed in I. Firstly, one must set up a physical basis and, after that, define a perturbative strategy. To overcome
the instabilities we must modify the wavefunctional ${\cal{Z}}_0(\eta_u)$ in Eq.~(\ref{2.34}) by assuming that:
\begin{equation}
\label{2.41}
 {{\cal{Z}}}_0[\eta_u] = \exp \left \{ -\frac{1}{4}  \int d \vec{x} \, d \vec{y}  \,\, \eta^a_{ui}(\vec{x}) \,  \left( G_u
\right)^{ab}_{ij}(\vec{x},\vec{y}) \,  \eta^b_{uj}(\vec{y})  \right\}     \; ,
\end{equation}
with
\begin{equation}
\label{2.42} 
\left( G_u \right)^{ab}_{ij}(\vec{x},\vec{y}) =  \sum_{\beta=u,v,w} \;  \sum_{N_{\beta}} \;  2 \,  \rho(N_{\beta}) \;
\phi^{a*}_i(N_{\beta};\vec{x}) \, \phi^{b}_j(N_{\beta};\vec{y})   \; ,
\end{equation}
where $ \rho(N_{\beta})$ are variational parameters. Note that the resulting wavefunctional  ${\cal{W}}_0[\eta]$ satisfies:
\begin{equation}
\label{2.43}
D^{ab}_i(\vec{x}) \, \frac{\delta{ \cal{W}}_0[\eta]}{\delta \eta^b_i(\vec{x})} \; = \; 0 \; .
\end{equation}
Starting from  ${\cal{W}}_0[\eta]$ one can obtain a basis  $\{ {\cal{W}}_n[\eta] \}$ of wavefunctional satisfying the Gauss constraints Eq.~(\ref{2.43})
by acting on ${\cal{W}}_0[\eta]$ with a suitable creation operator as defined in I, Appendix B. Starting from the orthogonal basis
 $\{ {\cal{W}}_n[\eta] \}$  we can set up a physical basis~\footnote{It is intended that the physical wavefunctionals  $ {\cal{G}}_n[\eta] $ are
 properly normalized, i.e.  $ < {\cal{G}}_n | {\cal{G}}_n > = 1 $.}: 
\begin{equation}
\label{2.44}
 {\cal{G}}_n[\eta] \; = \;        {\cal{W}}_n[\eta] \; \exp \left \{ \Gamma_n[\eta] \right \}  
\end{equation}
satisfying the Gauss law:
\begin{equation}
\label{2.45}
 \left[ D^{ab}_i(\vec{x}) \;  + \; g f^{acb} \,  \eta^c_i(\vec{x}) \right ] \,  \frac{\delta}{\delta \eta^b_i(\vec{x}) } \, 
{{\cal G}}_n[\eta] \; = \; 0 \; .
\end{equation}
In the spirit of considering perturbatively the quantum fluctuations over the background field one can solve Eq.~(\ref{2.45})
by writing:
\begin{equation}
\label{2.46}
 \Gamma_n[\eta]  \; = \; g \,    \Gamma_n^{(1)}[\eta]   \; + \; g^2 \, \Gamma_n^{(2)}[\eta] \; + \; ...
\end{equation}
Indeed, as shown in I, after inserting Eqs.~(\ref{2.44}) and (\ref{2.46}) into Eq.~(\ref{2.45}) one iteratively can determine the functional
$ \Gamma_n^{(1)}[\eta],  \Gamma_n^{(2)}[\eta], ... $. Subsequently, we can implement a perturbative expansion for the ground-state
energy by means of the well-known Brueckner-Goldstone formula (see, eg, Ref.~\cite{Fetter:1971}):
\begin{equation}
\label{2.47}
E  \; = \; < {\cal{G}}_0 | {\cal{H}} | {\cal{G}}_0 >  \; + \; < {\cal{G}}_0 | \,  {\cal{H}}_I \, \sum_{ k \ge 0} \; 
\left [ \frac{ {\cal{H}}_I}{ < {\cal{G}}_0 | {\cal{H}} | {\cal{G}}_0 > \, - {\cal{H}}_0} \right ]^k \;  |  {\cal{G}}_0 >_{connected}
\end{equation}
where:
\begin{equation}
\label{2.48}
  ({\cal{H}}_0)_{n m} \; =  \;  < {\cal{G}}_n | {\cal{H}} | \, {\cal{G}}_n >  \; \delta_{nm} \; ,
\end{equation}
\begin{equation}
\label{2.49}
({\cal{H}}_I)_{n m} \; =  \;  < {\cal{G}}_n | {\cal{H}} | \, {\cal{G}}_m  >  \; ( 1 \; - \; \delta_{nm} )  \; .
\end{equation}
In other words, the unperturbed Hamiltonian is given by the diagonal expectation values of the full Hamiltonian ${\cal{H}}$, while the
perturbations are the off-diagonal elements of   ${\cal{H}}$. If the ground-state wavefunctional  ${\cal{G}}_0[\eta]$ turns out to be close to
the true ground-state wavefunctional, then we have that ${\cal{H}}_I$ is a genuine small perturbation. \\
Since we shall work up to the second perturbative order, Eq.~(\ref{2.47}) reduces to:
\begin{equation}
\label{2.50}
E  \; = \; < {\cal{G}}_0 | {\cal{H}} | {\cal{G}}_0 >  \; + \;  \sum_{ n > 0} \;   \frac{1}{   ({\cal{H}})_{0 0 } \, - \,  ({\cal{H}})_{n n}} \; 
\left | < {\cal{G}}_n | \,  {\cal{H}} | {\cal{G}}_0 >  \right  |^2 \; \; .
\end{equation}
One further problem arises from the circumstance that the physical basis $\{ {\cal{G}}_n[\eta] \}$ is not orthogonal. Actually, one can transform
 the basis   $\{ {\cal{G}}_n[\eta] \}$ into an orthogonal basis through the L\"owdin's transformation~\cite{Lowdin:1950} (see I for further details).
Our aim is to evaluate the ground-state energy in the lowest order, i.e. in the one-loop approximation. However, we already remarked that
the dynamical condensation of the tachyonic modes forced us to consider the quartic self-coupling Hamiltonian ${\cal{H}}^{(4)}$ 
that is of order $g^2$ corresponding to a two-loop contribution to the ground-state energy. As a consequence, to respect the gauge
symmetry we must extend the calculation of the ground-state energy up to the second order in our perturbative scheme with
a variational physical basis. So that, our strategy is to evaluate the ground-state energy by means of Eq.~(\ref{2.50}) and, after
the stabilization of the tachyonic modes, we will retain only the lowest-order terms.
\section{\normalsize{The variational minimization of the vacuum energy }}
\label{S3}
The central issue of this Section is the calculation of the ground-state energy after the stabilization of the tachyonic modes.
In I we showed that the  Nielsen-Olesen modes were stabilized by the dynamical Bose-Einstein condensation leading to a peculiar
non-perturbative background field $\vec{u}^{\, a}(\vec{x})$. In the present case, we expect that the condensation of three different
kinds of tachyonic modes will induce three non-perturbative background fields   $\vec{u}^{\, a}(\vec{x}), a = 1,2$,  
$\vec{v}^{\, a}(\vec{x}), a = 4,5$ and  $\vec{w}^{\, a}(\vec{x}), a = 6,7$. It is convenient to introduce the SU(3) non-perturbative background
field  vector:
\begin{equation}
\label{3.1}
\vec{ U}(\vec{x})  \; = \;  
 \left\{ \begin{array}{ll}
 \;  \vec{u}^a(\vec{x})  \; \; &  a  \; = \; 1, \, 2 \;  
  \\
  \;  \vec{v}^a(\vec{x})  \; \; &  a  \; = \; 4, \, 5 \;  
  \\
  \;  \vec{w}^a(\vec{x})  \; \; &  a  \; = \; 6, \, 7 \;   
 \\
  \;  \; \; \; 0 \;  \; \; &  a  \; = \; 3, \, 8 \;   
\end{array}
    \right .  \;  \; \; .
\end{equation}
We may also introduce the $8 \times 8$ block-diagonal  SU(3) matrix:
\begin{equation}
\label{3.2}
G_{U ij}(\vec{x},\vec{y}) \; = \;
\begin{pmatrix}
 G_{u,ij}^{ab}(\vec{x},\vec{y})   &   &  &  &  \\
        &  0 &  &  &   \\
        &   &  G_{v,ij}^{cd}(\vec{x},\vec{y}) &  &     \\
          &   &  &  G_{w,ij}^{ef}(\vec{x},\vec{y})&     \\
                  &   &  &  &  0
\end{pmatrix}
\end{equation}
with 
\begin{equation}
\label{3.3} 
 G^{ab}_{u,ij}(\vec{x},\vec{y}) \;  =  \;  \sum_{p_2,p_3} \;  2 \,  \rho_u(p_2,p_3) \;
\phi^{a*}_{u i}(\vec{x}) \, \phi^{b}_{u j}(\vec{y})   \;  \;  , \;  \;  a, b \, = \, 1, 2 
\end{equation}
\begin{equation}
\label{3.4} 
 G^{cd}_{v,ij}(\vec{x},\vec{y}) \;   =  \;  \sum_{p_2,p_3} \;  2 \,  \rho_v(p_2,p_3) \;
\phi^{c*}_{v i}(\vec{x}) \, \phi^{d}_{v j}(\vec{y})    \;  \;  , \;  \;  c, d \, = \, 4, 5 
\end{equation}
\begin{equation}
\label{3.5} 
 G^{ef}_{w,ij}(\vec{x},\vec{y}) \;  =  \;  \sum_{p_2,p_3} \;  2 \,  \rho_w(p_2,p_3) \;
\phi^{e*}_{w i}(\vec{x}) \, \phi^{f}_{w j}(\vec{y})    \;  \;  , \;  \;  e, f \, = \, 6, 7 
\end{equation}
where we have explicated the indices $N_u, N_v$ and $N_w$ and the unstable-mode eigenvectors. We may also introduce
the  fluctuating SU(3) vector:
\begin{equation}
\label{3.6}
\vec{\eta}_U(\vec{x})  \; = \;  
 \left\{ \begin{array}{ll}
 \;  \vec{\eta}_u^{\,a}(\vec{x})  \; \; &  a  \; = \; 1, \, 2 \;  
  \\
  \;  \vec{\eta}_v^{\, a}(\vec{x})  \; \; &  a  \; = \; 4, \, 5 \;  
  \\
  \;  \vec{\eta}_w^{\, a}(\vec{x})  \; \; &  a  \; = \; 6, \, 7 \;   
 \\
  \;  \; \; \; 0 \;  \; \; &  a  \; = \; 3, \, 8 \;   
\end{array}
    \right.
\end{equation}
where
\begin{eqnarray}
\label{3.7}
   \vec{\eta}^{\,  a}_u(\vec{x}) = \sum_{p_2,p_3} \;  \; c_u(p_2,p_3)  \, \vec{\phi}^a_u(\vec{x})   \; .
\nonumber \\
  \vec{\eta}^{\, a}_{v}(\vec{x}) = \sum_{p_2,p_3} \;  \; c_v(p_2,p_3)  \, \vec{\phi}^a_v(\vec{x})   \; .
 \nonumber \\
  \vec{\eta}^{\, a}_{w}(\vec{x}) = \sum_{p_2,p_3} \;  \; c_w(p_2,p_3)  \, \vec{\phi}^a_w(\vec{x})   \; . 
\end{eqnarray}
Now, following I, we replace the wavefunctional  ${{\cal{Z}}}_0$, Eq.~(\ref{2.41}), with:
\begin{equation}
\label{3.8}
 {{\cal{Z}}}_0[\eta_U] = \exp \left \{ -\frac{1}{4}  \int d \vec{x} \, d \vec{y}  \,\, \left [ \eta^a_{Ui}(\vec{x}) \, - \,   U^a_{i}(\vec{x}) \right ]
 G_{U ij}^{ab}(\vec{x},\vec{y}) \, \left [ \eta^b_{Uj}(\vec{y}) \, - \,   U^b_{j}(\vec{y}) \right ]
  \right\}     \; .
\end{equation}
The main advantage in using this compact notation is that we can follow step by step the calculations presented in I for the SU(2) gauge theory.
The resulting calculations are rather involved and quite difficult to follow in details, nevertheless they can be borrowed quite easily from I
after replacing the SU(2) structure constants $\epsilon^{abc}$ with the SU(3) structure constants $f^{abc}$.
As a consequence, one finds that, in order to determine configurations which are able to stabilize the tachyonic modes, it is enough to evaluate
only the energy of the unstable sector. So that we have for the ground state energy:
\begin{equation}
\label{3.9}
 E \; =  \;  V \, \frac{H^2}{2} \; + \; E_S \; + \; E_U \; \; ,
\end{equation}
where the first term in the right hand site is the classical energy, the second the one-loop contributions due to the stable modes
given by Eq.~(\ref{2.40}), and finally:
\begin{equation}
\label{3.10}
 E_U \, =  \,  \frac{1}{8}  \int d \vec{x}  \,  G_{U ii}^{aa}(\vec{x},\vec{x}) \, + \, \frac{1}{2} 
 \int d \vec{x}  \; {\cal O}^{ab}_{ij}(\vec{x}) \left [  G_{U ij}^{ -1 \, ab}(\vec{x},\vec{x}) \, + \,  U^a_i(\vec{x})  U^b_j(\vec{x})
  \right  ]   \, + \,   E_U^{(4)}  
\end{equation}
with
\begin{eqnarray}
\label{3.11}
 E_U^{(4)}  \, = \,  \frac{g^2}{8} \epsilon_{ijk} \epsilon_{ij'k'} f^{abc} f^{ab'c'}     \int d \vec{x} 
\Bigg \{ U^b_j(\vec{x})  U^c_k(\vec{x}) U^{b'}_{j'}(\vec{x})  U^{c'}_{k'}(\vec{x}) \; + \;   \hspace{3 cm}
  \nonumber \\
 \bigg [   G_{U jj'}^{ -1 \, bb'}(\vec{x},\vec{x}) U^c_k(\vec{x})  U^{c'}_{k'}(\vec{x})  +          
 G_{U jk'}^{ -1 \, bc'}(\vec{x},\vec{x}) U^c_k(\vec{x})  U^{b'}_{j'}(\vec{x})  +   
   G_{U kj'}^{ -1 \, cb'}(\vec{x},\vec{x}) U^b_j(\vec{x})  U^{c'}_{k'}(\vec{x})  
   \nonumber \\
 \;   + \;   \; \;  permutations \; \;   \bigg ]    \Bigg  \}   \; . \hspace{6 cm}
\end{eqnarray}
From these equations we see that the ground state energy is a functional of $\rho_u(p_2,p_3)$,  $\rho_v(p_2,p_3)$, $\rho_u(p_2,p_3)$
and $\vec{u}^a(\vec{x})$, $\vec{v}^a(\vec{x})$, $\vec{w}^a(\vec{x})$. Our variational procedure amounts to minimize the vacuum energy with
respect to the variational parameters. To this end, we note that:
\begin{equation}
\label{3.12}
   \vec{u}^{\,  a}(\vec{x}) \; = \; \sum_{p_2,p_3} \;  \; b_u(p_2,p_3)  \, \vec{\phi}^a_u(\vec{x})   \; .
\end{equation}
As in I, we may assume that in Eq.~(\ref{3.12})   $b_u(p_2,p_3)$ factorizes. As a consequence we get:
\begin{eqnarray}
\label{3.13}
    u_k^{\pm }(\vec{x}) \; = \;  \frac{1}{\sqrt{2}} \; \bigg ( u_k^{1 }(\vec{x}) \; \pm \; i \,     u_k^{2}(\vec{x}) \bigg ) \; \; \;  \; \; 
\nonumber \\
  u_k^{\pm }(\vec{x}) \; = \;  \frac{f_u(x_3)}{\sqrt{2}} \, \left ( \begin{array}{c}  1 \\ \mp i \\ 0 \end{array} \right) \,
  g^u_{\pm}(x_1,x_2)
\end{eqnarray}
where  $f_u(x_3)$  is a real function and:
\begin{eqnarray}
\label{3.14}
  g^u_{+}(x_1,x_2)  \; = \;   \, \int_{- \infty}^{+ \infty} d p_2  \; b_u(p_2) \,  \frac{\exp ( i p_2 x_2)}{\sqrt{2 \pi}} \bigg ( \frac{gH}{\pi} \bigg )^{\frac{1}{4}}
\exp [ - \frac{gH}{2} (x_1 + \frac{p_2}{gH})^2 ] \; ,
  \nonumber \\
  g^u_{-}(x_1,x_2)  \; = \; \bigg [  g^u_{+}(x_1,x_2)  \bigg ]^*  \; . \hspace{8 cm}
\end{eqnarray}
Likewise, we get:
\begin{eqnarray}
\label{3.15}
    v_k^{\pm }(\vec{x}) \; = \;  \frac{1}{\sqrt{2}} \; \bigg ( v_k^{4 }(\vec{x}) \; \pm \; i \,     v_k^{5}(\vec{x}) \bigg ) \; \; \;  \; \; 
\nonumber \\
  v_k^{\pm }(\vec{x}) \; = \;  \frac{f_v(x_3)}{\sqrt{2}} \, \left ( \begin{array}{c}  1 \\ \mp i \\ 0 \end{array} \right) \,
  g^v_{\pm}(x_1,x_2) \; \; ,
\end{eqnarray}
and
\begin{eqnarray}
\label{3.16}
  g^v_{+}(x_1,x_2)  \; = \;   \, \int_{- \infty}^{+ \infty} d p_2  \; b_v(p_2) \,  \frac{\exp ( i p_2 x_2)}{\sqrt{2 \pi}} \bigg ( \frac{gH}{2 \pi} \bigg )^{\frac{1}{4}}
\exp [ - \frac{gH}{4} (x_1 + \frac{p_2}{\frac{gH}{2}})^2 ] \; ,
  \nonumber \\
  g^v_{-}(x_1,x_2)  \; = \; \bigg [  g^v_{+}(x_1,x_2)  \bigg ]^*  \; . \hspace{8 cm}
\end{eqnarray}
Finally:
\begin{eqnarray}
\label{3.17}
    w_k^{\pm }(\vec{x}) \; = \;  \frac{1}{\sqrt{2}} \; \bigg ( w_k^{6 }(\vec{x}) \; \pm \; i \,     w_k^{7}(\vec{x}) \bigg ) \; \; \;  \; \; 
\nonumber \\
  w_k^{\pm }(\vec{x}) \; = \;  \frac{f_w(x_3)}{\sqrt{2}} \, \left ( \begin{array}{c}  1 \\ \mp i \\ 0 \end{array} \right) \,
  g^w_{\pm}(x_1,x_2) \; \; ,
\end{eqnarray}
and
\begin{eqnarray}
\label{3.18}
  g^w_{+}(x_1,x_2)  \; = \;   \, \int_{- \infty}^{+ \infty} d p_2  \; b_w(p_2) \,  \frac{\exp ( i p_2 x_2)}{\sqrt{2 \pi}} \bigg ( \frac{gH}{2 \pi} \bigg )^{\frac{1}{4}}
\exp [ - \frac{gH}{4} (x_1 - \frac{p_2}{\frac{gH}{2}})^2 ] \; ,
  \nonumber \\
  g^w_{-}(x_1,x_2)  \; = \; \bigg [  g^w_{+}(x_1,x_2)  \bigg ]^*  \; . \hspace{8 cm}
\end{eqnarray}
One can easily check that  a necessary condition for our trial configurations could contribute to the ground-state energy density is:
\begin{equation}
\label{3.19}
\beta = \; u, v, w \; \; \;  \int_{- \frac{L}{2}}^{+ \frac{L}{2}} d x_1 \; d x_2  \; 
  g^{\beta}_{+}(x_1,x_2)  \;   g^{\beta}_{-}(x_1,x_2)  \; \sim \; \; L^2
\end{equation}
where $V = L^3$. Moreover, the minimum of the ground-state energy is attained for $|g^{\beta}_{+}(x_1,x_2)| = constant$ that leads to:
\begin{equation}
\label{3.20}
b_{\beta}(p_2) \; = \; \kappa_{\beta} \; \exp (i L p_2) \; \; \; , \; \; \;  \beta = \; u, v, w   
\end{equation}
for some real constants $\kappa_{\beta}$. Starting from Eqs.~(\ref{3.10}) and (\ref{3.11}), after using  Eqs.~(\ref{3.13})-(\ref{3.18}) and 
with some algebra, we get:
\begin{equation}
\label{3.21}
 E_U  \; =  \;   E_U^{(u)} \; +  \;  E_U^{(v)} \;  + \;   E_U^{(w)} 
\end{equation}
with:
\begin{eqnarray}
\label{3.22}
E_U^{(u)}  \; = \;   \frac{1}{2} \, V \, \frac{gH}{4 \pi^2}  \int_{- \infty}^{+ \infty}  d p_3  \left [ \rho_u(p_3) \;  + \; \frac{p_3^2 - gH}{\rho_u(p_3)} \right ] 
\; + \;  I^u_2 \;  \int_{- \frac{L}{2}}^{+ \frac{L}{2}}  d x_3 \;  f_u(x_3)  [ - \partial^2_3 \; - \; gH ] f_u(x_3) \;
\nonumber \\
 + \;   g^2 \, \frac{gH}{4 \pi^2}  \;  \int_{- \infty}^{+ \infty}  d p_3 \;  \frac{1}{\rho_u(p_3) }    I^u_2   \int_{- \frac{L}{2}}^{+ \frac{L}{2}} d x_3 \; f_u^2(x_3) \; 
\; + \;  \frac{g^2}{2} \, I^u_4 \;   \int_{- \frac{L}{2}}^{+ \frac{L}{2}} d x_3 \;  f_u^4(x_3) \;  , \hspace{3 cm} 
\end{eqnarray}
\begin{eqnarray}
\label{3.23}
E_U^{(v)}  \; = \;   \frac{1}{2} \, V \, \frac{gH}{8 \pi^2}  \int_{- \infty}^{+ \infty}  d p_3  \left [ \rho_u(p_3) \;  + \; \frac{p_3^2 - \frac{gH}{2}}{\rho_u(p_3)} 
\right ] 
\; + \;  I^v_2 \;  \int_{- \frac{L}{2}}^{+ \frac{L}{2}}  d x_3 \;  f_v(x_3)  [ - \partial^2_3 \; - \; \frac{gH}{2} ] f_v(x_3) \;
\nonumber \\
 + \;   g^2 \, \frac{gH}{8 \pi^2}  \;  \int_{- \infty}^{+ \infty}  d p_3 \;  \frac{1}{\rho_v(p_3) }   I^v_2   \int_{- \frac{L}{2}}^{+ \frac{L}{2}} d x_3 \;  f_v^2(x_3)  \; 
\; + \;  \frac{g^2}{2} \, I^v_4 \;   \int_{- \frac{L}{2}}^{+ \frac{L}{2}} d x_3 \;  f_v^4(x_3) \;  , \hspace{3 cm} 
\end{eqnarray}
\begin{eqnarray}
\label{3.24}
E_U^{(w)}  \; = \;   \frac{1}{2} \, V \, \frac{gH}{8 \pi^2}  \int_{- \infty}^{+ \infty}  d p_3  \left [ \rho_u(p_3) \;  + \; \frac{p_3^2 - \frac{gH}{2}}{\rho_w(p_3)} 
\right ] 
\; + \;  I^w_2 \;  \int_{- \frac{L}{2}}^{+ \frac{L}{2}}  d x_3 \;  f_w(x_3)  [ - \partial^2_3 \; - \; \frac{gH}{2} ] f_w(x_3) \;
\nonumber \\
 + \;   g^2 \, \frac{gH}{8 \pi^2}  \;  \int_{- \infty}^{+ \infty}  d p_3 \;  \frac{1}{\rho_w(p_3) }   I^w_2     \int_{- \frac{L}{2}}^{+ \frac{L}{2}} d x_3 \; f_w^2(x_3)\; 
\; + \;  \frac{g^2}{2} \, I^w_4 \;   \int_{- \frac{L}{2}}^{+ \frac{L}{2}} d x_3 \;  f_w^4(x_3) \;  . \hspace{3 cm} 
\end{eqnarray}
In the previous equations we have used the notations:
\begin{equation}
\label{3.25}
\beta = \; u, v, w \; \; \; I^{\beta}_2 \; = \;  \int_{- \frac{L}{2}}^{+ \frac{L}{2}} d x_1 \; d x_2  \; 
  g^{\beta}_{+}(x_1,x_2)  \;   g^{\beta}_{-}(x_1,x_2)  \;
\end{equation}
\begin{equation}
\label{3.26}
\beta = \; u, v, w \; \; \; I^{\beta}_4 \; = \;  \int_{- \frac{L}{2}}^{+ \frac{L}{2}} d x_1 \; d x_2  \; 
 [ g^{\beta}_{+}(x_1,x_2)  \;   g^{\beta}_{-}(x_1,x_2) ]^2  \; .
\end{equation}
Varying the vacuum energy functional with respect to $f_{\beta}(x_3)$ leads to the well-known kink 
equations~\cite{Coleman:1976,Rajaraman:1982,Manton:2004,Weinberg:2012}:
\begin{equation}
\label{3.27}
I^u_2 \, \left [ - \, \partial_3^2 \; - \; gH \right ]  f_u(x_3) \; + \;  g^2 \, I^u_4  \;  f_u^3(x_3)  \;  =  \; 0  \;   \; \; 
\end{equation}
\begin{equation}
\label{3.28}
I^v_2 \, \left [ - \, \partial_3^2 \; - \; \frac{gH}{2} \right ]  f_v(x_3) \; + \;  g^2 \,  I^v_4  \;  f_v^3(x_3)  \;  =  \; 0  \;   \; \; 
\end{equation}
\begin{equation}
\label{3.29}
I^w_2 \, \left [ - \, \partial_3^2 \; - \; \frac{gH}{2} \right ]  f_w(x_3) \; + \;  g^2 \,  I^v_4  \;  f_w^3(x_3)  \;  =  \; 0  \;  .
\end{equation}
Solving these last equations we obtain:
\begin{equation}
\label{3.30}
f_u(x_3) \; = \; \sqrt{\frac{gH}{g^2}}  \;  \sqrt{\frac{I^u_2}{I^u_4}} \; \tanh \bigg [ \sqrt{\frac{gH}{2}} \, (x_3 \, - \, \overline{x}_3) \bigg ] \;,
\end{equation}
\begin{equation}
\label{3.31}
f_v(x_3) \; = \; \sqrt{\frac{gH}{ 2 g^2}}  \;  \sqrt{\frac{I^v_2}{I^v_4}} \; \tanh \bigg [ \sqrt{\frac{gH}{4}} \, (x_3 \, - \, \overline{x}_3) \bigg ] \;,
\end{equation}
\begin{equation}
\label{3.32}
f_w(x_3) \; = \; \sqrt{\frac{gH}{ 2 g^2}}  \;  \sqrt{\frac{I^w_2}{I^w_4}} \; \tanh \bigg [  \sqrt{\frac{gH}{4}} \, (x_3 \, - \, \overline{x}_3) \bigg ] \; .
\end{equation}
After that, using:
\begin{equation}
\label{3.33}
\beta = \; u, v, w \; \;  \; \; \;  L \;  \frac{(I^{\beta}_2)^2}{I^{\beta}_4} \; = \; V \; \; ,
\end{equation}
we obtain:
\begin{eqnarray}
\label{3.34}
 E_U  \; =  \;  - \,   V \, \frac{H^2}{2} \; + \;   \frac{1}{2} \, V \, \frac{gH}{4 \pi^2}  \int_{- \infty}^{+ \infty}  d p_3  \left [ \rho_u(p_3) \;  + \; 
 \frac{p_3^2 + gH}{\rho_u(p_3)} \right ] 
 \nonumber \\
  \;  - \,   V \, \frac{H^2}{8} \; + \;   \frac{1}{2} \, V \, \frac{gH}{8 \pi^2}  \int_{- \infty}^{+ \infty}  d p_3  \left [ \rho_v(p_3) \;  + \; 
 \frac{p_3^2 + \frac{gH}{2}}{\rho_v(p_3)} \right ] 
\nonumber \\
 \; \;  - \,   V \, \frac{H^2}{8} \; +   \frac{1}{2} \, V \, \frac{gH}{8 \pi^2}  \int_{- \infty}^{+ \infty}  d p_3  \left [ \rho_w(p_3) \;  + \; 
 \frac{p_3^2 + \frac{gH}{2}}{\rho_w(p_3)} \right ]  \; .
\end{eqnarray}
Varying with respect to $ \rho_{\beta}(p_3)$ we finally get:
\begin{equation}
\label{3.35}
 \rho_{u}(p_3) \; = \; + \, \sqrt{p_3^2 \, + \, gH} \; \; , \; \;  \rho_{v}(p_3) \; = \;  \rho_{w}(p_3) \; = \; + \,  \sqrt{p_3^2 \, + \, \frac{gH}{2}}    
\end{equation}
leading to:
\begin{eqnarray}
\label{3.36}
 E_U  \; = \;  V \; \Bigg \{ \, - \,  \frac{H^2}{2} \; + \;  \frac{gH}{4 \pi^2}  \int_{- \infty}^{+ \infty}  d p_3 \,   \sqrt{p_3^2 + gH}  \hspace{1.5 cm}
 \nonumber \\
+ \;   2 \; \times \bigg [ \, - \,  \frac{H^2}{8} \; + \;  \frac{gH}{8 \pi^2}  \int_{- \infty}^{+ \infty}  d p_3 \,   \sqrt{p_3^2 + \frac{gH}{2}} \;   \bigg ] 
\; \;   \Bigg \}  \; .
\end{eqnarray}
A few comments are in order. Firstly, Eqs.~(\ref{3.30}), (\ref{3.31}) and (\ref{3.32}) manifest the non-analytic nature of the chromomagnetic kinks.
The origin of this non-analyticity can be understood if we look at the chromomagnetic field:
\begin{equation}
\label{3.37}
B^a_i(\vec{x})  \; = \; \frac{1}{2} \;  \epsilon_{ijk} \; \bigg [ \partial_j \, A^a_k(\vec{x}) \; - \; \partial_k \,  A^a_j(\vec{x}) \; + \; 
g \, f^{abc} \,      A^b_j(\vec{x}) \,  A^c_k(\vec{x}) \, \bigg ] \; . 
\end{equation}
The point is that the dynamical condensation of the tachyonic modes tries to screen as most as possible the constant Abelian background field
$H$ in such a way as to eliminate the instabilities. Now, an uniform classical chromomagnetic field can be realized by the Abelian term in
Eq.~(\ref{3.37}) or by the non-Abelian piece if $A^a_i \, \sim \frac{1}{\sqrt{g}}$. Indeed, in the next Section we will show that our chromomagnetic 
kink solutions lead to an almost complete cancellation of the external chromomagnetic field $H \delta^{a 3}$. On the other hand, the tachyonic
modes are frozen into the lowest Landau levels. As a consequence, these modes behave like (1+1)-dimensional charged scalar fields with negative
mass squared and stabilizing short-range interactions. This leads to the dynamical Bose-Einstein condensations with negative condensation energy
that look like a classical energy term as explicitly displayed by Eq.~(\ref{3.36}). \\
Let us now evaluate the ground state energy, Eq.~(\ref{3.9}), with $E_S$ given by Eq.~(\ref{2.40}) and $E_U$ by  Eq.~(\ref{3.36}). Following I,
where in Sect. VI  the relevant calculations are presented in details, after subtracting the ground state energy of the perturbative vacuum, we
get:
\begin{equation}
\label{3.38}
 \Delta E(gH)  \; = \;  V \; \Bigg \{ \, - \,  \frac{H^2}{4} \; + \;  \frac{(gH)^2}{32 \pi^2} \, \bigg [ \ln ( \frac{\Lambda^2}{gH} ) \; + \;  
 const \, \bigg ]  \;   \Bigg \}  \; 
\end{equation}
where $\Lambda$ is an ultraviolet cutoff and the actual value of the constant does not matter. Note that the presence of only logarithmic divergences
corroborates the gauge-invariance of our calculations. However, as noticed by R. P. Feynman~\cite{Feynman:1988}, the logarithmic divergences in
the ground-state energy are an inevitable consequence of the sensitivity of the variational procedure to the high frequencies. The fact that the 
coefficient of the logarithmic divergent term is positive strongly suggests that the stabilized ground state is not energetically favoured with
respect to the perturbative ground state. To better appreciate this last point, let us introduce a very high-energy scale $\Lambda_H$ defined by:
\begin{equation}
\label{3.39}
 \Delta E(gH = \Lambda_H^2)   \; = \;  0 \; \; .  
\end{equation}
Combining Eq.~(\ref{3.39}) with Eq.~(\ref{3.38}) we have:
\begin{equation}
\label{3.40}
\Delta \,  \varepsilon(gH) \; = \;  \frac{\Delta E(gH)}{V}   \; = \;  \frac{(gH)^2}{16 \pi^2} \, \ln \bigg ( \frac{\Lambda_H}{\sqrt{gH}} \bigg  ) \; 
\; , \; \;  \sqrt{gH} \; < \; \Lambda_H \; \; .  
\end{equation}
The physical meaning of the high-energy scale $\Lambda_H$ is that our variational ground-state wavefunctional is degenerate with the
perturbative vacuum at a very high scale. On the other hand, for large distances $d \gg \frac{1}{\Lambda_H}$, the perturbative vacuum should
be replaced by our variational vacuum wavefunctional. In this way our proposal for the QCD vacuum wavefunctional realizes at very short distances
the Bjorken's femptouniverse picture~\cite{Bjorken:1982}, while at larger distances the perturbative vacuum should be replaced by our
stabilized ground-state wavefunctional. However, there are obvious problems related to the fact that our variational wavefunctional, according
to Eq.~(\ref{3.40}), is not favoured energetically  nor it seems to have the needed ingredients to explain the observed colour and quark confinement
phenomena. Anyway, these last points will be further discussed more carefully in a later Section. 
\section{\normalsize{Vacuum chromodynamics}}
\label{S4}
In the previous Section we have implemented the variational procedure to overcome the one-loop instabilities. As a result we have
constructed the ground state wavefunctional that in the lowest-order approximation can be written as:
\begin{equation}
\label{4.1}
{\cal G}_0[A] =  {\cal N}   \exp \{ - \frac{1}{4} \int   d\vec{x} \, d\vec{y} \, 
 [ A^a_i(\vec{x}) - \bar{A}^a_i(\vec{x}) - U^a_i(\vec{x})  ] \,
G^{a b}_{i j} (\vec{x},\vec{y})  [ A^a_i(\vec{y}) - \bar{A}^a_i(\vec{y}) - U^a_i(\vec{y}) ] \} \, .
\end{equation}
where:
\begin{equation}
\label{4.2} 
 G^{a b}_{i j} (\vec{x},\vec{y})   =  
 \sum_{k=1}^{5}   \sum_{N_k} \;  2 \lambda^{\frac{1}{2}}(N_k) 
\phi^{a*}_i(N_k;\vec{x}) \, \phi^{b}_j(N_k;\vec{y})   
 +   \sum_{\beta = u,v,w} \sum_{p_2,p_3}  2 \,  \rho_{\beta}(p_2,p_3) 
\phi^{a*}_{\beta i}(\vec{x}) \, \phi^{b}_{\beta j}(\vec{y})    
\end{equation}
with $\rho_{\beta}(p_2,p_3)$ given by Eq.~(\ref{3.35}). In these approximations the scalar product of physical states is simply:
\begin{equation}
\label{4.3}
<{\cal G}_0 | {\cal G}_0 >  \; = \;   \int {\cal D} A \;  \;  {\cal G}_0^*[ A] \,  {\cal G}_0[ A] \;  
\delta [ D^{ab}_i(\vec{x}) \eta^b_i(\vec{x}) ] \; 
\end{equation}
where the Faddeev-Popov determinant has been englobed in the normalization constant. Presently, we are interested in evaluating the expectation
value of the field strength tensor $F^a_{\mu \nu}(x)$ on the ground-state wavefunctional, Eq.~(\ref{4.1}).
Evidently we have:
\begin{equation}
\label{4.4}
<{\cal G}_0 |\,  E^a_i(\vec{x}) \, | {\cal G}_0 >  \; = \;  <{\cal G}_0 |  \, i \, \frac{\delta}{\delta \, A^a_i(\vec{x})} \, | {\cal G}_0 >  \;  = \;  0 \; \; .
\end{equation}
On the other hand, it is easy to see that the chromomagnetic fields:
\begin{equation}
\label{4.5}
  B^a_i(\vec{x})  \; = \;  <{\cal G}_0 | \, \frac{1}{2} \, \epsilon_{ijk}  [ \partial_j \, A^a_k(\vec{x}) \, - \, \partial_k \,  A^a_j(\vec{x})  +  
g \, f^{abc} \,      A^b_j(\vec{x}) \,  A^c_k(\vec{x})  ] \,  | {\cal G}_0 > 
\end{equation}
have a non-zero  expectation value. In fact, one obtains:
\begin{equation}
\label{4.6}
 B^a_i(\vec{x})   =  \frac{1}{2} \, \epsilon_{ijk}  \bigg \{  \partial_j [ \bar{A}^a_k(\vec{x}) \, + U^a_k(\vec{x}) ]
   - \, \partial_k \, [ \bar{A}^a_j(\vec{x}) \, + U^a_j(\vec{x}) ]  +  
g \, f^{abc} \,   [ \bar{A}^a_k(\vec{x}) \, + U^a_k(\vec{x}) ] [ \bar{A}^a_k(\vec{x}) \, + U^a_k(\vec{x}) ] \bigg \} \, .
\end{equation}
Before addressing into the concrete calculations of the ground-state chromomagnetic field, we need to better characterize the peculiar
structure of the background fields generated by the dynamical condensation of the tachyonic modes. As we have shown before, the background
field  $\vec{U}^{\, a}(\vec{x})$ is composed by three background fields  $\vec{u}^{\, a}(\vec{x})$,  $\vec{v}^{\, a}(\vec{x})$ and 
$\vec{w}^{\, a}(\vec{x})$ corresponding to the condensation of  the three different unstable modes, respectively. Since the structure of these last background fields is quite similar, for definitiveness we shall focus on $\vec{u}^{\, a}(\vec{x})$. According to the previous Section we need to consider
only the background field  $\vec{u}^{\, +}(\vec{x})$, Eq.~(\ref{3.13}):
\begin{equation}
\label{4.7}
u_k^{+ }(\vec{x}) \; = \;  \frac{f_u(x_3)}{\sqrt{2}} \, \left ( \begin{array}{c}  1 \\ - i \\ 0 \end{array} \right) \,
  g^u_{+}(x_1,x_2)
\end{equation}
with:
\begin{equation}
\label{4.8}
  g^u_{+}(x_1,x_2)  \; = \;   \, \int_{- \infty}^{+ \infty} d p_2  \; b_u(p_2) \,  \frac{\exp ( i p_2 x_2)}{\sqrt{2 \pi}} \bigg ( \frac{gH}{\pi} \bigg )^{\frac{1}{4}}
\exp [ - \frac{gH}{2} (x_1 + \frac{p_2}{gH})^2 ] \; .
\end{equation}
Now, taking into account Eq.~(\ref{3.20}) we can write:
\begin{equation}
\label{4.9}
  g^u_{+}(x_1,x_2)   =   \int_{- \infty}^{+ \infty} d p_2  \; \kappa_u  \, \exp (i L p_2) \,   \frac{\exp ( i p_2 x_2)}{\sqrt{2 \pi}} 
  \bigg ( \frac{gH}{\pi} \bigg )^{\frac{1}{4}} \exp [ - \frac{gH}{2} (x_1 + \frac{p_2}{gH})^2 ] \; .
\end{equation}
Evidently we have:
\begin{eqnarray}
\label{4.10}
  g^u_{+}(x_1,x_2)    g^u_{-}(x_1,x_2)  =   \int_{- \infty}^{+ \infty} d p_2  \, d p'_2  \; \kappa_u^2   \;  \exp [i L (p_2 - p'_2)]  \; 
   \frac{\exp [ i (p_2 - p'_2) x_2]}{2 \pi} 
   \nonumber \\
  \bigg ( \frac{gH}{\pi} \bigg )^{\frac{1}{2}} \;  \exp [ - \frac{gH}{2} (x_1 + \frac{p_2}{gH})^2 ] \;  \exp [ - \frac{gH}{2} (x_1 + \frac{p'_2}{gH})^2 ]  \; .
\end{eqnarray}
The function within the integrals is absolutely summable, so that by the Riemann-Lebesgue lemma in the thermodynamical limit $L \rightarrow \infty$
survives only the term with $p_2 = p'_2$. so that we are left with:
\begin{equation}
\label{4.11}
  g^u_{+}(x_1,x_2)    g^u_{-}(x_1,x_2)  =   \int_{- \infty}^{+ \infty} d p_2  \; \frac{\kappa_u^2}{2 \pi}    \;
  \bigg ( \frac{gH}{\pi} \bigg )^{\frac{1}{2}} \;  \exp [ - gH  (x_1 + \frac{p_2}{gH})^2 ] \;  =  \;  \kappa_u^2 \;  \frac{gH}{2 \pi}  \; .
\end{equation}
Since $g^u_- = g^u_+*$, we can write:
\begin{equation}
\label{4.12}
  g^u_{\pm}(x_1,x_2)  \; = \;    | g^u_{+}(x_1,x_2) | \; \exp [ \pm  \, i \,  \delta^u(x_1,x_2) ]   \;  = \; 
  \kappa_u \; \sqrt{ \frac{gH}{2 \pi}} \,  \exp [ \pm  \, i \,  \delta^u(x_1,x_2) ]   \; .
\end{equation}
Moreover, the same arguments lead to the conclusion that in the thermodynamical limit  $g^u_{\pm}(x_1,x_2)  \rightarrow 0$. So that we must admit
that the phase   $\delta^u(x_1,x_2)$ is rapidly varying so that  $\exp [ \pm  \, i \,  \delta^u(x_1,x_2) ]$ average to zero, while
 $| g^u_{+}(x_1,x_2) |$ stays finite. This last result is not so strange for, as we have repeatedly stressed, the induced background fields
 are generated by the dynamical condensation of the tachyonic modes by quantum fluctuations. Nevertheless, these evanescent background
 fields are able to give a non-zero contributions to physical observables that are not sensitive to the quantum phases $\delta^u$, $\delta^v$
 and $\delta^w$. \\
 Let us, now, turn on the calculation of the chromomagnetic field, Eq.~(\ref{4.6}). Evidently, we have:
\begin{equation}
\label{4.13}
 B^3_i(\vec{x}) \;   = \;  \epsilon_{ijk} \;  \partial_j   \bar{A}^a_k(\vec{x}) \;  + \; 
\frac{g}{2}  \, \epsilon_{ijk} \,  f^{3bc} \,  U^b_j(\vec{x}) \, U^c_k(\vec{x}) \; ,
\end{equation}
or
\begin{eqnarray}
\label{4.14}
 B^3_i(\vec{x}) \;   = \; \delta_{i 3} \, H \; + \;
   \frac{g}{2}  \, \epsilon_{ijk} \,  f^{312} \,  [ u^1_j(\vec{x}) \, u^2_k(\vec{x})  \;  - \;  u^2_j(\vec{x}) \, u^1_k(\vec{x}) ] 
   \hspace{5.1 cm}
\nonumber \\
+ \;   \frac{g}{2}  \, \epsilon_{ijk} \,  f^{345} \,  [ v^4_j(\vec{x}) \, v^5_k(\vec{x})  \;  - \;  v^5_j(\vec{x}) \, v^4_k(\vec{x}) ]  \; 
+ \;  \frac{g}{2}  \, \epsilon_{ijk} \,  f^{367} \,  [ w^6_j(\vec{x}) \, w^7_k(\vec{x})  \;  - \;  w^7_j(\vec{x}) \, w^6_k(\vec{x}) ] \; . \; \;  \; \;  \; 
\end{eqnarray}
Now a straightforward calculation shows that:
\begin{equation}
\label{4.15}
   \frac{g}{2}  \, \epsilon_{ijk} \,  f^{312} \,  [ u^1_j(\vec{x}) \, u^2_k(\vec{x})  \;  - \;  u^2_j(\vec{x}) \, u^1_k(\vec{x}) ]  \; = \; 
   \delta_{i 3} \,  i \, g \;  [ u^+_1(\vec{x}) \, u^-_2(\vec{x})  \;  - \;  u^-_1(\vec{x}) \, u^+_2(\vec{x}) ]   \;  \; ,
\end{equation}
that leads to:
\begin{equation}
\label{4.16}
   \frac{g}{2}  \, \epsilon_{ijk} \,  f^{312} \,  [ u^1_j(\vec{x}) \, u^2_k(\vec{x})  \;  - \;  u^2_j(\vec{x}) \, u^1_k(\vec{x}) ]  \; = \; 
   - \, \delta_{i 3} \,  g \;  f^2_u(x_3) \,   g^u_{+}(x_1,x_2)    g^u_{-}(x_1,x_2)    \;  \; .
\end{equation}
Further, considering that $f^{345} = - f^{367} = \frac{1}{2}$ one can check that the contributions due to   $\vec{v}^{\, a}(\vec{x})$ and 
$\vec{w}^{\, a}(\vec{x})$ cancel out. So that we are left with:
\begin{equation}
\label{4.17}
 B^3_i(\vec{x}) \;   = \;   \delta_{i 3}  \, [ H \; -  g \,  f^2_u(x_3) \,   g^u_{+}(x_1,x_2)    g^u_{-}(x_1,x_2) ] \; .
\end{equation}
Finally, after using Eqs.~(\ref{4.11}) and (\ref{3.30}) we obtain:
\begin{equation}
\label{4.18}
 B^3_i(\vec{x}) \;   = \;   \delta_{i 3}  \,  H \; \Bigg \{ 1 \,  - \,   \tanh^2  \bigg [ \sqrt{\frac{gH}{2}} \, (x_3 \, - \, \overline{x}_3) \bigg ]   \Bigg \} \; \;  
\end{equation}
or better:
\begin{equation}
\label{4.19}
 B^3_i(\vec{x}) \;   = \;   \delta_{i 3} \; \frac{ H}{  \cosh^2  \bigg [ \sqrt{\frac{gH}{2}} \, (x_3 \, - \, \overline{x}_3) \bigg ] } \; \; .
\end{equation}
These last equations clearly show that the induced background field  $\vec{u}^{\, a}(\vec{x})$ screens almost completely the external
Abelian chromomagnetic field $H \delta^{a 3}$ such that the resulting vacuum chromomagnetic field is localized on the kink-plane
$x_3 = \overline{x}_3$. Interestingly enough, the screening of the external Abelian chromomagnetic field resemble closely what happens in
type II superconductors where an external magnetic field is screened into narrow flux tubes by means of the creation of Abrikosov
vortices. \\
With similar calculations one can evaluate the other components of the chromomagnetic field. Here we merely display the final results:
\begin{equation}
\label{4.20}
 B^1_i(\vec{x}) \;   = \;  - \,  \delta_{i 3}  \,  \frac{H}{2}  \;  \tanh^2  \bigg [ \sqrt{\frac{gH}{4}} \, (x_3 \, - \, \overline{x}_3) \bigg ]  \;
 \exp [ - \frac{gH}{2} \, x_1^2 ] \; \; ,  
\end{equation}
\begin{equation}
\label{4.21}
 B^2_i(\vec{x}) \;   = \;   0  \; \; ,  
\end{equation}
\begin{equation}
\label{4.22}
 B^4_i(\vec{x}) \;   = \;   \delta_{i 3}  \,  \frac{H}{\sqrt{3}}  \;  \tanh \bigg [ \sqrt{\frac{gH}{2}} \, (x_3 \, - \, \overline{x}_3) \bigg ]  \;
 \tanh \bigg [ \sqrt{\frac{gH}{4}} \, (x_3 \, - \, \overline{x}_3) \bigg ]  \;
  \exp [ - \frac{3}{4} \, gH \,  x_1^2 ] \; \; ,  
\end{equation}
\begin{equation}
\label{4.23}
 B^5_i(\vec{x}) \;   = \;   0  \; \; ,  
\end{equation}
\begin{equation}
\label{4.24}
 B^6_i(\vec{x}) \;   = \;   \delta_{i 3}  \,  \frac{H}{\sqrt{3}}  \;  \tanh \bigg [ \sqrt{\frac{gH}{2}} \, (x_3 \, - \, \overline{x}_3) \bigg ]  \;
 \tanh \bigg [ \sqrt{\frac{gH}{4}} \, (x_3 \, - \, \overline{x}_3) \bigg ]  \;
  \exp [ - \frac{gH}{2} \,  x_1^2 ] \; \; ,  
\end{equation}
\begin{equation}
\label{4.25}
 B^7_i(\vec{x}) \;   = \;   0  \; \; ,  
\end{equation}
\begin{equation}
\label{4.26}
 B^8_i(\vec{x}) \;   = \;  - \,  \delta_{i 3}  \,  \frac{\sqrt{3}}{2}  \, H \;  \tanh^2  \bigg [ \sqrt{\frac{gH}{4}} \, (x_3 \, - \, \overline{x}_3) \bigg ]  \; \;  . 
\end{equation}
The last point we would like to discuss in the present Section is that our stabilized ground-state wavefunctional implies the presence of
chromomagnetic charges according to:
\begin{equation}
\label{4.27}
D^{a b}_i(\vec{x}) \,  B^b_i(\vec{x}) \;   =  \;  \rho^a_M(\vec{x}) \; \; .
\end{equation}
Combining Eq.~(\ref{4.27}) with Eqs.~(\ref{4.19}) - (\ref{4.26}) we obtain:
\begin{equation}
\label{4.28}
 \rho^1_M(\vec{x}) \;   = \;  -   \, H \,  \frac{\sqrt{gH}}{2}  \; \frac{ \tanh \bigg [ \sqrt{\frac{gH}{4}} \, (x_3 \, - \, \overline{x}_3) \bigg ] }{
 \cosh^2  \bigg [ \sqrt{\frac{gH}{4}} \, (x_3 \, - \, \overline{x}_3) \bigg ] } \;  \exp [ - \frac{gH}{2} \, x_1^2 ] \; \;  ,  
\end{equation}
\begin{equation}
\label{4.29}
 \rho^3_M(\vec{x}) \;   = \;  - \sqrt{2}  \, H \,  \sqrt{gH}  \; \frac{ \tanh \bigg [ \sqrt{\frac{gH}{2}} \, (x_3 \, - \, \overline{x}_3) \bigg ] }{
 \cosh^2  \bigg [ \sqrt{\frac{gH}{2}} \, (x_3 \, - \, \overline{x}_3) \bigg ] }  \; \;  ,  
\end{equation}
\begin{equation}
\label{4.30}
 \rho^4_M(\vec{x})    =   \frac{H}{\sqrt{3}}  \,  \frac{\sqrt{gH}}{2}  \; \Bigg \{ \frac{ \tanh \bigg [ \sqrt{\frac{gH}{4}} \, (x_3 \, - \, \overline{x}_3) \bigg ] }{
 \cosh^2  \bigg [ \sqrt{\frac{gH}{2}} \, (x_3 \, - \, \overline{x}_3) \bigg ] } \,   + \,   
 \frac{ \tanh \bigg [ \sqrt{\frac{gH}{2}} \, (x_3 \, - \, \overline{x}_3) \bigg ] }{
 \cosh^2  \bigg [ \sqrt{\frac{gH}{4}} \, (x_3 \, - \, \overline{x}_3) \bigg ] } 
 \Bigg \}  
  \exp [ - \frac{ 3 gH}{4} \, x_1^2 ] \;  ,  
\end{equation}
\begin{equation}
\label{4.31}
 \rho^6_M(\vec{x})    =   \frac{H}{\sqrt{3}}  \,  \frac{\sqrt{gH}}{2}  \; \Bigg \{  \sqrt{2} \frac{ \tanh \bigg [ \sqrt{\frac{gH}{4}} \, (x_3 \, - \, \overline{x}_3) \bigg ] }{
 \cosh^2  \bigg [ \sqrt{\frac{gH}{2}} \, (x_3 \, - \, \overline{x}_3) \bigg ] } \,   + \,   
 \frac{ \tanh \bigg [ \sqrt{\frac{gH}{2}} \, (x_3 \, - \, \overline{x}_3) \bigg ] }{
 \cosh^2  \bigg [ \sqrt{\frac{gH}{4}} \, (x_3 \, - \, \overline{x}_3) \bigg ] } 
 \Bigg \}  
  \exp [ - \frac{ gH}{2} \, x_1^2 ] \;  ,  
\end{equation}
\begin{equation}
\label{4.32}
 \rho^8_M(\vec{x}) \;   = \;  -  \; \frac{\sqrt{3}}{2}   \, H \,  \sqrt{gH}  \; \frac{ \tanh \bigg [ \sqrt{\frac{gH}{4}} \, (x_3 \, - \, \overline{x}_3) \bigg ] }{
 \cosh^2  \bigg [ \sqrt{\frac{gH}{4}} \, (x_3 \, - \, \overline{x}_3) \bigg ] }  \; \;  ,  
\end{equation}
\begin{equation}
\label{4.33}
 \rho^2_M(\vec{x}) \;   = \;  \rho^5_M(\vec{x}) \;   =   \;    \rho^7_M(\vec{x})  \;   =  \;  0  \;    \; .
\end{equation}
Later on we shall see that these chromomagnetic  charges will play a fundamental role in the formation of the chromoelectric flux tube generated
by static colour sources.
\section{\normalsize{The QCD vacuum wavefunctional }}
\label{S5}
In the previous Sections we have been able to implement the stabilization of the one-loop instabilities. Accordingly, we have seen that to the lowest order
the trial wavefunctional can be written as:
\begin{equation}
\label{5.1}
{\cal G}_0[{\tilde{\eta}}] \; = \;   {\cal N} \;  \exp{ \left [  - \frac{1}{4} \int d\vec{x} \, d\vec{y} \;  {\tilde{\eta}}^a_i(\vec{x}) \,
G^{a b}_{i j} (\vec{x},\vec{y})  \, {\tilde{\eta}}^b_j(\vec{y})  \right ]}
\end{equation}
where:
\begin{equation}
\label{5.2}
{\tilde{\eta}}^a_i(\vec{x})  \; = \;   A^a_i(\vec{x}) \; - \;   \bar{A}^a_i(\vec{x})  \; -  \;  U^a_i(\vec{x}) \; 
\end{equation}
and $G^{a b}_{i j} (\vec{x},\vec{y}) $ given by Eq.~(\ref{4.2}). Moreover,  the scalar product between physical states reduces to:
\begin{equation}
\label{5.3}
<{\cal G} | {\cal F} >  \; = \;   \int {\cal D} {\tilde{\eta}} \;  \;  {\cal G^*}[ {\tilde{\eta}}] \,  {\cal F}[ {\tilde{\eta}}]
\end{equation}
where the functional integrations are restricted to transverse gauge field fluctuations and the harmless Faddeev-Popov determinant
has been englobed in the normalization constant. Our aim is to set up a wavefunctional for the QCD vacuum. Evidently, the trail wavefunctional
Eq.~(\ref{5.1}) is not good enough. Firstly, we showed in the preceding Section that the chromomagnetic fields have a non-zero value on this wavefunctional. More importantly, the vacuum energy associated to ${\cal G}_0$ is greater than the perturbative vacuum energy. Since the 
proposal that the states with a constant chromomagnetic field could lower the vacuum energy~\cite{Savvidy:1977,Matinyan:1978,Pagels:1978},
there is a widespread conviction in the literature that, even after stabilization of the tachyonic unstable modes, these states are energetically
favoured. On the contrary, we have shown that a full quantum-mechanical treatment of the tachyonic modes taking into account the severe
constraints due to the gauge symmetry leads to a stabilized state that increases the vacuum energy for both SU(2) and SU(3) gauge theories. \\
Let us look closely to the structure of the induced background field $U^a_i(\vec{x})$. The dynamical condensation of the unstable modes lead to
the non-perturbative background fields $U^a_i(\vec{x})$ characterized by the kink-profile functions $f_u(x_3)$, $f_v(x_3)$ and  $f_w(x_3)$.
These last functions vanish at the kink plane  $x_3 = \overline{x}_3$. 
Evidently, for   $x_3 = \overline{x}_3$ the background field is given by the Abelian background field:
\begin{equation}
\label{5.4}
\bar{A}^a_i(x_1,x_2,\overline{x}_3) \; = \;  \delta^{a,3} \,  \delta_{i,2} \, x_1 H  \;  .
\end{equation}
Therefore,  on the kink plane we are in the same situation as in the seminal Feynman's paper on the qualitative
behaviour of Yang-Mills theory in (2+1)-dimensions~\cite{Feynman:1981}. We, now, reproduce  in SU(3) the gauge transformations used 
by Feynman, but in SU(2). We write for a SU(3) unitary matrix:
\begin{equation}
\label{5.5}
\Lambda(\vec{x})  \; = \;  \exp [ -i g \,  \theta^a(\vec{x}) \frac{\lambda^a}{2} ]  \;  ,
\end{equation}
where  $\lambda^a$ are the Gell-Mann matrices. We further set:
\begin{equation}
\label{5.6}
\theta^a(x_1, x_2) \;  = \;   \theta^a(x_1, x_2, \overline{x}_3) \; ,
\end{equation}
and introduce:
\begin{equation}
\label{5.7}
\bar{A}_i(x_1,x_2) \; = \; \frac{\lambda^a}{2} \,  \bar{A}_i^a(x_1,x_2)  \; = \;  \, \frac{\lambda^3}{2}\,  \delta_{i,2} \, x_1 H  \;  .
\end{equation}
Under a planar gauge transformation we can write:
\begin{equation}
\label{5.8}
g \bar{A}'_i (x_1,x_2) \; = \;  \Lambda(x_1,x_2) \,  g \bar{A}_i (x_1,x_2) \,  \Lambda^{-1}(x_1,x_2) \; - \, i \,
\partial_i  \Lambda(x_1,x_2)  \,   \Lambda^{-1}(x_1,x_2) \;  .
\end{equation}
Now, we  divide the $x_3 = \overline{x}_3$ plane into $L_D \times L_D$ square domains,  with $L_D \ll L$. Further, we label  the domains with an integer
$m$ as follows:
\begin{equation}
\label{5.9}
x_1 \; = \;  \frac{L_D}{2 \pi} \, \xi  \; \; ,  \; \;  x_2 \; = \;  \frac{L_D}{2 \pi} \, \eta \; \; ,
\end{equation}
\begin{eqnarray}
\label{5.10}
\nonumber
m = 0 \; \;  \; \; \; \; \;  \; \; \;  \; \; 0 \le x_1 \le L_D \; \; \; \; \; \; \; \; 0 \; \le \xi \; \le 2 \pi   \\
\nonumber  
m = 1 \; \;  \;  \; \; \;  L_D  \le x_1 \le 2 \, L_D \;    \; \;   \; \; \; 2 \, \pi  \; \le \xi \; \le 4 \pi  
 \\
 . \;  .  \;  .  \; .  \hspace{7 cm}
\end{eqnarray}
So that we have:
\begin{equation}
\label{5.11}
 \xi  \; =   \;  2 \, \pi \, m \; + \; \beta \; \; , \; \;  0 \; \le \; \beta
 \; \le 2 \, \pi \; . 
\end{equation}
We do not need the residual of $x_2$ since only the integer $m$ matter here. \\
Writing:
\begin{equation}
\label{5.12}
\Lambda(x_1,x_2)  \; = \;  \Lambda_1(x_1,x_2) \, \Lambda_2(x_1,x_2)  \;  ,
\end{equation}
with:
\begin{equation}
\label{5.13}
\Lambda_1(x_1,x_2)  \; = \;    \exp \{ - i  \, \frac{\eta}{2} \, [  \cos (\frac{\beta}{2})  \, \lambda^3  \; + \;  
 \sin (\frac{\beta}{2})  \, \lambda^2  ] \}   \;  ,
\end{equation}
\begin{equation}
\label{5.14}
\Lambda_2(x_1,x_2)  \; =  \;  \exp \{ - i  \, \frac{\ell_D^2}{2 \pi} \, \eta \,  ( m \, + \, \frac{1}{2} ) \, \lambda^3  \}  \; ,
\end{equation}
$\ell_D$ being the dimensionless domain linear size, i.e. $L_D = \ell_D \, a_H$ with the magnetic length $a_H \simeq \frac{1}{\sqrt{gH}}$, 
it is easy to check that:
\begin{equation}
\label{5.15}
\Lambda_1(x_1,x_2)  \; =  \;  \cos (\frac{\eta}{2})  \; - \; i \,  \sin (\frac{\eta}{2}) \, [ \cos (\frac{\beta}{2}) \, \lambda^3 \; + \;  
 \sin (\frac{\beta}{2}) \, \lambda^2 ]  \;  
\end{equation}
and
\begin{equation}
\label{5.16}
\Lambda_2(x_1,x_2)  \; =  \;   \cos [ (m \, + \, \frac{1}{2}) \, \frac{\ell_D^2}{2 \pi} \, \eta ] \; - \; i \, \, \lambda^3
  \sin [ (m \, + \, \frac{1}{2}) \, \frac{\ell_D^2}{2 \pi} \, \eta ]  \;   .
\end{equation}
The gauge transformed vector potential can be easily evaluated. We have:
\begin{equation}
\label{5.17}
g \bar{A}'_2 (x_1,x_2) \; = \; gH \, x_1 \;  \Lambda(x_1,x_2) \, \frac{\lambda^3}{2}  \,  \Lambda^{-1}(x_1,x_2) \; - \, i \,
\partial_2 \Lambda(x_1,x_2)  \,   \Lambda^{-1}(x_1,x_2) \;  .
\end{equation}
Now, a straightforward calculation leads to:
\begin{equation}
\label{5.18}
 gH \, x_1 \;  \Lambda(x_1,x_2) \, \frac{\lambda^3}{2}  \,  \Lambda^{-1}(x_1,x_2) \; = \;
 \;  \frac{L_D}{2 \pi} \, gH \, (2 \, \pi \, m \; + \; \beta)  \;   \Lambda_1(x_1,x_2) \, \frac{\lambda^3}{2}  \,  \Lambda_1^{-1}(x_1,x_2) \; ,
\end{equation}
\begin{equation}
\label{5.19}
 - \, i \, \partial_2 \Lambda(x_1,x_2)  \,   \Lambda^{-1}(x_1,x_2) \;  = \;
 - \, i \,  \frac{2 \pi}{L_D} \,  \frac{\partial}{\partial \eta}  \Lambda_1  \,   \Lambda_1^{-1} \; - \;
 \Lambda_1 \, [  \frac{\ell_D^2}{L_D} \, (m\, + \, \frac{1}{2})  \frac{\lambda^3}{2} ]  \, \Lambda_1^{-1}  \; .
\end{equation}
Combining Eqs.~(\ref{5.18}) and (\ref{5.19}) we see that the terms depending on $m$ cancel out. Thus, we are left with:
\begin{equation}
\label{5.20}
g \bar{A}'_2 (x_1,x_2) \, = \, \Lambda_1\, \left \{ \left [\sqrt{gH} \ell_D (\frac{\beta}{2 \pi} + \frac{1}{2}) -
 \frac{4 \pi}{\ell_D} \sqrt{gH} \cos \frac{\beta}{2} \right ] \,  \frac{\lambda^3}{2}  \,  - \, 
\frac{4 \pi}{\ell_D} \sqrt{gH} \sin \frac{\beta}{2} \,  \frac{\lambda^2}{2}  
 \right \}  \Lambda_1^{-1} \;  .
\end{equation}
This last quantity varies periodically but remains of the order $\sqrt{gH}$ over the entire area $L^2$. Moreover, we have:
\begin{equation}
\label{5.21}
g \bar{A}'_1 (x_1,x_2) \; = \;  - \;  i \, \partial_1 \Lambda(x_1,x_2)  \,   \Lambda^{-1}(x_1,x_2) \;  .
\end{equation}
After some algebra we find:
\begin{equation}
\label{5.22}
g \bar{A}'_1(x_1,x_2)  =  -  \frac{2 \pi}{\ell_D} \, \sqrt{gH} \,
 \left \{  \sin \frac{\eta}{2} \left [- \sin \frac{\beta}{2}  \frac{\lambda^3}{2} + \cos \frac{\beta}{2}  \frac{\lambda^2}{2} \right ] \cdot
\left [\cos \frac{\eta}{2} + i \sin \frac{\eta}{2} (  \cos \frac{\beta}{2} \lambda^3 + \sin \frac{\beta}{2}\lambda^2)  \right ] \,  
 \right \}   \;  .
\end{equation}
Also $g \bar{A}'_1(x_1,x_2)$ does not depend on $m$, so it is periodic over the square domains and of order $\sqrt{gH}$.
Let us consider, now, the chromomagnetic field $B^a_i$. We have:
\begin{equation}
\label{5.23}
g B'_3 (x_1,x_2) \; = \; gH  \;  \Lambda(x_1,x_2) \, \frac{\lambda^3}{2}  \,  \Lambda^{-1}(x_1,x_2) \; = \;
gH  \;  \Lambda_1(x_1,x_2) \, \frac{\lambda^3}{2}  \,  \Lambda_1^{-1}(x_1,x_2) \; .
\end{equation}
From Eq.~(\ref{5.15}) we see that also the gauge-transformed chromomagnetic field does not depend on $m$ and, therefore,
it is periodic with period $L_D$. Now, Feynman pointed out that the vector potential $A^a_i(x_1,x_2)$ may vary independently
in the various domains. Indeed, one must avoid to insist on correlations which are not required by the potential energy interaction terms. As
discussed in Ref.~\cite{Feynman:1981}, one can vary the vector potential $A'_i$ in a given domain to become zero while in all the other
domains it stayed the same. It turns out that no potential energy barrier arises to prevent the given domain from behaving independently
of the other domains. In our case, however, aside of the Abelian vector potential, we must also consider the non-Abelian vector potentials induced
by the vacuum condensation of the tachyonic modes. In addition, we are dealt with three spatial dimensions instead of two dimensions 
discussed in Ref.~\cite{Feynman:1981}.
Now, we would like to show that indeed there are no strong potential energy barriers that prevent the chromomagnetic domains to behave independently. 
To see this let us consider the contributions of the unstable modes to the vacuum energy, Eq.~(\ref{3.21}). After taking into account the 
kink equations Eqs.~(\ref{3.27}), (\ref{3.28})  and (\ref{3.29}), we rewrite Eqs.~(\ref{3.21}) - (\ref{3.24}) as:
\begin{equation}
\label{5.24}
E_U  \; = \; E_U^{(u)}  \; + \;   E_U^{(v)}  \; + \;  E_U^{(w)}  
\end{equation}
where:
\begin{eqnarray}
\label{5.25}
E_U^{(u)}  \; = \;   \frac{1}{2} \, V \, \frac{gH}{4 \pi^2} , \int_{- \infty}^{+ \infty}  d p_3 \left [ \rho_u(p_3) \;  + \; \frac{p_3^2 - gH}{\rho_u(p_3)} \right ] 
\; - \;  \frac{g^2}{2} \, I^u_4 \; \int_{- \frac{L}{2}}^{+ \frac{L}{2}}  d x_3 \;  f_u^4(x_3) \; 
\nonumber \\
+ \; g^2 \, \frac{gH}{4 \pi^2}  \;  \int_{- \infty}^{+ \infty}  d p_3  \;  \frac{1}{\rho_u(p_3) }  \;      I^u_2 \;  \int_{- \frac{L}{2}}^{+ \frac{L}{2}}  d x_3 \;  f_u^2(x_3) \; \; ,
\hspace{3.5 cm}
\end{eqnarray}
\begin{eqnarray}
\label{5.26}
E_U^{(v)}  \; = \;   \frac{1}{2} \, V \, \frac{gH}{8 \pi^2}  \int_{- \infty}^{+ \infty}  d p_3  \left [ \rho_v(p_3) \;  + \; \frac{p_3^2 - \frac{gH}{2}}{\rho_v(p_3)} \right ] 
\; - \;  \frac{g^2}{2} \, I^v_4 \;  \int_{- \frac{L}{2}}^{+ \frac{L}{2}}  d x_3 \;  f_v^4(x_3) \; 
\nonumber \\
+ \; g^2 \, \frac{gH}{8 \pi^2}  \;  \int_{- \infty}^{+ \infty}  d p_3  \;  \frac{1}{\rho_v(p_3) }  \;      I^v_2 \;  \int_{- \frac{L}{2}}^{+ \frac{L}{2}}  d x_3 \;  f_v^2(x_3) \;  \; ,
\hspace{3.5 cm}
\end{eqnarray}
\begin{eqnarray}
\label{5.27}
E_U^{(w)}  \; = \;   \frac{1}{2} \, V \, \frac{gH}{8 \pi^2}  \int_{- \infty}^{+ \infty}  d p_3 \left [ \rho_w(p_3) \;  + \; \frac{p_3^2 - \frac{gH}{2}}{\rho_w(p_3)} \right ] 
\; - \;  \frac{g^2}{2} \, I^w_4 \;  \int_{- \frac{L}{2}}^{+ \frac{L}{2}}  d x_3 \;  f_w^4(x_3) \; 
\nonumber \\
+ \; g^2 \, \frac{gH}{8 \pi^2}  \;  \int_{- \infty}^{+ \infty}  d p_3  \;  \frac{1}{\rho_w(p_3) }  \;      I^w_2 \;  \int_{- \frac{L}{2}}^{+ \frac{L}{2}}  d x_3 \;  f_w^2(x_3) \;  \; .
\hspace{3.5 cm}
\end{eqnarray}
Moreover, it is useful to recall that we have the following kink equations:
\begin{eqnarray}
\label{5.28}
I^u_2 \, \left [ - \, \partial_3^2 \; - \; gH \right ]  f_u(x_3) \; + \;  g^2 \, I^u_4  \;  f_u^3(x_3)  \;  =  \; 0  \;   \; \; 
\nonumber \\
I^u_2 \; = \;  \int_{- \frac{L}{2}}^{+ \frac{L}{2}} dx_1 \, dx_2 \;  g^u_+(x_1,x_2) \, g^u_-(x_1,x_2) \; \; \; \; \;  \; \; \; \; \; \; 
\nonumber \\
I^u_4 \; =  \int_{- \frac{L}{2}}^{+ \frac{L}{2}}  dx_1 \, dx_2 \;  [ g^u_+(x_1,x_2) \, g^u_-(x_1,x_2)]^2  \; \; , \; \;  \; \; \; \; 
\end{eqnarray}
\begin{eqnarray}
\label{5.29}
I^v_2 \, \left [ - \, \partial_3^2 \; - \; \frac{gH}{2} \right ]  f_v(x_3) \; + \; g^2 \, I^v_4  \;  f_v^3(x_3)  \;  =  \; 0  \;   
\nonumber \\
I^v_2 \; = \;  \int_{- \frac{L}{2}}^{+ \frac{L}{2}}  dx_1 \, dx_2 \;  g^v_+(x_1,x_2) \, g^v_-(x_1,x_2) \; \; \; \; \;  \; \; \; \; \; 
\nonumber \\
I^v_4 \; =  \int_{- \frac{L}{2}}^{+ \frac{L}{2}} dx_1 \, dx_2 \;  [ g^v_+(x_1,x_2) \, g^v_-(x_1,x_2)]^2  \; \; , \; \;  \; \;  \;
\end{eqnarray}
\begin{eqnarray}
\label{5.30}
I^w_2 \, \left [ - \, \partial_3^2 \; - \; \frac{gH}{2} \right ]  f_w(x_3) \; + \; g^2 \, I^w_4  \;  f_w^3(x_3)  \;  =  \; 0  \;   
\nonumber \\
I^w_2 \; = \;  \int_{- \frac{L}{2}}^{+ \frac{L}{2}} dx_1 \, dx_2 \;  g^w_+(x_1,x_2) \, g^w_-(x_1,x_2) \; \; \; \; \;  \; \; \; \; \; 
\nonumber \\
I^w_4 \; =  \int_{- \frac{L}{2}}^{+ \frac{L}{2}}  dx_1 \, dx_2 \;  [ g^w_+(x_1,x_2) \, g^w_-(x_1,x_2)]^2  \; \; . \; \;  \; \;  \;
\end{eqnarray}
We said that to stabilize the one-loop instabilities the induced gauge vector potential $U^a_i(\vec{x})$ must satisfy the constraints:
\begin{equation}
\label{5.31}
\int^{\frac{L}{2}}_{- \frac{L}{2}} \, d x_3 \;  f^2(x_3) \; \sim \; L \; \; , \; \; \int^{\frac{L}{2}}_{- \frac{L}{2}} \, d x_3  \; f^4(x_3) \; \sim \; L
\end{equation}
together with:
\begin{equation}
\label{5.32}
I_2 \; \sim \; L^2 \; \; , \; \; I_4 \; \sim \; L^2 \; \; .
\end{equation}
Moreover, the minimum of the vacuum energy is attained when:
\begin{equation}
\label{5.33}
\frac{(I^u_2)^2}{I^u_4}  \; =  \; \frac{(I^v_2)^2}{I^v_4}  \; =  \; \frac{(I^w_2)^2}{I^w_4}  \; =  \; L^2  \; \; .
\end{equation}
It is evident that the kink solutions of Eqs.~(\ref{5.28}),  (\ref{5.29}) and (\ref{5.30}) satisfy the above constraints. On the other hand,
the most general solution of the kink equations are given by the dilute multi-kink solutions:
\begin{equation}
\label{5.34}
f^u(x_3)  \; =  \; \sqrt{ \frac{gH}{g^2} \, \frac{I^u_2}{I^u_4}} \; \;  \prod_{n=1}^{N_{kink}}    sign (x_3 \, - \, x_n) \; ,
\end{equation}
and the analogous expressions for $f^v$ and $f^w$. In Eq.~(\ref{5.34}) the hyperbolic tangent function has been approximated by the step
function and $N_{kink} = \frac{L}{L_D}$. We note that  the width of the kink plane  is of order
of the magnetic length  $a_H $  such that  for distances greater  than the magnetic length  
the kink-profile functions reduce to a constant. Since $L_D$ is the distance between kinks, the dilute approximation is assured
if $\ell_D \gg 1$. A more stringent condition on $\ell_D$ comes from the fact that we must allow for unstable modes. This leads to:
\begin{equation}
\label{5.35}
\ell_D \; \gtrsim  \;   \sqrt{2} \, \pi \simeq 4.4 \; \; . 
\end{equation}
In this way, we have divided the spatial volume $V$ into slices of thickness $L_D$. The kink planes are assumed to be at the middle of the
slice. Moreover, we have seen that the kink planes  can be further divided into squares of linear size $L_D$. As a consequence, the
whole volume $V$ turns out to be divided into cubic domains of volume $L_D^3$. Since we are assuming $L_D \,  \gg \, a_H$,
Eqs.~(\ref{5.31}), (\ref{5.32})  and (\ref{5.33}) are still valid after replacing $L$ with $L_D$. Therefore the vacuum energy can be written
as:
\begin{equation}
\label{5.36}
\Delta E(gH) \; \simeq \; N_D \, V_D \; \frac{(gH)^2}{16 \, \pi^2} \ln \left ( \frac{\Lambda_H}{\sqrt{gH}} \right ) \; \; ,
\end{equation}
where:
\begin{equation}
\label{5.37}
 V_D \;  = \; L_D^3 \; \; , \; \; N_D \; = \; \frac{V}{V_D}  \; \; .
\end{equation}
Interestingly enough,  we can rewrite the vacuum functional Eq.~(\ref{5.1}) as:
\begin{equation}
\label{5.38}
{\cal G}_0[{\tilde{\eta}}] \; = \;   {\cal N} \; \prod_{n} \;  \exp{ \left [  - \frac{1}{4} \int_{D_n}  d\vec{x} \, d\vec{y} \;  {\tilde{\eta}}^a_i(\vec{x}) \,
G^{a b}_{i j} (\vec{x},\vec{y})  \, {\tilde{\eta}}^b_j(\vec{y})  \right ]}
\end{equation}
where the integer $n$ labels the $N_D$ domains. It follows, then, that we can look at the quantum state Eq.~(\ref{5.38}) as a collection
of $N_D$ cubic domains with approximately the same energy:
\begin{equation}
\label{5.39}
\Delta E_D(gH) \; \simeq  \;  V_D \; \frac{(gH)^2}{16 \, \pi^2} \ln \left ( \frac{\Lambda_H}{\sqrt{gH}} \right ) \; \; .
\end{equation}
Moreover, the calculations of the chromomagnetic fields extend also to a single domain. So that our state resemble closely a 
ferromagnetic substance~\footnote{For an excellent discussion on the physics of ferromagnetism, see Ref.~\cite{Kittel:1949}.} 
where, however, all the domains have the same magnetic moment oriented in the same direction.
We recall that the domain chromomagnetic fields are directed along the perpendicular to the kink plane with strength that depends on
the chromomagnetic condensate $\sqrt{gH}$. \\
We would like to push a little further the analogy with ferromagnetism. Ferromagnetic materials are paramagnetic but show drastically
different behaviour since, below the Curie temperature, they show spontaneous magnetization. P. Weiss~\cite{Weiss:1907}
was able to explain the principal aspects of ferromagnetism by postulating the existence of a molecular field and the existence of
domain structure. It is now known that the origin of the molecular field lies in quantum-mechanical exchange forces.
The explanation of the origin of domain structure as a natural consequence of the various contributions due to the exchange,
anisotropy and magnetic energies was given by Landau and Lifshitz~\cite{Landau:1935}.
The direction of magnetization of different domain need not necessarily be parallel such that the resultant magnetization
vanishes. When an external magnetic field is applied, the domains rotate to align their magnetic moment with the field
direction leading to a non-zero magnetization. Two domains magnetized in different directions are separated by
a transition layer, called Bloch wall~\cite{Bloch:1932}, that, in general, has a certain amount of energy associated with it
characterized by the energy per unit area. \\
Returning to our wavefunctional, we have seen in Sect.~\ref{S4} that this quantum state has a non-zero expectation value of the
chromomagnetic field. On the other hand, we found that the gauge system can be thought  as made of $N_D$ cubic domains.
The non-zero expectation value of the chromomagnetic field arises from the implicit assumption that all the domains are 
oriented in the same direction. However, following the Feynman's suggestion that one must be careful not to insist on
correlations which are not required by the potential energy, we should check if, indeed, there are huge potential barriers 
preventing a given domain from behaving independently from the other domains. To this end, let us consider, firstly, 
two adjacent domains. Let us suppose, now, to revert the direction of the chromomagnetic field in one domain.
From the calculations presented in the previous Section, it is easy to see that $gB \rightarrow - gB$ amounts to
$g^u_+(x_1,x_2) \rightarrow g^u_-(x_1,x_2)$ for the u-kink, and the same for the v- and w-kinks.
From  Eqs.~(\ref{5.24}) - (\ref{5.27}) it is evident that these changes do not vary the domain energy. In other words,
reverting the direction of the chromomagnetic field in a given domain does not cost in energy.
As a consequence, we can change the sign of $gB$ independently without affecting the vacuum energy. For instance,
we may arrange the domain chromomagnetic field in a three-dimensional checkerboard order so that the chromomagnetic
fields averaged over distances much greater than $L_D$ vanish. This could make the wavefunctional Eq.~(\ref{5.38}) a better
candidate for the QCD vacuum. In addition, we may also rotate the chromomagnetic field in a given domain without changing
the vacuum energy since Eq.~(\ref{5.39}) shows that the domain energy depends only on the strength of the chromomagnetic
condensate $\sqrt{gH}$. However, there is a transition layer  that separates adjacent domains with chromomagnetic fields
pointing in different directions were we need to smoothly connect the domain background field to the one in the adjacent domain.
Evidently, this will require a certain expenditure of energy to establish a boundary layer. On the other hand, this energy will
depend on $| \vec{\nabla} (\bar{A}^a_i + U^a_i)|^2$ and on the area of the boundary layer. Observing that the domain energy scales as
$L_D^3$, while the surface energy of the boundary layer (Bloch wall) grows as $L_D^2$, we see that no potential energy barriers 
arise to prevent domains from behaving independently once that $L_D \gg a_H$. \\
Before proceeding further, it is useful to pause for summarizing our results. We have considered the SU(3) pure gauge theory
in presence of a constant Abelian chromomagnetic background field. We found that in the one-loop approximation
there are three different kinds of unstable modes. We performed a full quantum-mechanical variational calculation leading
to the dynamical condensation of the tachyonic modes that, in turns, generated a new non-perturbative background field.
Starting from the multi-kink structure of the induced background fields and following the Feynman's argumentations on the
effects of the gauge symmetry  on the ground state wavefunctional, we concluded that the stabilized vacuum wavefunctional 
can be thought as a collection of independent chromomagnetic domains that can be rewritten more explicitly as:
\begin{equation}
\label{5.40}
{\cal G}_0[A] =  {\cal N} \; \prod_{n} \;  \exp \{ - \frac{1}{4} \int_{D_n}  d\vec{x} \, d\vec{y} \, 
 [ A^a_i(\vec{x}) - \bar{A}^a_i(\vec{x}) - U^a_i(\vec{x})  ] \,
G^{a b}_{i j} (\vec{x},\vec{y})  [ A^a_i(\vec{y}) - \bar{A}^a_i(\vec{y}) - U^a_i(\vec{y}) ] \} \, .
\end{equation}
In  Eq.~(\ref{5.40}) the background field in a given domain is intended to be directed along an arbitrary spatial direction. Moreover, 
due to the local gauge symmetry, we can also orientate the domain background field in an arbitrary color direction. As a consequence,
the functional measure is given by:
\begin{equation}
\label{5.41}
{\cal{D}}  \, A \; = \; \prod_{n} \left ( {\cal{D}} \; A \right )_n
\end{equation}
where the functional integrations over the domain $D_n$ involve gauge potential vector fields transverse with respect
to the domain background fields averaged over all the allowed spatial and color directions. 
To ensure the translational invariance of the ground state wavefunctional all the domains must be characterized by the same
average chromomagnetic condensate that, henceforth, will be denoted by $\sqrt{gH_0}$.
We may conclude, thus, that our ground state wavefunctional describes the quantum vacuum as a disordered chromomagnetic
condensate. Therefore, we are led to suppose  that the wavefunctional Eq.~(\ref{5.40}), being gauge and translational invariant and having the
energy that scales with the spatial volume, should be a good candidate for the QCD vacuum, at least for large distances
$\sqrt{gH_0} \ll \Lambda_H$. However,  Eq.~(\ref{5.36}) shows that the energy of our wavefunctional is greater with
respect to the perturbative ground state. In other words, our vacuum wave functional is not energetically favoured. 
Nevertheless, having pictured the gauge system as a collection of $N_D$ independent domains, the number of gauge
field configurations accounted for by the wavefunctional   Eq.~(\ref{5.40})  is easily estimated as:
\begin{equation}
\label{5.42}
N_D ! \; \simeq \; \exp \{ N_D \,  \ln N_D  \}  \; .
\end{equation}
This allows us to introduce the configurational entropy:
\begin{equation}
\label{5.43}
\exp \{ S_{conf} \} \; \simeq \; N_D ! \; \simeq \; \exp \{ N_D \,  \ln N_D  \}  \; ,
\end{equation}
or:
\begin{equation}
\label{5.44}
 S_{conf}  \; \simeq \;  \frac{V}{L_D^3} \;   \ln N_D    \; .
\end{equation}
So that  the configurational entropy scales with the volume. This means that, even thought our vacuum wavefunctional has greater energy with
respect to the perturbative vacuum wavefunctional:
\begin{equation}
\label{5.45}
{\cal G}^{pert}_0[A] =  {\cal N} \;   \exp \bigg \{ -  \int  \,  d\vec{x} \, d\vec{y} \; \; 
 \frac{\vec{B}^a(\vec{x}) \, \cdot \,  \vec{B}^a(\vec{y})}{4 \, \pi^2 \; |\vec{x} \, - \vec{y}|^2 }  \, \bigg \}  \;  ,
\end{equation}
the number of gauge field configurations that realize that vacuum wavefunctional is large enough to span a set of finite measure in the functional space of physical states. On the contrary, the perturbative vacuum spans a zero-measure set of gauge field configurations~\footnote{An enlightening 
discussion, albeit within the path integral approach, can be found in the S. Coleman's lecture~\cite{Coleman:1985}. See, also, the
discussion in Ref.~\cite{Teper:1979}.}. More precise statements will be addressed later on. Therefore, the transition from the perturbative vacuum
to our variational vacuum wavefunctional can be thought as a order-disorder quantum phase transition analogous to the
Berezinskii-Kosterlitz-Thouless (B-K-T) 
phase transition~\cite{Berezinskii:1971,Berezinskii:1972,Kosterlitz:1972,Kosterlitz:1973,Kosterlitz:1974,Kosterlitz:2016}.
As is well known, in the B-K-T phase transitions the increase of the entropy due to the unbinding of topological excitations overcomes the energy 
barrier, leading to the decrease of the free energy F = E - TS, T being the temperature.
We can come in closer analogy with the thermodynamics of the B-K-T phase transitions by introducing the domain fugacity:
\begin{equation}
\label{5.46}
z_D \; = \; \left ( \kappa \, \mu \, L_D \right )^3  
\end{equation}
where $\mu$ is an energy scale and $\kappa$ a constant that will be specified later on.
If we have $N_D$ domains, we may define the entropy as:
\begin{equation}
\label{5.47}
\exp \{ S_D \} \; = \; z_D^{N_D} \; = \; \exp \{ N_D \,  \ln z_D  \}  \; = \;  \exp \{ \frac{V}{V_D} \,  \ln z_D  \}  \;  .
\end{equation}
We have already shown that the energy needed to create $N_D$ domains is:
\begin{equation}
\label{5.48}
\Delta E(gH_0) \; \simeq \; N_D \, V_D \; \frac{(gH_0)^2}{16 \, \pi^2} \ln \left ( \frac{\Lambda_H}{\sqrt{gH_0}} \right ) \; \; .
\end{equation}
Since we are dealing with a quantum phase transition the temperature T is zero. To introduce the free energy the role of the temperature is
played by the energy scale $\mu$.
Accordingly, we may introduce the quantum free energy:
\begin{equation}
\label{5.49}
F_D \; = \;  \Delta E \;  -  \; \mu \, S_D
\end{equation}
or
\begin{equation}
\label{5.50}
\frac{F_D}{N_D}  \; \simeq \;  V_D \,  \frac{(gH_0)^2}{16 \, \pi^2} \ln \left ( \frac{\Lambda_H}{\sqrt{gH_0}} \right )  \; - \; 
3 \, \mu \, \ln \left (  \kappa \, \mu \, L_D \right )  \; .
\end{equation}
At the phase transition the free energy vanishes and the fugacity becomes $z_D = 1$.  From Eq.~(\ref{5.50}) we see that $z_D =1$ for
$\mu = \mu^*$ where:
\begin{equation}
\label{5.51}
V_D \,  \frac{(gH_0)^2}{16 \, \pi^2} \ln \left ( \frac{\Lambda_H}{\sqrt{gH_0}} \right )  \; \simeq  \; 
3 \, \mu^* \, \ln \left (  \kappa \, \mu^* \, L_D \right )  \; .
\end{equation}
Neglecting logarithmic corrections we get:
\begin{equation}
\label{5.52}
\mu^* \; \simeq \; V_D \,  \frac{(gH_0)^2}{48 \, \pi^2}   \; \simeq  \; 
  \frac{\ell_D^3}{48 \, \pi^2}  \;  \sqrt{gH_0}  \; ,
\end{equation}
and
\begin{equation}
\label{5.53}
\kappa \; \simeq \;   \frac{48 \, \pi^2}{\ell_D^4} \; . 
\end{equation}
From Eqs.~(\ref{5.35}) and (\ref{5.53}) we infer that $\kappa \sim O(1)$, and Eq.~(\ref{5.52}) can be rewritten as:
\begin{equation}
\label{5.54}
\mu^* \; \simeq  \;  \frac{\sqrt{gH_0}}{\ell_D}  \; .
\end{equation}
At the quantum phase transition there is a proliferation of domains without variation of the free energy for adding one more
domain the  increase of the entropy compensates the energy variation. \\
To conclude the present Section, our variational perturbative approach aimed to stabilize  the SU(3) pure gauge theory in presence
of an Abelian constant background field lead to the conclusion that the QCD vacuum at large scales behaves like a disordered
chromomagnetic condensate. Our results picture   the quantum phase transition from the perturbative vacuum to the confining
QCD vacuum as a order-disordered transition driven by the proliferation of chromomagnetic domains. 
Even thought these results look promising, it remains to check if the proposed QCD vacuum 
wavefunctional does display the known physical properties of the confining physical vacuum.
\section{\normalsize{Color confinement, flux tubes and Meissner effects }}
\label{S6}
Let ${\cal{O}}[A]$ be a generic physical observable, then we define the vacuum expectation value as usual:
\begin{equation}
\label{6.1}
 <{ \cal{O}}[A] >  \;  = \; \int  {\cal{D}}  A \;  {\cal{G}}_0^*[A] \;  {\cal{O}}[A]  \; {\cal{G}}_0[A]  \; ,
\end{equation}
where the vacuum functional is given by Eq.~(\ref{5.40}) and the functional integrations are performed according to Eq.(\ref{5.41}).
The normalization constant is fixed by:
\begin{equation}
\label{6.2}
 \int  {\cal{D}}  A \;  {\cal{G}}_0^*[A] \;   {\cal{G}}_0[A]  \;  = \; 1  \; .
\end{equation}
Evidently, if we consider a local coloured observable $ {\cal{O}}^{a,b,..}[A]$, then:
\begin{equation}
\label{6.3}
 <{ \cal{O}}^{a,b,..}[A] >  \;  = \; 0  \;
\end{equation}
in accordance with the Elitzur's theorem~\cite{Elitzur:1975}. In particular, we have:
\begin{equation}
\label{6.4}
 < B^a_i(\vec{x})>  \;  = \;  < E^a_i(\vec{x})>  \; = \; 0  \; .
\end{equation}
So that only colorless local observables have a non-zero expectation  value on the vacuum wavefunctional   ${\cal{G}}_0[A]$.
For instance, we have that  $< (B^a_i(\vec{x}))^2> $ and   $ < (E^a_i(\vec{x}))^2>$  are different from zero. Indeed, we can write:
\begin{eqnarray}
\label{6.5}
 <(gB^a_i(\vec{x}))^2 >  \;  = \; \int  {\cal{D}}  A \;  {\cal{G}}_0^*[A] \; \big [ gB^a_i(\vec{x}) \big ]^2  \; {\cal{G}}_0[A]  \;  \; \; 
 \nonumber \\
= \;   \int  {\cal{D}}  A \;  {\cal{G}}_0^*[A] \; \left \{ \frac{1}{V_D} \int_{D(\vec{x})} d \vec{x} \; (gB^a_i(\vec{x}))^2  \right \} \; {\cal{G}}_0[A] 
\; \; 
\end{eqnarray}
where the spatial integration is over the volume of the domain $D(\vec{x})$ containing $\vec{x}$.
In Eq.~(\ref{6.5}) we used the traslational invariance of the vacuum functional. Using Eqs.~(\ref{5.40}) and (\ref{5.41}) we have (aside
of the normalization constant):
\begin{eqnarray}
\label{6.6}
 <(gB^a_i(\vec{x}))^2 >  \;  = \;  \int  \left [ {\cal{D}}  A \right ]_{D(\vec{x})} \;   \left \{ \frac{1}{V_D} \int_{D(\vec{x})} d \vec{x} \;
  (gB^a_i(\vec{x}))^2  \right \}  \; \times \;
 \nonumber \\ 
  \exp \{ - \frac{1}{2} \int_{D(\vec{x})} d\vec{x} \, d\vec{y} \, 
 [ A^a_i(\vec{x}) - \bar{A}^a_i(\vec{x}) - U^a_i(\vec{x})  ] 
G^{a b}_{i j} (\vec{x},\vec{y})  [ A^a_i(\vec{y}) - \bar{A}^a_i(\vec{y}) - U^a_i(\vec{y}) ] \} \, .
\end{eqnarray}
From the results of Sect.~\ref{S4} it is straightforward to ascertain that:
\begin{equation}
\label{6.7}
 <(gB^a_i(\vec{x}))^2 >  \;  \simeq  \; (gH_0)^2 \; \; .
\end{equation}
Analogous calculations can be performed to evaluate   $ < (E^a_i(\vec{x}))^2>$ that, however, turns out to be negligible small with respect
to the chromomagnetic contributions. As a consequence, our vacuum wavefunctional has a non-trivial gluon condensate:
\begin{equation}
\label{6.8}
 <\frac{\alpha}{\pi} F^a_{\mu \nu}(\vec{x})   F_a^{\mu \nu}(\vec{x})  >  \;  \simeq \; \frac{1}{2 \pi^2} \;   <(gB^a_i(\vec{x}))^2 >  \;  
 \simeq  \frac{(gH_0)^2}{2 \pi^2} \; . 
\end{equation}
Later on we shall see that for QCD $\sqrt{gH_0} \simeq 1.0$ GeV, so that we reach an estimate of the gluon condensate:
\begin{equation}
\label{6.9}
 <\frac{\alpha}{\pi} F^a_{\mu \nu}  F_a^{\mu \nu} >  \;  \simeq \; \left ( 0.1 \; {\text{GeV}} \right )^4   \;  
\end{equation}
that is in reasonable agreement with phenomenological estimates (see, eg, Table 1 in Ref.~\cite{Narison:2018}) and the direct
determination on the lattice~\cite{Bali:2014}. \\
The vacuum wavefunctional does not admit long-range color correlations. For instance, let us consider the two point correlation function
 $ < gB^a_i(\vec{x}) \,  gB^a_i(\vec{y})>$.  Evidently, for $| \vec{x} - \vec{y}| > L_D$ we have:
\begin{equation}
\label{6.10}
< gB^a_i(\vec{x}) \,  gB^a_i(\vec{y})>   \;  \simeq \;  < gB^a_i(\vec{x}))> \; <  gB^a_i(\vec{y})> \; = \; 0 \; .
\end{equation}
In general, the lack of long-range color correlations can be easily understood. In fact, our vacuum functional basically is made of
chromomagnetic domains completely decorrelated. Thereby, the quantum vacuum does not share the color coherence needed
to propagate an arbitrary color disturbance over distances larger than $L_D$. Finally, the lowest excited states are gapped.
Indeed, as pointed out by Feynman in Ref.~\cite{Feynman:1981}, when a system can be considered as made of approximately
independent parts, the lowest excitation energy is the excitation of one of the parts. Evidently, the lowest excitation energy of a single
chromomagnetic domain is of order $\frac{1}{L_D}$, so that for the energy gap we get:
\begin{equation}
\label{6.11}
\Delta    \;  \simeq \;  \frac{\sqrt{gH_0}}{\ell_D} \; \simeq \mu^*   \; .
\end{equation}
The absence of long-range color correlations together with the presence of a finite energy gap for the low-lying excitations ensures
that our vacuum wavefunctional satisfies the color confinement criterion. Moreover, from phenomenological considerations we
infer that $L_D \sim 1.0$ fm. Since we already anticipated that the chromomagnetic condensate strength is about 1.0 GeV,
we are lead to estimate $\ell_D \sim 5$, leading to a mass gap $\Delta \sim 10^2$~MeV. \\
Nevertheless,  this is not enough to conclude that our vacuum wavefunctional is good enough to capture the relevant physical
properties to describe the large-distance dynamics of the QCD vacuum.
The crucial point is that we must be also able to explain the quark confinement physics. Since we are still dealing with
the pure gauge theory, the quark confinement problem amounts to demonstrate that a static quark-antiquark pair
interact via a confining linear potential:
\begin{equation}
\label{6.12}
V_{conf} \; = \sigma \; R
\end{equation}
for distances $R$ large enough. In Eq.~(\ref{6.12}) $\sigma$ is the string tension that is related to the Regge 
slope~\cite{Necco:2002}:
\begin{equation}
\label{6.13}
 \sqrt{\sigma } \; \simeq \; 464 \; {\text MeV} \; .
\end{equation}
Actually, a great deal of numerical evidences showed that a static quark-antiquark pair interacts by means of a linear
potential for distances above about 0.5 fm. Moreover, the linear potential is almost completely due to the chromoelectric fields
that are longitudinal, namely oriented along the line connecting the static color sources. However, a first-principle theoretical
explanation of this phenomenon is still lacking. Therefore, the current understanding of quark confinement is mainly based
on models of the QCD vacuum. We intend to show that our theoretical proposal for the QCD vacuum allows to gain a vivid
picture on the formation of the chromoelectric flux tube between static color charges and, in addition, to determine the color
structure and the transverse profile of the flux-tube chromoelectric fields.\\
The presence of static color   charges modify the Gauss's law constraints as:
\begin{equation}
\label{6.14}
 \left \{ i \, D^{ab}(\vec{x}) \, \frac{\delta}{\delta \, \eta^b(\vec{x})}  - Q^a \left [  \delta(\vec{x} - \vec{x}_Q) \, - \,  \delta(\vec{x} - \vec{x}_{\bar{Q}})
 \right ]  \right \} {\cal G}[A] = 0 \; ,
\end{equation}
where we are assuming the presence of a static quark charge $Q^a$ at $\vec{x}_Q$ and a static antiquark charge $\bar{Q}^a$
at $\vec{x}_{\bar{Q}}$. We have seen that the ground-state wavefunctional $ {\cal G}_0[A]$ does not allow color disturbances
to propagate over large distances. On the other hand, the Gauss  law imposes that the chromoelectric flux originating from
the source $Q^a$ must reaches the sink $\bar{Q}^a$. As a consequence, it is necessary to modify  $ {\cal G}_0[A]$ into a
wavefunctional $ {\cal G}[A]$ that, indeed, satisfies the Gauss  law constraints Eq.~(\ref{6.14}). 
Evidently, we need a wavefunctional with energy as low as possible and, at the same time, we must ensure a region around the static
color charges with restored color coherence  so that color disturbances are allowed to spread over large distances.
The most obviously way is to create  bags of perturbative vacuum (false vacuum) around the color sources. Indeed, by assuming
that by quantum fluctuations a chromomagnetic domain may evaporate, one gains energy $\Delta E_D(gH_0)$, Eq.~(\ref{5.39}),
and loses  configurational entropy such that, according to Eq.~(\ref{5.51}), there is no appreciable variation of the vacuum
free energy. Evidently, the amplitude for quantum fluctuations to tunnel into the perturbative vacuum is of order
${\cal{A}} \sim \exp ( - \Delta E_D \, R ) \sim \exp ( - \sqrt{gH_0} \,  R )$ with $R = | \vec{x}_Q - \vec{x}_{\bar{Q}}|$.
We already said that $\sqrt{gH_0} \sim 1.0$ GeV, so that the tunnelling probability is sizeable for distances $R \sim a_H \sim 10^{-1}$ fm.
However,  extensive numerical simulations of quenched QCD on the lattice demonstrated unequivocally the formation
of the squeezed chromoelectric flux tube  for distances well above $R \sim 1.0$ fm. Since for $R \gg a_H$ the tunnelling 
probability is vanishingly small, we need an alternative mechanism to explain the formation of the chromoelectric flux tubes
around static color charges separated by large distances. As a matter of fact, an alternative way to generate long-range
color coherence is to polarize the chromomagnetic domains such that the chromomagnetic fields point in the same direction.
\begin{figure}
%
\begin{center}
\includegraphics[width=0.7\textwidth,clip]{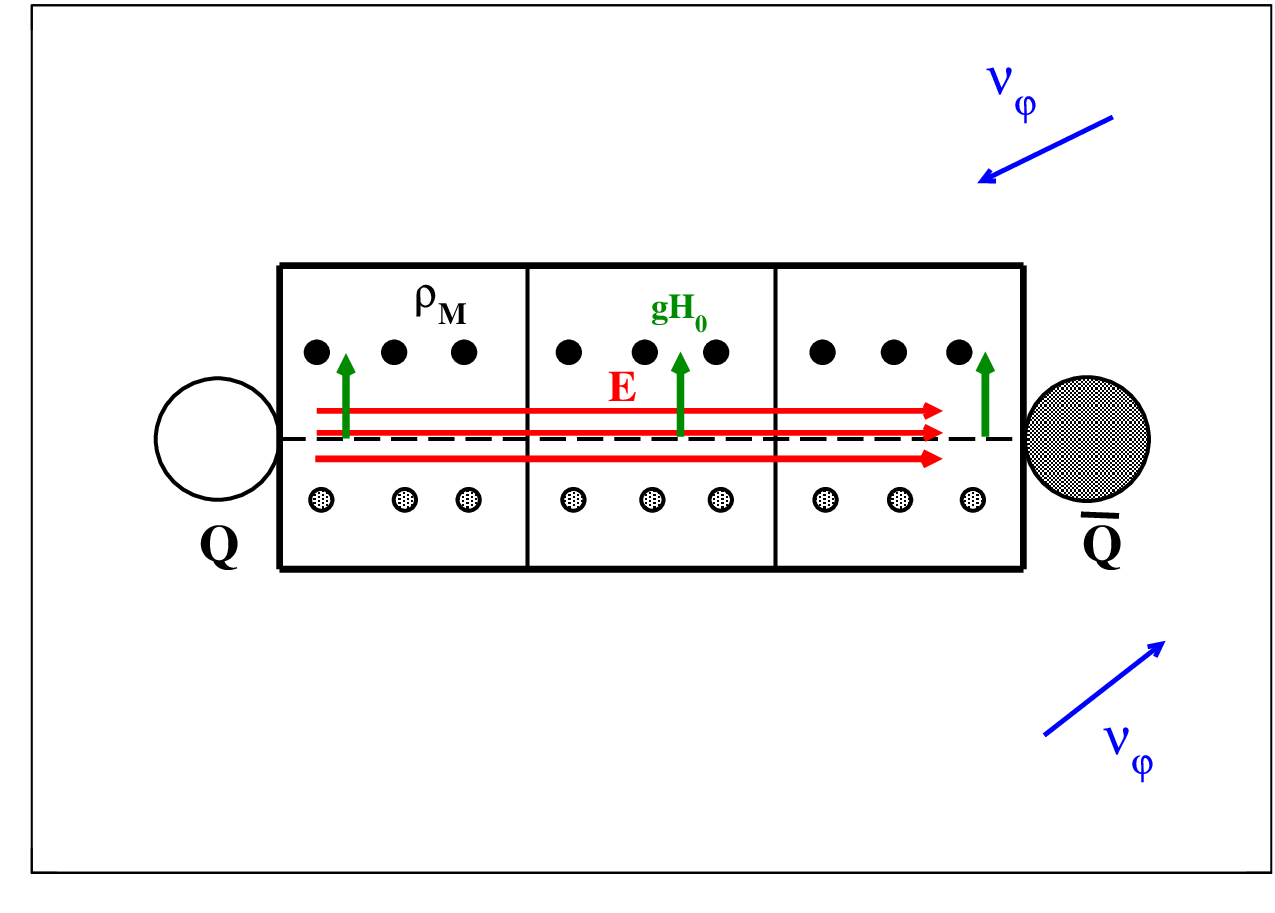}
\end{center}
\vspace{-0.5cm}
\caption{\label{Fig1} 
(Color online) Pictorial representation of the domain polarization due to a static quark-antiquark pair.}
\end{figure}
More precisely, we must admit that in a spatial region comprising the static color charges the chromomagnetic domains share the
same kink plane with the domain chromomagnetic fields pointing in the same direction transverse to the kink plane.
It should be clear that the polarization of the domains does not vary the energy of the wavefunctional but it increases the free energy
since we lost configurational entropy. To minimize the vacuum free energy the polarization region must have the spatial  volume as
small as possible. Due to the symmetry of the problem, the polarization volume is a cylinder with the symmetry axis coincident with the 
line jointing   the static color charges  and with  transverse sectional  area  of order $A_T \lesssim L_D^2$.
The cylinder symmetry axis must lie on the common kink plane of the polarized chromomagnetic domains. This allows the polarized domains
to rotate rigidly around the symmetry axis such that the increase of the vacuum free energy is reduced as much as possible.
As a consequence, we are led to a modified vacuum wavefunctional as schematically illustrated in Fig.~\ref{Fig1}. \\
We noted in Sect.~\ref{S4} that the vacuum chromomagnetic fields lead to the presence of chromomagnetic charges whose density was given by:
\begin{equation}
\label{6.15}
\vec{D}^{ab}(\vec{x}) \, \cdot  \, \vec{B}^b(\vec{x}) \; = \; \rho^a_M(\vec{x}) \; .
\end{equation}
These chromomagnetic charges give rise to a chromomagnetic current density:
\begin{equation}
\label{6.16}
\vec{J}^{a}_M(\vec{x})  \; = \; \vec{v}_{\phi} \, \rho^a_M(\vec{x}) \; 
\end{equation}
where $\vec{v}_{\phi}$ is the rotational velocity of the polarized domains. Now, let us introduce cylindrical coordinate system
$(x_{\ell}, x_t, \phi)$ where $x_{\ell}$ is the coordinate along the symmetry axis, $x_t$ the distance from the axis and $\phi$
the azimuthal angle. Evidently:
\begin{equation}
\label{6.17}
 \vec{v}_{\phi} \; = \;  v_{\phi}  {\hat{\phi}}    \; , 
\end{equation}
so that the azimuthal chromomagnetic current will give rise to a Lorentz force which tends to squeeze the chromoelectric fields
of the static color charges into a narrow structure directed along the longitudinal direction ${\hat{x_{\ell}}}$. To do this, however, we need
a chromomagnetic current density that is almost uniform along the flux tube.  From the results in Sect.~\ref{S4} we see that this
requirement is satisfied only by the Abelian component of the chromomagnetic charge density $\rho^3_M(\vec{x})$ and
$\rho^8_M(\vec{x})$. Evidently, we have:
\begin{equation}
\label{6.18}
\vec{J}^{\; 3}_M(\vec{x})  \; = \; \vec{v}_{\phi} \, \rho^3_M(\vec{x}) \;  \;  , \; \; \vec{J}^{\; 8}_M(\vec{x})  \; = \; \vec{v}_{\phi} \, \rho^8_M(\vec{x}) \; ,
\end{equation}
with:
\begin{equation}
\label{6.19}
  g \rho^3_M(\vec{x}) \;  \simeq \;   - \, \sqrt{2} \, (gH_0)^{\frac{3}{2}} \;  \, \frac{ \tanh \left ( \sqrt{\frac{gH_0}{2}} \, x_t  \right ) }
 {\cosh^2  \left ( \sqrt{\frac{gH_0}{2}} \, x_t  \right )    }   \;  \; ,
\end{equation}
\begin{equation}
\label{6.20}
 g  \rho^8_M(\vec{x})  \; \simeq \;    -  \, \frac{\sqrt{3}}{2}  \, (gH_0)^{\frac{3}{2}} \;  \, \frac{ \tanh \left ( \sqrt{\frac{gH_0}{4}} \, x_t  \right ) }
 {\cosh^2  \left ( \sqrt{\frac{gH_0}{4}} \, x_t  \right )    }    \;  \; .
\end{equation}
Observing that the chromomagnetic current densities belong to the maximal Abelian SU(3)  subgroup, the equations relating
the chromomagnetic currents to the chromoelectric fields are given by the Ampere law (see, for instance, 
Refs.~\cite{Jackson:1999,Felsager:1983}):
\begin{equation}
\label{6.21}
  -  \;  \vec{\nabla} \; \times \vec{E}^a(\vec{x}) \; = \; \vec{J}^{a}_M(\vec{x})     \;  \; .
\end{equation}
From Eqs.~(\ref{6.18}),  (\ref{6.19}),  (\ref{6.20}) and    (\ref{6.21}) we get:
\begin{equation}
\label{6.22}
\vec{E}^a(\vec{x})  \;  \simeq \; E^a(x_t)  \; \hat{x}_{\ell}   \;  \; ,
\end{equation}
\begin{equation}
\label{6.23}
 E^a(x_t)  \;  \simeq \; \delta^{a 3} \, E^3(x_t) \; + \;  \delta^{a 8} \, E^8(x_t) \;
\end{equation}
with:
\begin{equation}
\label{6.24}
 \frac{d}{d x_t} \, g E^3(x_t) \; \simeq \;   v_{\phi} \;  g  \rho^3_M(x_t) \;  \; ,
\end{equation}
\begin{equation}
\label{6.25}
 \frac{d}{d x_t} \, g E^8(x_t) \; \simeq \;   v_{\phi} \; g   \rho^8_M(x_t) \;  \; .
\end{equation}
One easily obtains:
\begin{equation}
\label{6.26}
 g E^3(x_t) \; \simeq \;   v_{\phi} \;   gH_0 \;  \, \frac{1}{\cosh^2  \left ( \sqrt{\frac{gH_0}{2}} \, x_t  \right )  }    \;  \; ,
\end{equation}
\begin{equation}
\label{6.27}
 g E^8(x_t) \; \simeq \; \frac{\sqrt{3}}{2} \,   v_{\phi} \;   gH_0 \;  \, \frac{1}{\cosh^2  \left ( \sqrt{\frac{gH_0}{4}} \, x_t  \right )  }    \;  \; .
\end{equation}
It is worth pausing to briefly recap on the origin of the chromomagnetic charge density that plays a fundamental role in
the formation and structure of the chromoelectric flux tube between far apart static color charges.
From the results presented in Sect.~\ref{S4} it should be evident that the chromomagnetic charge densities originated from
the kink structures that, in turns, come from the condensation of the tachyonic unstable modes. As discussed in details in I and in 
Sects.~\ref{S2} and \ref{S3} of the present paper, the spin couplings of the gauge vector fields to the chromomagnetic 
background field generate negative mass squared terms in the lowest Landau levels. These instabilities drive the dynamical condensation of 
the tachyonic modes that were stabilized by the short-range repulsive interactions due to the positive quartic self-couplings of
the gauge vector bosons. Albeit our calculations are based on a perturbative approach by means of a variational basis, we already
remarked that the results presented in I for the SU(2) gauge theory and for SU(3) in the present paper have been corroborated by 
non-perturbative lattice numerical simulations of non-Abelian gauge theories in presence of external background
 fields~\cite{Cea:1991,Cea:1993,Cea:1997d,Cea:1998,Cea:1999b}.  \\
The static quark-antiquark system is confined through the generation of the chromoelectric flux tube for which the quark and antiquark
act as source and sink. Moreover, since the chromoelectric fields in the flux tube do not depend on the longitudinal coordinate
$x_{\ell}$, we see that the quark and the antiquark at large separation distances are confined by a linearly rising potential,
Eq.~(\ref{6.12}), with string tension given by the energy per unit length stored in the flux-tube chromoelectric fields:
\begin{equation}
\label{6.28}
\sigma \; \simeq \; \frac{1}{2} \; \int d \vec{x}_t \left \{  [  gE^3(x_t) ]^2 \;+ \;  [  gE^8(x_t) ]^2   \right \} \;  \; .
\end{equation}
A straightforward calculation gives:
\begin{equation}
\label{6.29}
\sqrt{\sigma}  \; \simeq \;  \sqrt{ 5 \pi J_1} \; v_{\phi} \; \sqrt{gH_0} \; , 
\end{equation}
where:
\begin{equation}
\label{6.30}
J_1\; =  \;  \int_0^{\infty} dz  \; \frac{z}{\cosh^4 z} \; \simeq \;  0.295431 \; \; . 
\end{equation}
We may also introduce the radius $R_T$ of the transverse section of the flux tube as:
\begin{equation}
\label{6.31}
R_T\; \simeq \;  \; \frac{  \int d \vec{x}_t  \; |\vec{x}_t| \; \left \{  [  gE^3(x_t) ]^2 \;+ \;  [  gE^8(x_t) ]^2   \right \}  }{
\int d \vec{x}_t   \;  \left \{  [  gE^3(x_t) ]^2 \;+ \;  [  gE^8(x_t) ]^2   \right \} }
 \;  \; .
\end{equation}
We easily obtain:
\begin{equation}
\label{6.32}
R_T\; \simeq \;  \; \frac{  2 \, \sqrt{2} \, + \, 6 }{5 } \; \frac{J_2}{J_1} \; \frac{1}{\sqrt{gH_0}} \; \; ,
\end{equation}
where:
\begin{equation}
\label{6.33}
J_2\; =  \;  \int_0^{\infty} dz  \; \frac{z^2}{\cosh^4 z} \; \simeq \;  0.214978  \; \; . 
\end{equation}
An alternative  definition of the transverse radius is given by $w$ where:
\begin{equation}
\label{6.34}
w^2 \; \simeq \;  \; \frac{  \int d \vec{x}_t  \; |\vec{x}_t|^2  \;  gE_{\ell}(x_t)  }{
 \int d \vec{x}_t  \;  gE_{\ell}(x_t)  }  \;  \; ,
\end{equation}
with
\begin{equation}
\label{6.35}
 gE_{\ell}(x_t)  \;  =  \;  gE^3(x_t) \; + \;    gE^8(x_t)  \; \; .
\end{equation}
Performing the integrals one gets:
\begin{equation}
\label{6.36}
w^2 \; \simeq \;  \frac{9}{4} \; \frac{ 1 \, + \, 2 \sqrt{3}}{1 \, + \, \sqrt{3}} \; \frac{\zeta(3)}{\ln 2} \; \frac{1}{gH_0} \; \; .
\end{equation}
It turns out that:
\begin{equation}
\label{6.37}
w \; \simeq \;  2.0 \; R_T \;  \; .
\end{equation}
According to Eqs.~(\ref{6.26}) and (\ref{6.27}) the transverse profile of the flux-tube chromoelectric fields  depends only on the strength
of the chromomagnetic condensate $\sqrt{gH_0}$, while the azimuthal velocity fixes the chromoelectric field normalization.
In principle, we can determine these two parameters by looking at observations. However, due to quark confinement we can only compare
with theoretical experiments such as the non-perturbative numerical simulations of QCD on the lattice.
In a series of paper~\cite{Cea:2016,Cea:2017,Baker:2019,Baker:2020,Baker:2022} it was investigated by lattice Monte Carlo simulations of both SU(3) pure gauge theory and (2+1)-flavor QCD at almost the physical point some properties
of the chromoelectric flux tube at zero temperature generated by a static quark-antiquark pair. More precisely, these distributions were obtained
from lattice measurements of the connected correlators between a plaquette and a Wilson loop. Indeed, the connected correlator provides
a lattice definition of a gauge-invariant field strength tensor $gF_{\mu \nu}$ generated by the static color sources. The Wilson loop
connected to the plaquette generates the static-quark color fields which point in an unknown direction $n^a$ in color space. The Schwinger
lines connecting the Wilson loop to the plaquette perform the parallel transport of the color direction $n^a$ from the Wilson loop
to the plaquette, so that:
\begin{equation}
\label{6.38}
gF_{\mu \nu}(x)  \;  = \;  gF^a_{\mu \nu}(x) \, \cdot \, n^a  \;  \; .
\end{equation}
Equation~(\ref{6.38})  is an inevitable consequence of the gauge invariance of the lattice operator and its linear dependence on the
color fields in the continuum limit has been explicitly tested in Ref.~\cite{Cea:2016} (see Fig.~3 there).
Remarkably, extensive lattice simulations showed that, far from the sources, the flux tube is almost completely formed by the
longitudinal chromoelectric field $gE_{\ell}$ which is constant along the flux tube and decreases rapidly in the transverse
direction $x_t$. Introducing (here $\beta$ is the lattice gauge coupling):
\begin{equation}
\label{6.39}
E_{\ell}(x)  \;  = \;  \sqrt{\frac{\beta}{6}} \;  \, gE_{\ell}(x)   \;  \; ,
\end{equation}
the formation of the longitudinal chromoelectric field $E_{\ell}(x)$ was interpreted as the dual Meissner effect within the dual
superconductor mechanism of quark confinement. Accordingly, the lattice data for the longitudinal chromoelectric field were
analyzed by exploiting a variational model for the magnitude of the normalized order parameter of an isolated vortex in type II
superconductors advanced in Ref.~\cite{Clem:1975}. As a consequence, the transverse distribution of the chromoelectric flux tube
were described according to:
\begin{equation}
\label{6.40}
E_{\ell}(x_t)  \;  = \;  \frac{\varphi}{2 \pi} \; \frac{1}{\lambda \, \xi_v} \;  
\frac{K_0(\sqrt{x_t^2 \, + \, \xi_v^2)}}{K_1(\frac{\xi_v}{\lambda}) }    \;  \; ,
\end{equation}
where  $\xi_v$   is a variational  core-radius parameter, $\lambda$ is the penetration length and $K_n(x)$ is the modified Bessel function
of order $n$. Moreover, the so-called Ginzburg-Landau parameter $\kappa_{GL}$ can be obtained by:
\begin{equation}
\label{6.41}
\kappa_{GL} \;  = \;  \frac{\lambda}{\xi} \; = \; \sqrt{2} \; \,  \frac{\lambda}{\xi_v} \;  \left [
1 \; - \; \frac{K_0^2(\frac{\xi_v}{\lambda}) }{K_1^2(\frac{\xi_v}{\lambda}) } 
 \right ]^{\frac{1}{2}} \; \; , 
\end{equation}
$\xi$ being the coherence length. 
\\
It resulted that  the phenomenological law Eq.(\ref{6.40}) allowed to track very well the transverse profile of the longitudinal
chromoelectric field giving support to the dual superconductor mechanism of quark confinement. On the other hand, our attempt
to unveil from first principles the structure of the large-scale QCD vacuum led us  to a completely different picture for the
formation of the color flux tube generated by static sources. According to our previous discussion, the chromoelectric flux tube
is mainly composed by the Abelian component $gE^3$ and $gE^8$. Therefore, we can safely assume that:
\begin{equation}
\label{6.42}
  gE^a(x) \, \cdot \, n^a  \;  \simeq  \;  gE^3(x) \; + \; gE^8(x)
\end{equation}
where  $gE^3(x)$ and   $gE^8(x)$ are explicitly given by Eqs.~(\ref{6.26}) and (\ref{6.27}). So  that we have for the measured chromoelectric
longitudinal field:
\begin{equation}
\label{6.43}
  E_{\ell}(x_t) \; \simeq \;   v_{\phi} \;   gH_0 \; \left \{   \, \frac{1}{\cosh^2  \left ( \sqrt{\frac{gH_0}{2}} \, x_t  \right )  }    \;  + \; 
 \frac{\sqrt{3}}{2} \;  \frac{1}{\cosh^2  \left ( \sqrt{\frac{gH_0}{4}} \, x_t  \right )  }  
 \right \}  \;  \; .
\end{equation}
\begin{figure}
\begin{center}
\includegraphics[width=0.55\textwidth,clip]{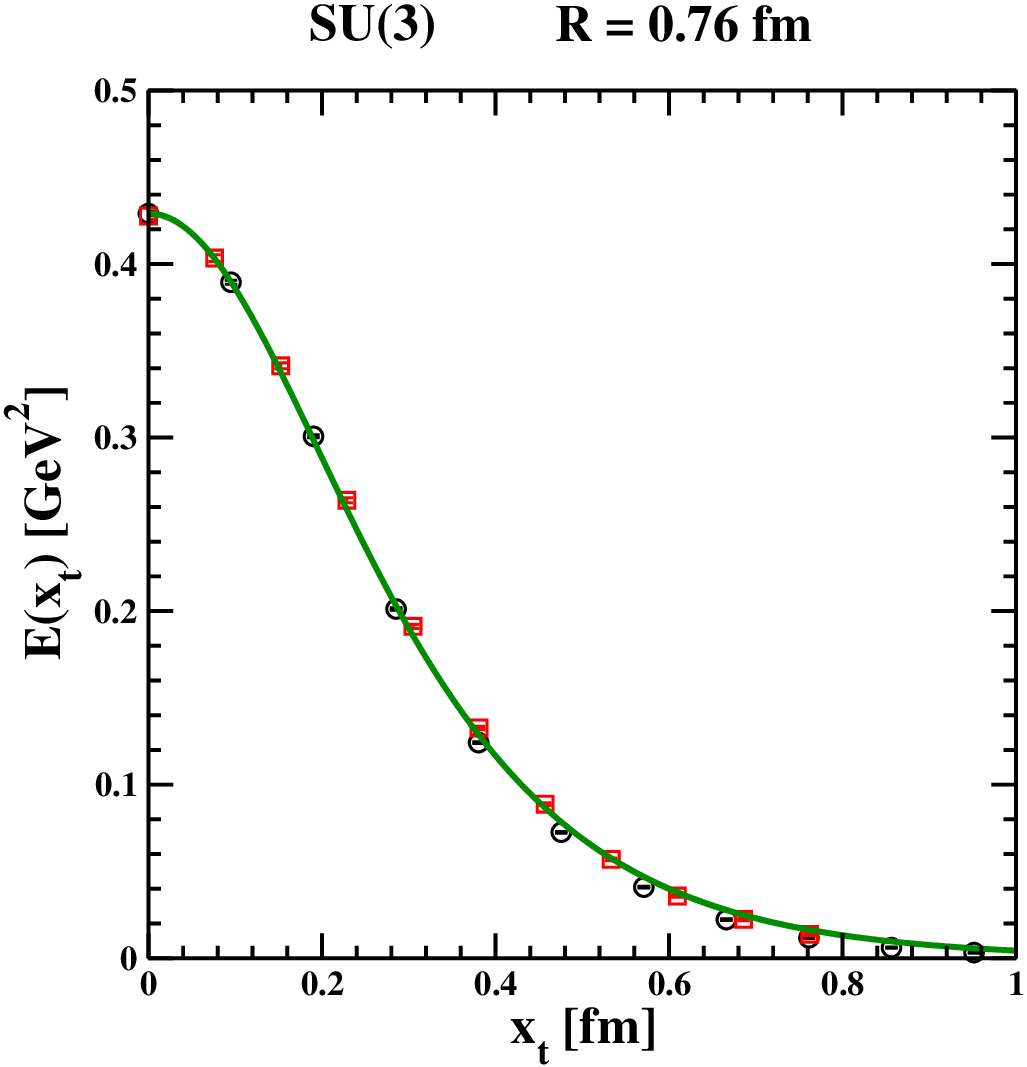}
\end{center}
\caption{\label{Fig2} 
(Color online) The longitudinal chromoelectric field versus the transverse distance for SU(3) at $\beta = 6.050$ (circles) and $\beta = 6.195$
(squares). The data have been taken from Fig.~3, left panel, of Ref.~\cite{Cea:2017}. The continuous line is Eq.~(\ref{6.43}) with
parameters given by Eq.~(\ref{6.44}).}
\end{figure}
In Eq.~(\ref{6.43}) we considered that $\sqrt{\frac{\beta}{6}} \approx 1$ and that the smearing procedure needed to extract the physical
informations from the measured lattice connected correlator leads to an effective non-perturbative finite renormalization of the
field strength tensor. As a consequence, in Eq.~(\ref{6.43}) it is intended the presence of an unknown renormalization constant
that, however, affects only the value of the longitudinal chromoelectric field at $x_t=0$.
A remarkable consequence of Eq.~(\ref{6.43}) is that the transverse profile of the longitudinal chromoelectric field depends only on the
vacuum chromomagnetic condensate strength $\sqrt{gH_0}$. For comparison, the Clem's ansatz Eq.~(\ref{6.40}) needs two
parameters $\lambda$ and $\xi_v$ to track the transverse profile. \\
We have contrasted Eq.~(\ref{6.43}) to several lattice measurements of the longitudinal chromoelectric field and, surprisingly, we found
that Eq.~(\ref{6.43}) is able to reproduce the transverse profiles of $E_{\ell}(x_t)$ quite well. To better appreciate this last point,
in Fig.~\ref{Fig2} we display the behaviour of the longitudinal chromoelectric field as a function of the transverse distance for two
different values of the gauge coupling for the SU(3) pure gauge theory as reported in Fig.~3, left panel, of Ref.~\cite{Cea:2017}.
In Ref.~\cite{Cea:2017} the lattice data for $E_{\ell}(x_t)$ were nicely fitted by the Clem's ansatz Eq.~(\ref{6.40}). Likewise, we
find that the transverse profile of the longitudinal chromoelectric field can be accounted for by using Eq.~(\ref{6.43}) with
(see Fig.~\ref{Fig2}):
\begin{equation}
\label{6.44}
  v_{\phi}  \; \simeq \;  0.21 \; \; , \; \; \sqrt{gH_0} \; \simeq \; 1.05 \; \text{GeV} \; \; .  
\end{equation}
However, in Refs.~\cite{Baker:2019,Baker:2020} it was enlighten the presence of an effective Coulomb-like chromoelectric field
$\vec{E}^C$ associated with the static quark sources. Indeed, we have said that it is conceivable by quantum tunnelling the
formation of bubbles of perturbative vacuum around the static color sources with a linear size of order the chromomagnetic length.
So that near the static charges there is a perturbative Coulomb chromoelectric field that, however, may penetrate at larger
distances due to the restoration of color coherence along the flux tube. Therefore, the measured chromoelectric field
$\vec{E}$ can be written as~\cite{Baker:2019,Baker:2020}:
\begin{equation}
\label{6.45}
  \vec{E} \;  =  \; \vec{E}^{NP} \; + \;  \vec{E}^{C}  \;  \;  
\end{equation}
with the non-perturbative chromoelectric field $\vec{E}^{NP}$ being purely longitudinal. In other words, $\vec{E}^{NP}$ must be
identified with the confining field of the QCD flux tube. Following Ref.~\cite{Baker:2020}, in Fig.~\ref{Fig3}, left panel we display the
transverse profile of the non-perturbative chromoelectric field for the SU(3) pure gauge theory corresponding to a source distance 
of $R \simeq 0.85$ fm. The transverse profile is consistent with Eq.~(\ref{6.43}) with (solid line in Fig.~\ref{Fig3}):
\begin{equation}
\label{6.46}
v_{\phi}  \; \simeq \;  0.185 \; \; , \; \; \sqrt{gH_0} \; \simeq \; 1.0 \; \text{GeV} \; \; , \; \; \sqrt{\sigma} \; \simeq \; 398.5 \; \text{MeV} \; \; . 
\end{equation}
In Eq.~(\ref{6.46}) the string tension has been evaluated by means of Eq.~(\ref{6.29}). Note that the value of the string tension in 
Eq.~(\ref{6.46}) is smaller with respect to the accepted value Eq.~(\ref{6.13}). This is to be expected for we have already noticed that there is a 
non-perturbative renormalization constant due to the smearing procedure that affects the measured chromoelectric field.
Such a renormalization constant can be easily evaluated from  the ratio of the estimate of the string tension in Eq.~(\ref{6.46})
to the reference value given by Eq.~(\ref{6.13}). \\
We saw that the connected correlators allow to extract the gauge-invariant flux-tube field strength tensor $F_{\mu \nu}$.
In Ref.~\cite{Baker:2020}  from the  field strength tensor it was constructed a stress energy-momentum tensor $T_{\mu \nu}(x)$
having the Maxwell form. In fact, in the Appendix A of Ref.~\cite{Baker:2020} the energy-momentum tensor $T_{\mu \nu}$
was considered as a function of the field strength tensor $F_{\mu \nu}$ characterizing the color flux tube. Assuming that the field strength
tensor points in a single color direction parallel to the color direction of the static sources, it follows that the energy-momentum tensor
lies in the same single color direction and, therefore, it has the Maxwell form~\cite{Baker:2020}:
\begin{equation}
\label{6.47}
T_{\mu \nu}(x)   \; = \;  F_{\mu \alpha}(x)   F_{\alpha \nu}(x) \;  - \; \frac{1}{4} \,    g_{\mu \nu}  \,  F_{\alpha \beta}(x)   F_{\alpha \beta}(x) \; \; .
\end{equation}
\begin{figure}
\begin{center}
\includegraphics[width=0.48\textwidth,clip]{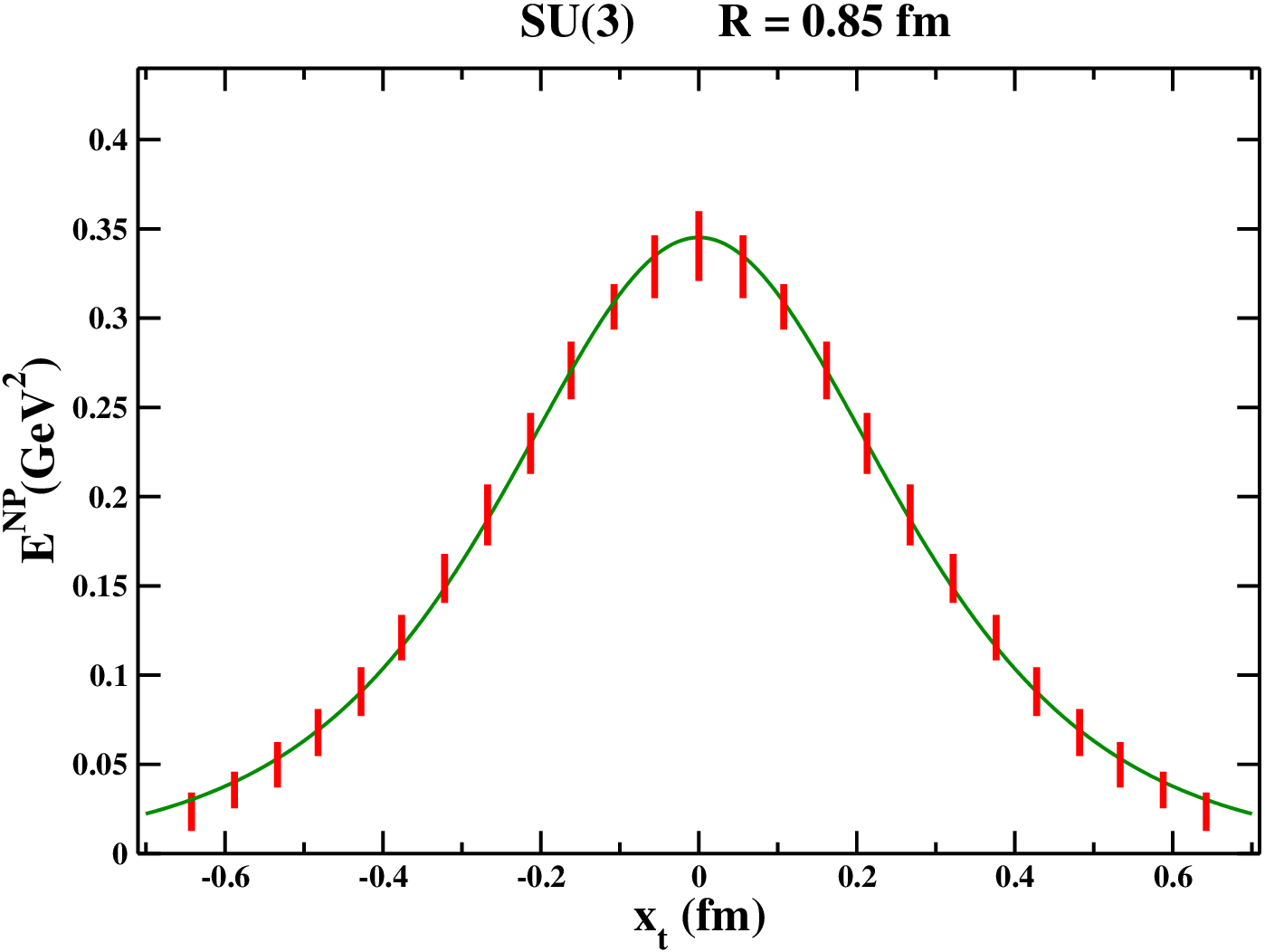}
\; 
\includegraphics[width=0.47\textwidth,clip]{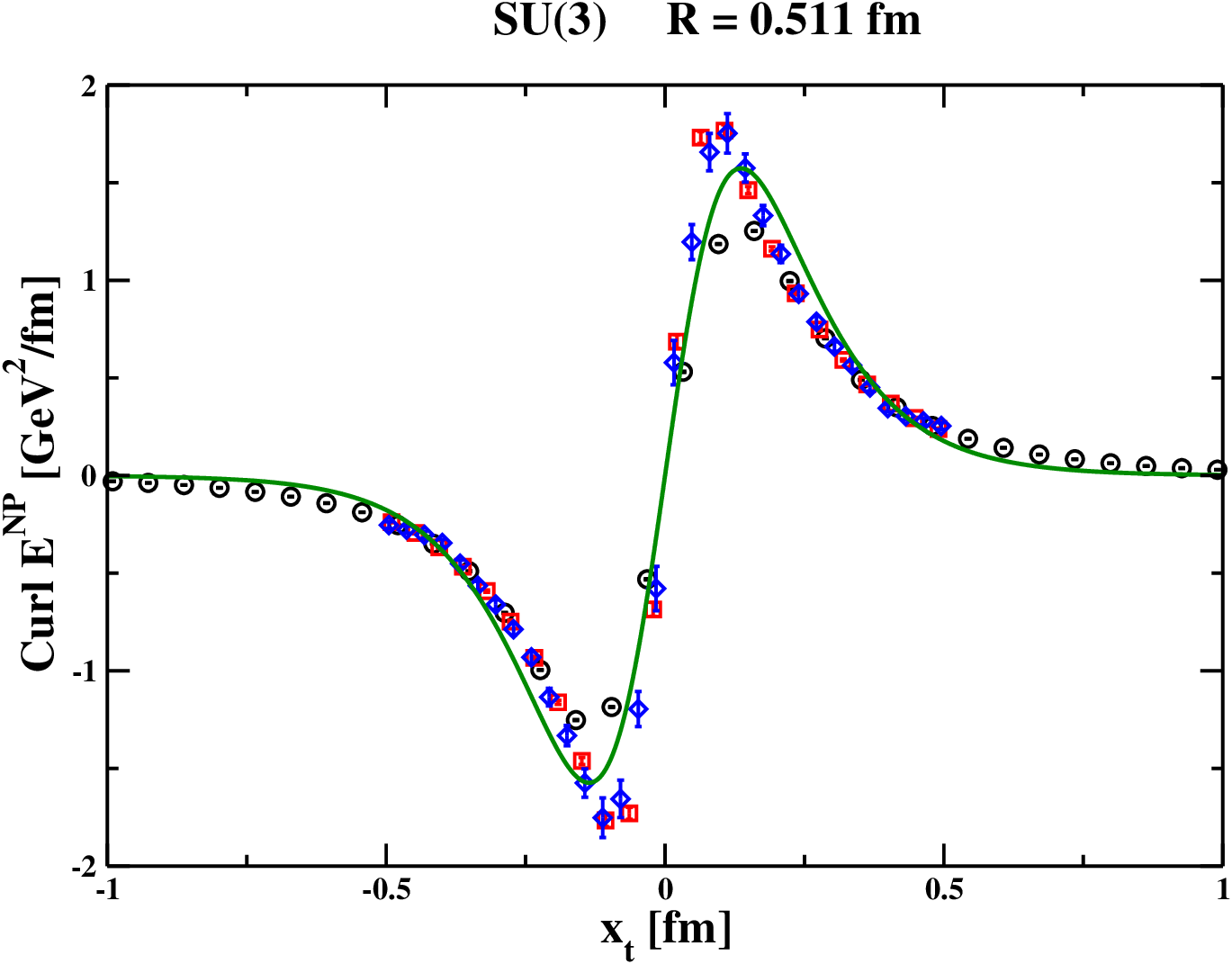} 
\end{center}
\caption{\label{Fig3} 
(Color online) Left panel. Transverse distribution of the non-perturbative flux-tube chromoelectric field. The vertical bars
mark the values of the chromoelectric field as the longitudinal coordinate varies along the flux tube. The data have been
taken from Fig.~5 in Ref.~\cite{Baker:2020}. The solid line is the fit to the data using Eq.~(\ref{6.46}).
Right panel. The transverse distribution of the rotational of the non-perturbative chromoelectric field for three different
valuues of the gauge coupling, $\beta = 6.240$ (circles), $\beta = 6.544$ (squares) and $\beta = 6.769 $ (diamonds).
The data have been taken from Fig.~5 in Ref.~\cite{Baker:2022}. The solid line is the chromomagnetic current given by 
Eq,~(\ref{6.49}) by assuming $\sqrt{gH_0} \simeq 1.5$ GeV. }
\end{figure}
However, we have shown that the field strength tensor is mainly composed by the Abelian color directions $a = 3, 8$. Nevertheless, the
Abelian nature of the flux-tube field strength tensor fully justify the Maxwell construction proposed in Ref.~\cite{Baker:2020}.
More recently, in Ref.~\cite{Baker:2022} it was presented for the first time the evidence of a solenoidal chromomagnetic current 
responsible for the formation of the longitudinal chromoelectric field. The chromomagnetic current density was defined by the
Ampere law:
\begin{equation}
\label{6.48}
\vec{J}_{mag}(\vec{x})   \; =   \;  \vec{\nabla} \; \times \; \vec{E}(\vec{x})    \; =  \;   \vec{\nabla} \; \times \vec{E}^{NP}(\vec{x})   \; ,
\end{equation}
where we used $ \vec{\nabla} \, \times \, \vec{E}^C(\vec{x})  = 0$. Comparing Eq.~(\ref{6.48}) with our Eq.~(\ref{6.21}) we
infer that:
\begin{equation}
\label{6.49}
\vec{J}_{mag}(\vec{x})   \; =   \; - \left [  \vec{J}^3_{M}(\vec{x}) +  \vec{J}^8_{M}(\vec{x})  \right ]   
\end{equation}
with $ \vec{J}^{3,8}_{M}(\vec{x})$ given by Eqs.~(\ref{6.18}), (\ref{6.19}) and (\ref{6.20}).
In Fig.~\ref{Fig3}, right panel, we report the chromomagnetic current, Eq.~(\ref{6.48}), for three different values of the gauge coupling evaluated at the transverse plane at a distance $\frac{R}{4}$ from the source. The data have been reproduced from Fig.~5 in Ref.~\cite{Baker:2022}.
Since the distance between the static sources is rather small ($R \simeq 0.511$ fm) the data  seem to display some noticeable scattering
 probably due to systematic effects arising from the subtraction of the Coulomb field from the measured flux-tube chromoelectric field.
 Nevertheless, our theoretical chromomagnetic current seem to track reasonably well the lattice data. \\
In our opinion, the essential Abelian nature of the chromomagnetic currents and the ensuing flux-tube chromoelectric fields are at
the heart of the so-called Abelian dominance observed in several lattice simulations~\cite{Ripka:2004,Greensite:2011,Kondo:2015}.
Indeed, the dual superconductivity scenario is realized by adopting a gauge-fixing procedure analogous to the unitary gauge where
a suitable matrix-valued operator is diagonalized leaving unfixed the maximal Abelian subgroup U(1)$\times$U(1) whose
generators belong to the Cartan subalgebra of the SU(3) gauge group. In these gauges the given gauge theory can be thought of as
an essentially Abelian gauge theory. Our previous results should make evident that the observed Abelian dominance is not
the cause but it is a consequence of the structure of the confining quantum vacuum. \\
It is interesting to  also check if our theoretical transverse shape of the flux-tube chromoelectric field is consistent with numerical
studies in QCD with maximal Abelian gauge fixing.
\begin{figure}
\begin{center}
\includegraphics[width=0.50\textwidth,clip]{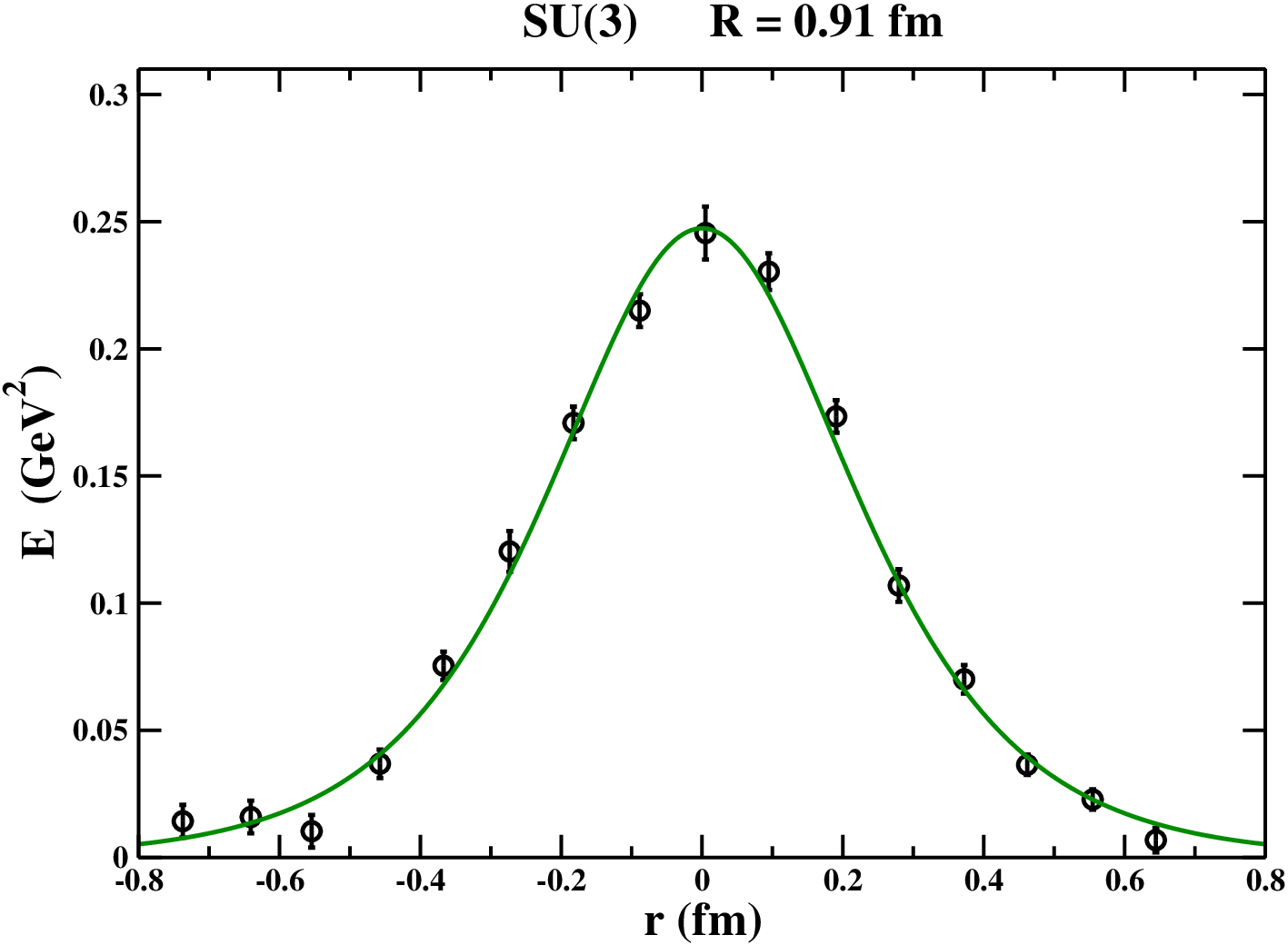}
\; \; \; \;
\includegraphics[width=0.35\textwidth,clip]{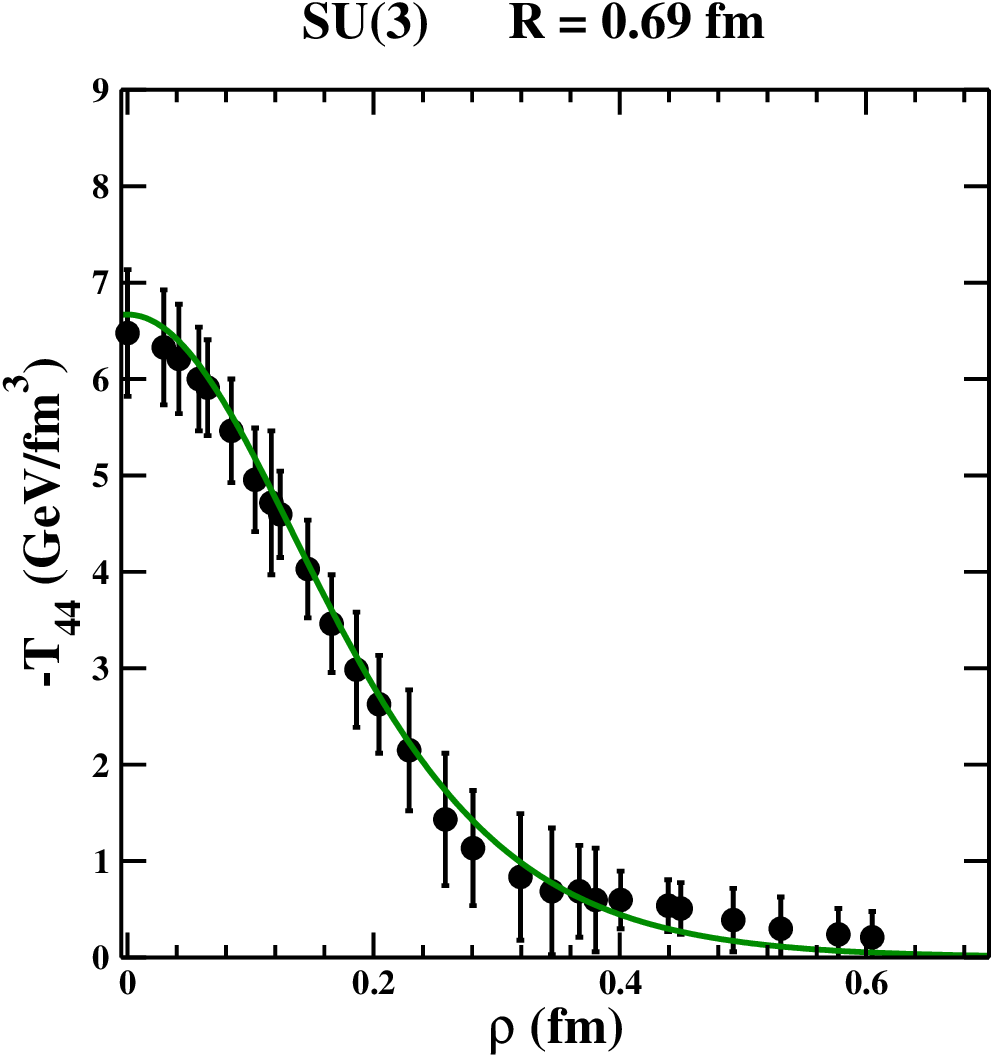} 
\end{center}
\caption{\label{Fig4} 
(Color online) Left panel.  Transverse distribution of the Abelian chromoelectric field generated by a static quark-antiquark
pair at distance $R \simeq 0.91$ fm in quenched QCD in the maximal Abelian gauge. The data have been taken form Fig.~15
in Ref.~\cite{Bornyakov:2004}. The continuous  line is  Eq.~(\ref{6.43}) with   parameters given by Eq.~(\ref{6.50}).
Right panel. The transverse distribution at the mid transverse plane of the energy density around a static quark-antiquark
pair. The data have been extracted from Fig.~3, panel (b), in Ref.~\cite{Yanagihara:2019}. The continuous line corresponds to
Eq.~(\ref{6.51}) with parameters given by  Eq.~(\ref{6.52}). }
\end{figure}
As a matter of fact,  in Fig.~\ref{Fig4}, left panel, we report the profile of the Abelian flux tube chromoelectric field (in physical units)
in quenched QCD in the maximal Abelian gauge as displayed in Fig.~15 of Ref.~\cite{Bornyakov:2004} together with our theoretical transverse 
profile, Eq.~(\ref{6.43}) with:
\begin{equation}
\label{6.50}
v_{\phi}  \; \simeq \;  0.102 \; \; , \; \; \sqrt{gH_0} \; \simeq \; 1.14 \; \text{GeV} \; \; . 
\end{equation}
We can see that, also in this case, our theoretical expectations are able to track quite well the lattice data. \\
As the last check, we focus, now, on several numerical studies of the pure gauge SU(3) theory on the the structure of the static quark
flux tube implemented with correlators between Wilson loops and plaquettes. In this case, in general, the lattice observables
give informations on the square of the flux-tube field strength tensor. For definitiveness, we have considered the quite recent
studies on the spatial distribution of the stress energy-momentum tensor  around a static quark-antiquark color sources
reported in Ref.~\cite{Yanagihara:2019}. The authors of Ref.~\cite{Yanagihara:2019} investigated the spatial distribution of the stress
energy-momentum tensor $T_{\mu \nu}$ around a static quark-antiquark pair in the SU(3) pure gauge theory. To smooth the
gauge field configurations it was employed the Yang-Mills gradient flow (for a recent review see, eg, Ref.~\cite{Schindler:2022}
and references therein). The spatial distribution of the static color source energy-momentum tensor was obtained by measuring
the correlators of the conserved renormalized stress tensor~\cite{Suzuki:2013} and a Wilson loop. It is worth mentioning that in 
Appendix A of Ref.~\cite{Baker:2020} it was shown that the Maxwell energy-momentum tensor built from the field strength tensor
characterizing the SU(3) flux tube resulted  to be in satisfying agreements with the results presented in  Ref.~\cite{Yanagihara:2019}. 
For a more quantitative comparison we looked at the transverse distribution of the flux-tube energy density ${\cal{E}}(x) = - \, T_{44}(x)$
measured at the mid transverse plane and displayed in Fig.~3 of Ref.~\cite{Yanagihara:2019}.
According to our results we should have:
\begin{equation}
\label{6.51}
- \,  T_{4 4} \; \simeq \; \frac{1}{2} \; \left \{  [  gE^3(r) ]^2 \;+ \;  [  gE^8(r)  ]^2   \right \} \;  
\end{equation}
where $r$ is the transverse distance in the cylindrical coordinate system. The fit of Eq.~(\ref{6.51}) to the lattice data returned (see
Fig.~\ref{Fig4}, right panel):
\begin{equation}
\label{6.52}
v_{\phi}  \; \simeq \;  0.20 \; \; , \; \; \sqrt{gH_0} \; \simeq \; 1.10 \; \text{GeV} \; \; , \; \; \sqrt{\sigma} \; \simeq \; 474 \; \text{MeV} \; .
\end{equation}
Again, we see that the lattice data are in quite good agreement with theoretical expectations. Furthermore, we note that the estimate of
 the string tension in Eq.~(\ref{6.52}) is in accordance with the large $R$ behaviour of the quark-antiquark force, displayed 
 in Fig.~4 of Ref.~\cite{Yanagihara:2019}, as directly determined from the Wilson loops and the renormalized energy-momentum
 tensor as well as with the string tension reference value Eq.~(\ref{6.13}), signalling that the renormalized lattice energy-momentum
 tensor does not need further non-perturbative renormalization since it is expected  to have a smooth continuum limit~\cite{Suzuki:2013}. \\
 To summarize, we have shown that our picture for the formation of a chromoelectric squeezed flux tube generated by a static quark-antiquark
 pair in the SU(3) pure gauge theory turned out to be in reasonable agreement with lattice outcomes from different collaborations.
 Our previous discussion leads to the following estimate for the strength of the vacuum chromomagnetic condensate in the pure gauge
 SU(3) theory: 
\begin{equation}
\label{6.53}
 \sqrt{gH_0} \; \simeq \; 1.0 \; - \; 1.1 \;  \text{GeV} \; \;  \; \;  \; \;  \text{SU(3)} \; .
\end{equation}
From this last equation we may estimate the flux-tube transverse radius, Eq.~(\ref{6.32}), $R_T \, \simeq \, 0.2$ fm, or the width,
Eq.~(\ref{6.37}), $w \, \simeq  \, 0.4$ fm in agreement with several lattice determinations. \\
Up to now we have  dealt with only the SU(3) gluon fields and completely neglected the dynamical role of the fermion fields. Although this
point will be addressed in more details in the following Section, it is, nevertheless, worthwhile to ascertain how dynamical quarks affect
the structure of the chromoelectric fields generated by a static quark-antiquark pair. Indeed, we looked at lattice studies in full QCD on
the flux-tube color fields and in Fig.~\ref{Fig5} we present the transverse distribution of the longitudinal flux-tube chromoelectric
field as obtained  by three different lattice group.
\begin{figure}
\begin{center}
\includegraphics[width=0.26\textwidth,clip]{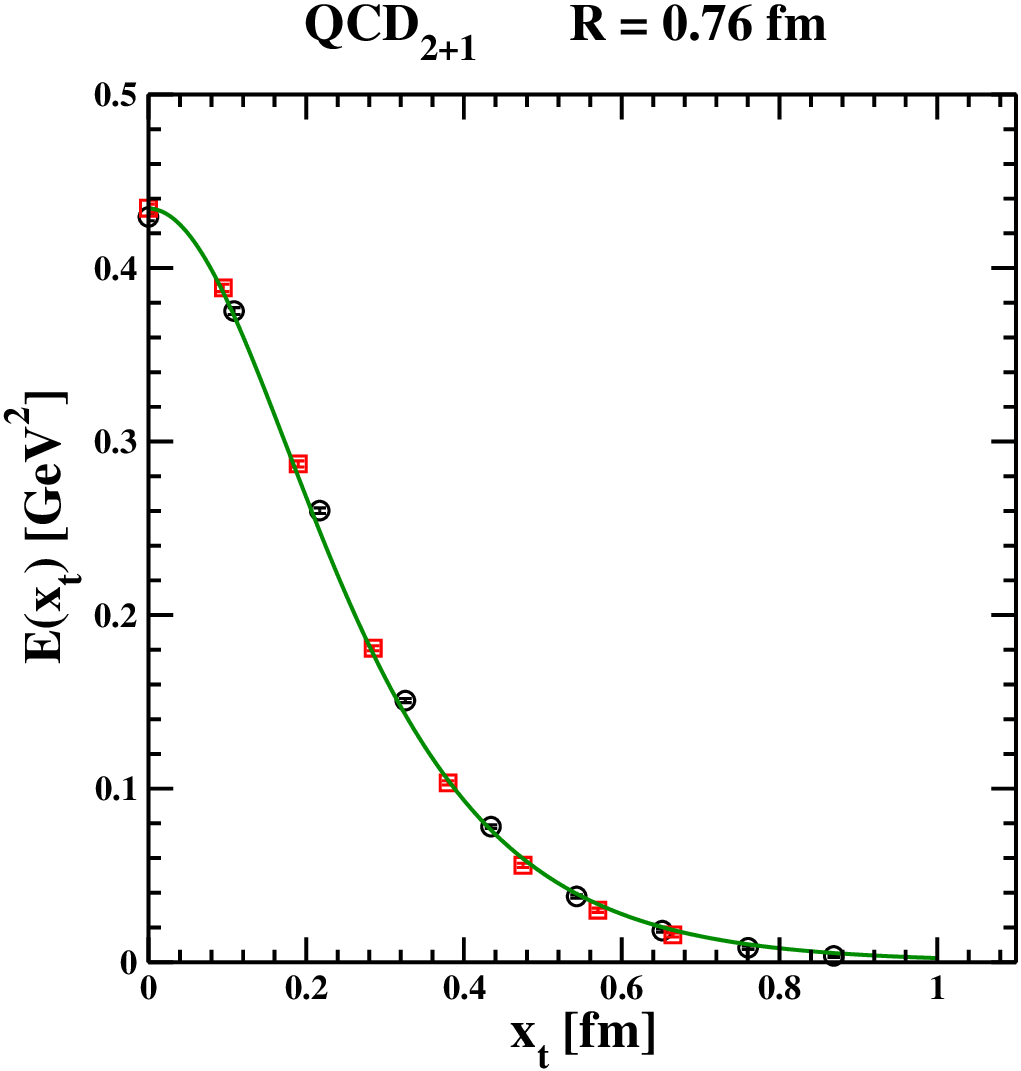}
\includegraphics[width=0.35\textwidth,clip]{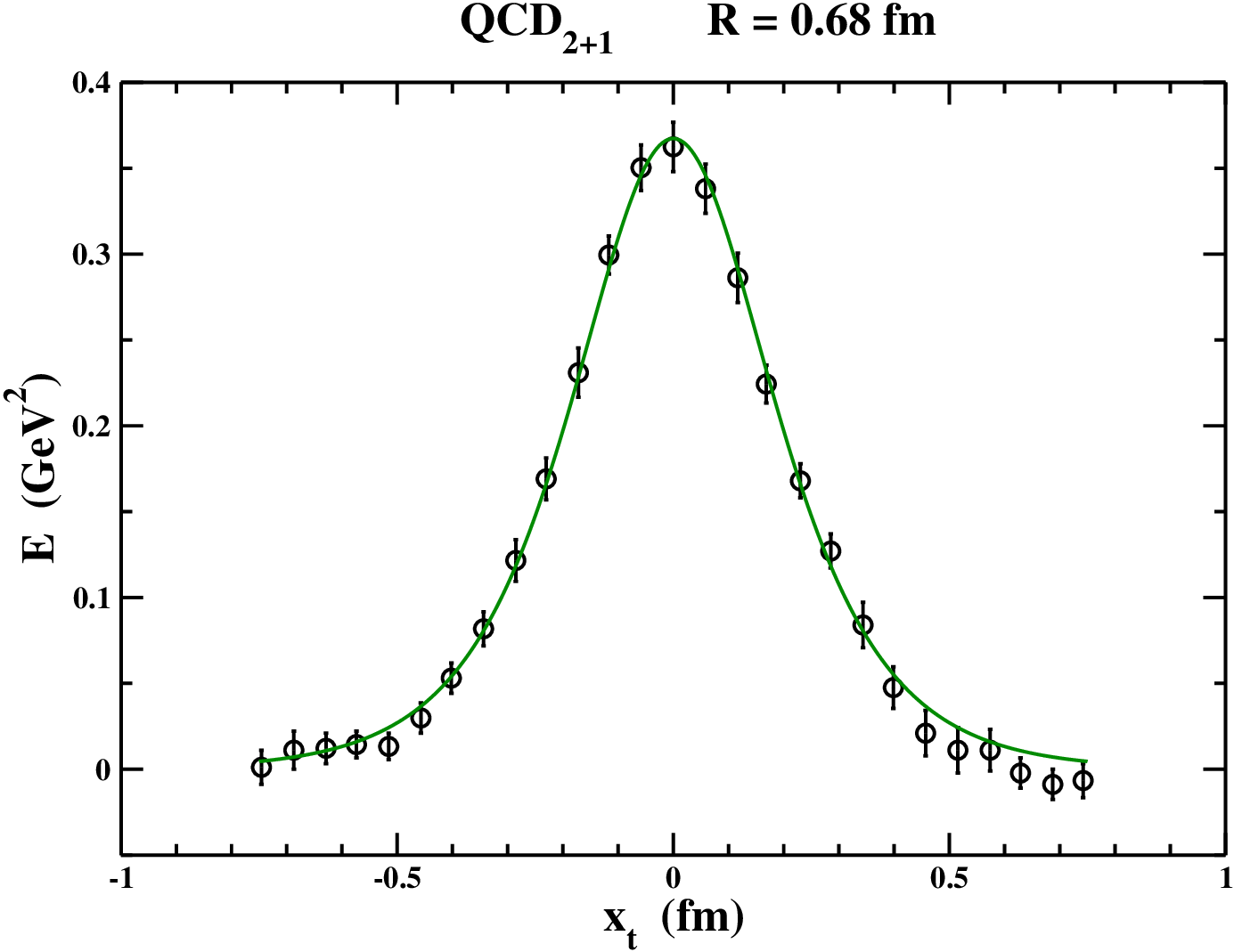} 
\includegraphics[width=0.35\textwidth,clip]{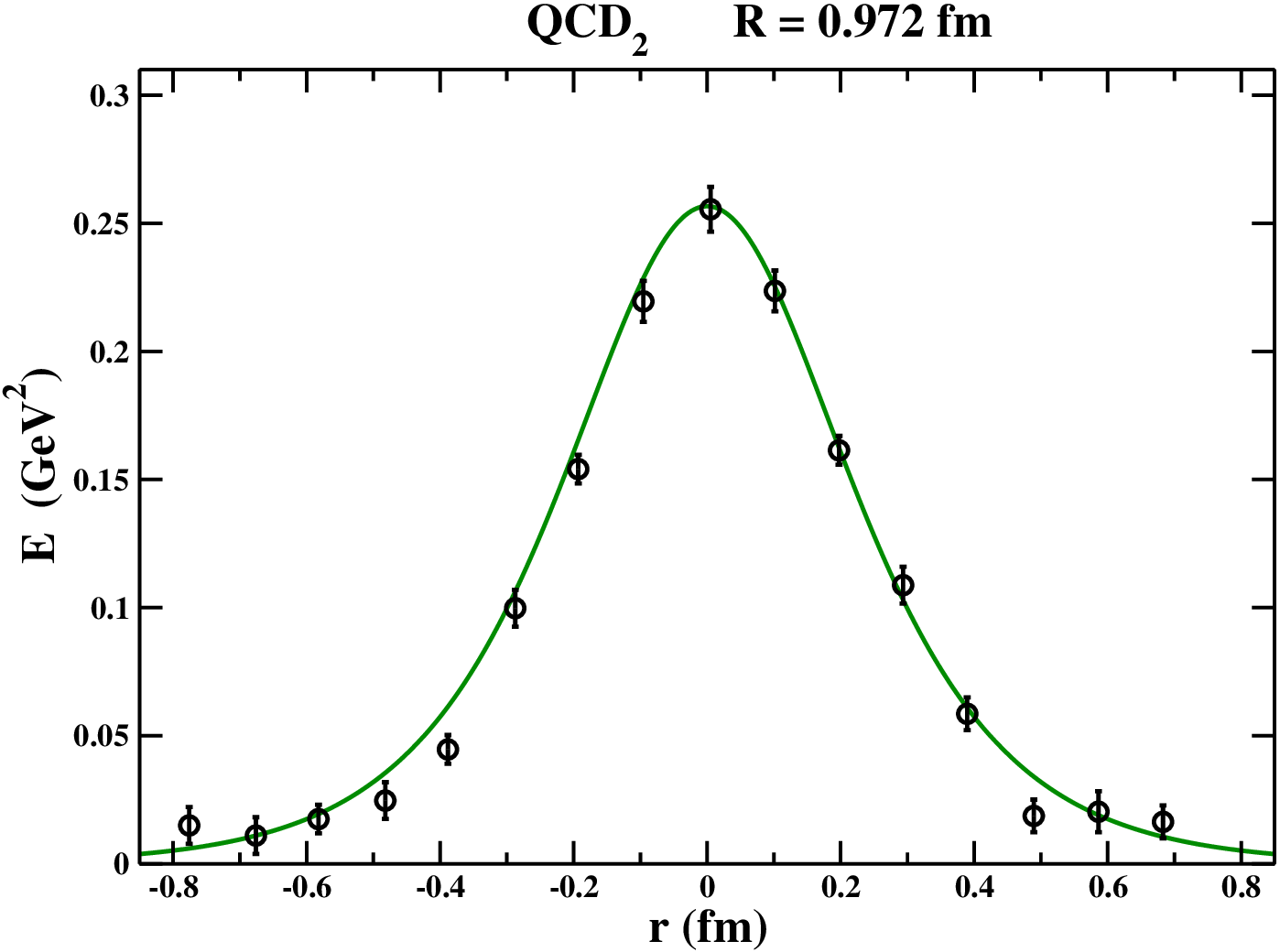} 
\end{center}
\caption{\label{Fig5} 
(Color online) The transverse shape of the flux-tube longitudinal chromoelectric field in full QCD. The continuous lines are qualitative fits
of the lattice data to our Eq.~(\ref{6.43}). }
\end{figure} 
Firstly, in Fig.~\ref{Fig5}, left panel, it is displayed  the behaviour of the longitudinal chromoelectric field versus the transverse distance
as obtained from two different values of the gauge coupling, $\beta = 6.743$ (circles) and $\beta = 6.885$ (squares), for (2+1)-flavours
QCD with HISQ/Tree action~\cite{Cea:2017}. The data was obtained on the line of constant physics by adjusting the gauge coupling and 
the bare quark  masses so as to keep the strange quark mass fixed at the physical point and the light quark masses corresponding
to a pion mass $m_{\pi} \simeq 160$ MeV. The data  have been extracted from Fig.~3, right panel, in Ref.~\cite{Cea:2017}. The continuous line
corresponds to Eq.~(\ref{6.43}) with:
\begin{equation}
\label{6.54}
v_{\phi}  \; \simeq \;  0.17 \; \; , \; \; \sqrt{gH_0} \; \simeq \; 1.17 \; \text{GeV} \; \;  .
\end{equation}
In Fig.~\ref{Fig5}, middle panel,  we report the transverse profile of the flux-tube chromoelectric field reported in Fig.~14 of Ref.~\cite{DElia:2021},
where the flux tube properties were investigated by numerical lattice simulations for full QCD with (2+1)-flavours of stoud-improved staggered
fermions with physical quark masses (pion mass $m_{\pi} \simeq 135$ MeV). In this case we found (continuous line in Fig.~\ref{Fig5}):
\begin{equation}
\label{6.55}
v_{\phi}  \; \simeq \;  0.108 \; \; , \; \; \sqrt{gH_0} \; \simeq \; 1.35 \; \text{GeV} \; \;  .
\end{equation}
Finally, in Fig.~\ref{Fig5}, right panel, we consider the transverse distribution of the flux-tube chromoelectric field for QCD in the maximal 
Abelian gauge with two degenerate non-perturbatively improved Wilson fermions with a rather heavy mass 
$m_q \simeq 100$ MeV~\cite{Bornyakov:2004}. Fitting the lattice data to our Eq.~(\ref{6.43}) we get:
\begin{equation}
\label{6.56}
v_{\phi}  \; \simeq \;  0.104 \; \; , \; \; \sqrt{gH_0} \; \simeq \; 1.15 \; \text{GeV} \; \;  .
\end{equation}
It is remarkable that, in accordance with common expectations, the inclusion of dynamical quarks does not substantially modify the color
structure of the flux tube. The only effect seems to be a small increase of the chromomagnetic condensate strength:
\begin{equation}
\label{6.57}
 \sqrt{gH_0} \; \simeq \; 1.2 \; - \; 1.3 \;  \text{GeV} \; \;  \; \;  \; \;  \text{QCD} \; .
\end{equation}
Our quantum vacuum functional leads us to reach a clear picture on the flux-tube physics. Indeed, the squeezing of the chromoelectric fields
generated by a static quark-antiquark pair is due to chromomagnetic currents belonging to the maximal Abelian subgroup of SU(3) whose
transverse distribution only depends  on the strength of the vacuum chromomagnetic condensate. It is remarkable that the resulting flux-tube
chromoelectric fields are obtained by solving the quite simple Ampere law, Eq.~(\ref{6.21}). Evidently, the qualitative
agreement of our theoretical expectations to several lattice measurements leads us to believe that we are on the right path. Undoubtedly, 
the knowledge of the color structure and distribution of the flux-tube has several interesting phenomenological consequences.
Here we restrict ourself to the comparison with the seminal paper by Casher, Neuberger and Nussinov~\cite{Casher:1979} that lies at the basis
of several high-energy physics Monte Carlo codes.
In Ref.~\cite{Casher:1979} quark confinement is assumed to be generated by the formation of chromoelectric flux tubes with almost
uniform energy density. By employing the approximations where the chromoelectric field which develops between the quarks is assumed
to be an almost classical field, the authors of Ref.~\cite{Casher:1979} found that the flux-tube chromoelectric field is pratically longitudinal.
Moreover, to estimate the production of quark pairs in the flux tubes it is assumed that the longitudinal chromoelectric field is  Abelian:
\begin{equation}
\label{6.58}
 g E^a  \;  \simeq \; \delta^{a 3} \, g  E^3 \; + \;  \delta^{a 8} \, g E^8 \; .
\end{equation}
Using the Gauss law and:
\begin{equation}
\label{6.59}
 \frac{1}{2} \; A_T \; \sum_{a=1}^{8} \left ( E^a \right )^2 \; \simeq \sigma \;  
\end{equation}
where $A_T \simeq \pi R_T^2$ is the cross sectional area, one gets:
\begin{equation}
\label{6.60}
 \frac{1}{2} \; g E^3  \;  \simeq \; \frac{3}{2} \; \sigma \; \; , \; \;   g E^8 \; \simeq \; \frac{1}{\sqrt{3}} \;  g E^3 \; .
\end{equation}
Within the above framework, the quark dynamics into the flux tube may be approximately described by the Dirac equation in presence
of a classical almost uniform longitudinal chromoelectric field given by a diagonal matrix in color space~\cite{Casher:1979}:
\begin{equation}
\label{6.61}
\left [ - \; i \, \slashed \partial \; - \; m_{eff} \; - \;  2 \; i \; \sigma \, z \; 
\begin{pmatrix}
   1      &   &  \\
         & - \frac{1}{2} &  \\
    &  &  -  \frac{1}{2}
\end{pmatrix}
\right ] \; \psi \; = 0 \; \; ,
\end{equation}
where $m_{eff}$ is the effective quark mass and $z$ is our longitudinal coordinate $x_{\ell}$. After that, the dynamical process of quark pair
creation may be described rather well by the well-known Schwinger mechanism~\cite{Schwinger:1951,Brezin:1970}. \\
It is  remarkable that within our approach we have confirmed the Abelian structure of the flux-tube chromoelectric fields. More precisely,
the chromoelectric fields that confine the static quark-antiquark pair is longitudinal and it is almost completely formed by the $a = 3,8$ color
directions. However, if we assume an  uniform flux-tube, instead of Eq.~(\ref{6.60}) we obtain:
\begin{equation}
\label{6.62}
 g E^8(x_t = 0) \; \simeq \; \frac{2}{\sqrt{3}} \;  g E^3(x_t = 0) \; ,
\end{equation}
and 
\begin{equation}
\label{6.63}
g E^3(x_t = 0) \;  \simeq \; v_{\phi} \, g H_0 \; \simeq \; \frac{\sigma}{5 \pi J_1 v_{\phi}} \; \simeq \; 1.08 \, \sigma \; ,
\end{equation}
where we used Eq.~(\ref{6.29}) and assumed $v_{\phi} \simeq 0.2$. As a consequence we are led to the following Dirac equation:
\begin{equation}
\label{6.64}
\left [ - \; i \, \slashed \partial \; - \; m_{eff} \; - \;  1.08  \; i \; \sigma \, z \; 
\begin{pmatrix}
   1      &   &  \\
         & - \frac{1}{3} &  \\
    &  &  -  \frac{2}{3}
\end{pmatrix}
\right ] \; \psi \; = 0 \; \; .
\end{equation}
In addition, it is interesting to note that we may go beyond the approximations adopted in Ref.~\cite{Casher:1979} for we have an explicit expression
for the transverse distribution of the flux-tube chromoelectric fields. In fact, the transverse profile of the chromoelectric fields allows, in
principle, to determine the transverse distribution of the quark-antiquark pairs produced by the Schwinger mechanism ( for a good
account see, eg, Ref.~\cite{Wong:1994} and references therein),  however  such matter goes beyond the aim of the present paper. \\
Let us conclude this rather lengthy  Section by discussing the remarkable color Meissner effect~\cite{Cea:2003,Cea:2005,Cea:2006,Cea:2007}
that led us to picture the QCD vacuum like a chromomagnetic condensate. In order to investigate non-perturbatively the quantum vacuum
structure in Refs.~\cite{Cea:1997a,Cea:1997b,Cea:1997c}, by means of the so-called Schr\"odinger functional,  it was introduced a 
gauge-invariant effective action for gauge systems in external static background fields. In particular, it was enlightened that the deconfinement
temperature depends on the strength of an external Abelian chromomagnetic field in both  the pure SU(3) gauge 
theory~\cite{Cea:2003,Cea:2005,Cea:2006} and QCD with two degenerate staggered dynamical quarks~\cite{Cea:2007}.
Actually, it was ascertained that the deconfinement temperature decreases when the strength of the applied field is increased and eventually
goes to zero. 
Indeed, in Fig.~\ref{Fig6} we display the critical temperature of deconfinement versus the strength of the applied chromomagnetic field
for SU(3) (left panel) and QCD (right panel). For the pure gauge theory the data have been taken from Fig.~4 in Ref.~\cite{Cea:2005}
taking into account that $T_c \, = \, 300(7)$ MeV~\cite{Teper:1998}, while for QCD with two dynamical staggered quarks with a rather
large mass (pion mass $m_{\pi}$ around 500 MeV) the data have been extracted from Fig.~7 of Ref.~\cite{Cea:2007} assuming
$T_c~ \simeq~170$~MeV. From Fig.~\ref{Fig6} we infer that in both cases the deconfinement critical temperature decreases
linearly with $\sqrt{gH}$. In fact, the lattice data can be fitted to Eq.~(\ref{1.1}) giving  (see the continuous lines in Fig.~\ref{Fig6}):
\begin{figure}
\begin{center}
\includegraphics[width=0.45\textwidth,clip]{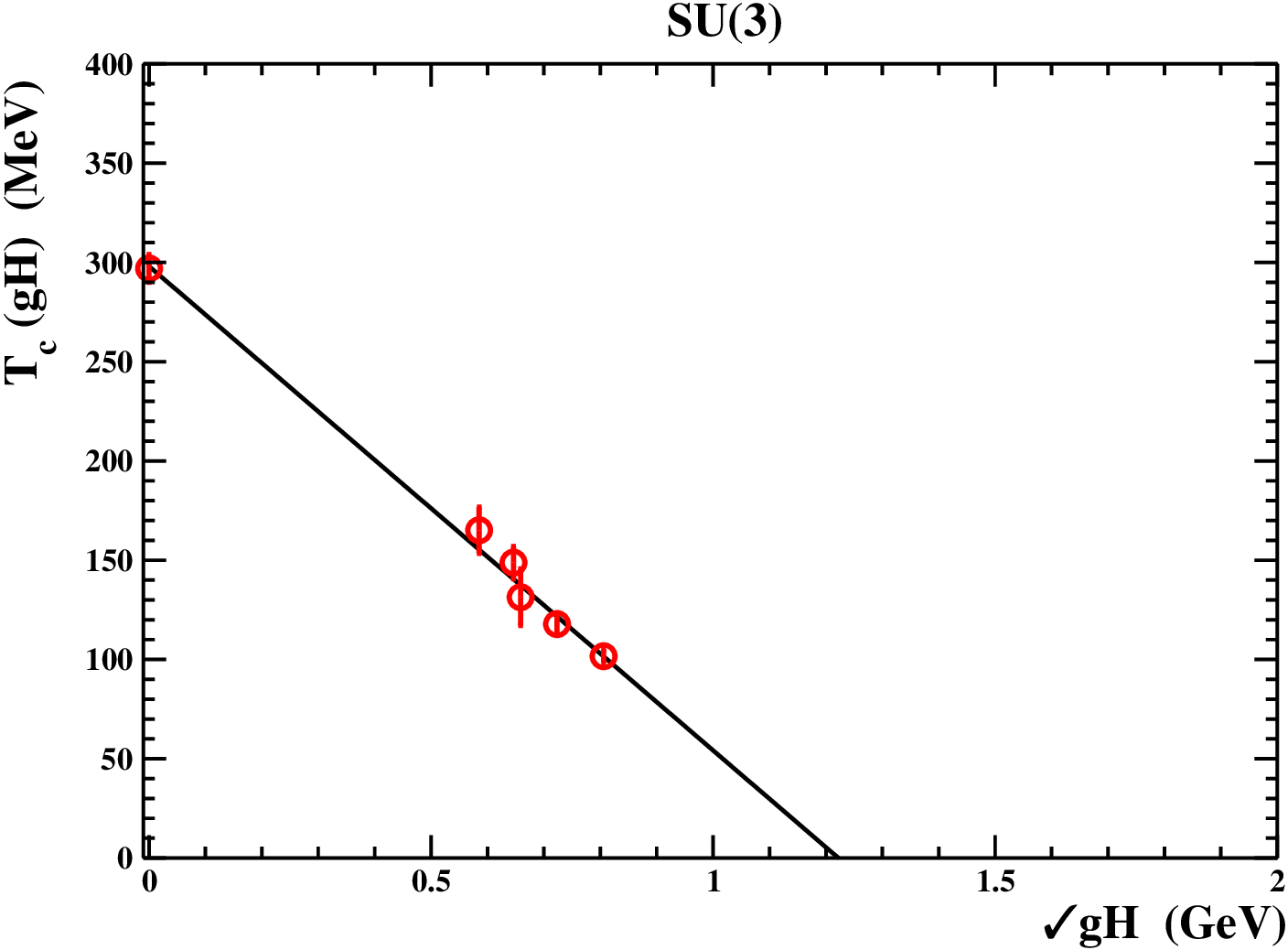}
\; \; 
\includegraphics[width=0.45\textwidth,clip]{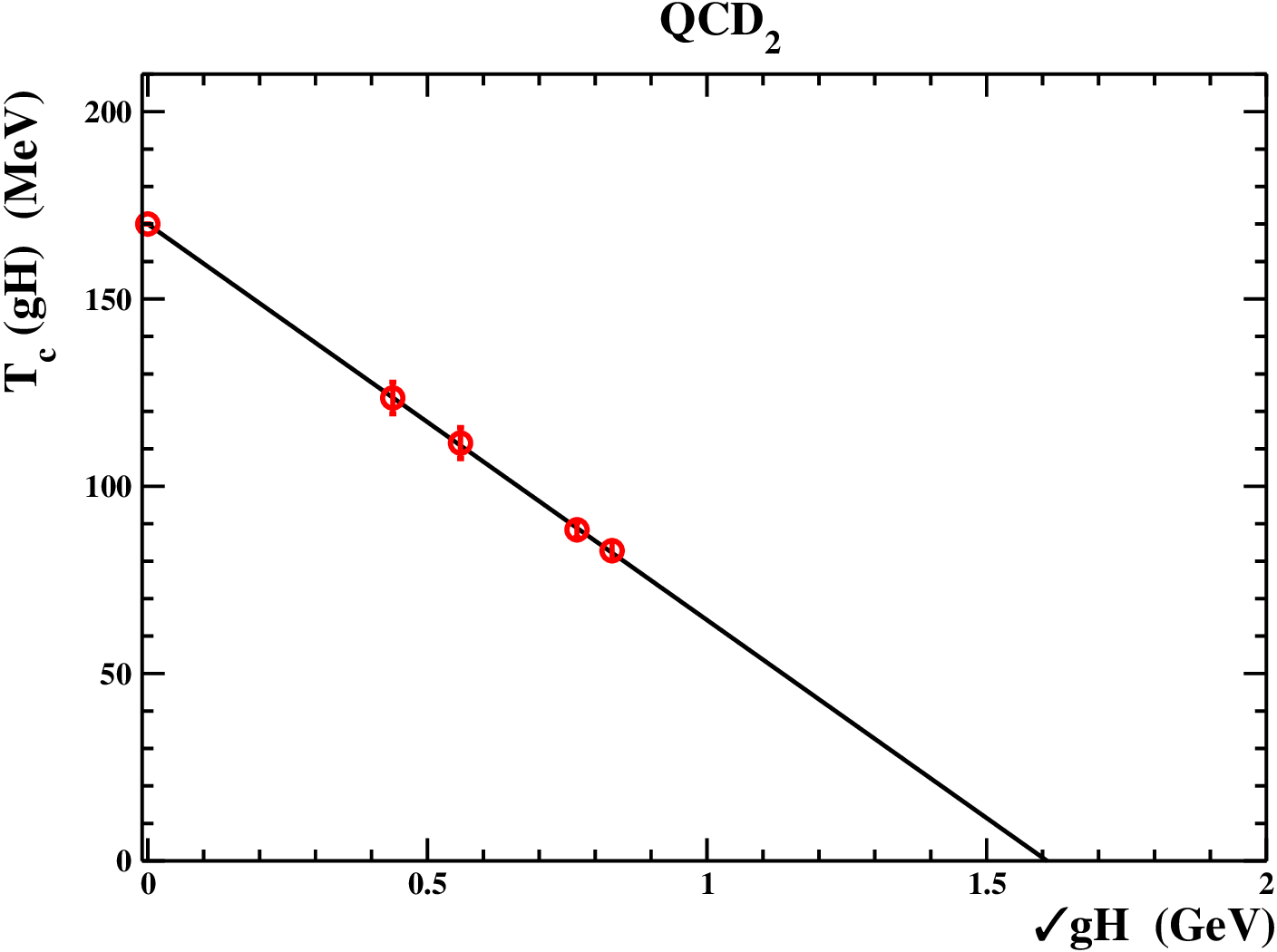} 
\end{center}
\caption{\label{Fig6} 
(Color online) Behaviour of the deconfinement critical temperature versus the strength of the external chromomagnetic firld
for SU(3) pure gauge theory (left panel) and  QCD with two degenerate staggered quarks  (right panel). The continuous lines  are
the best fits of the lattice data to Eq.~(\ref{1.1}).}
\end{figure}
\begin{equation}
\label{6.65}
T_c(0) \; = \;  298(10) \; \text{MeV} \; \; ,  \; \; \sqrt{gH_c} \; = \; 1.22(4)   \;  \text{GeV} \; \;  \; \;  \; \;  \text{SU(3)} \; ,
\end{equation}
\begin{equation}
\label{6.66}
T_c(0) \; = \;  170(12) \; \text{MeV} \; \; , \; \;  \sqrt{gH_c} \; = \; 1.61(14)   \;  \text{GeV} \; \;  \; \;  \; \;  \text{QCD} \; .
\end{equation}
From a dynamical point of view these peculiar behaviours of the critical temperatures point to the prevalent role of the chromomagnetic length. 
Actually, we may now  try to interpret the color Meissner effect within our picture of the confining vacuum wavefunctional. Firstly, we note
that the gauge systems was probed with external fields whose strengths $\sqrt{gH}$   was smaller than $\sqrt{gH_0}$, 
Eqs.~(\ref{6.53}) and (\ref{6.57}). So that it is conceivable that the main effects of the applied chromomagnetic fields are to polarize the
chromomagnetic domains such that the vacuum chromomagnetic condensate tends to lie in the same direction as the external
field. As a consequence, we see that the polarization effects cause  an increase of the effective domain size leading to a decrease of
the mass gap $ \Delta \sim  \frac{1}{L_D}$. Since on dimensional ground the deconfinement temperature $T_c$ is of order
of the mass gap, we may explain the observed decreases of the critical temperature $T_c(gH)$. If this interpretation is correct,
then at the critical strength $\sqrt{gH_c} \sim \sqrt{gH_0}$ the gauge system undergoes a quantum deconfinement phase transition.
However, the deconfined phase is far from being similar to a weakly interacting gas of quarks and gluons, but it should resemble 
the quantum Hall liquid of the condensed matter physics. A clear signature of this would  come from the measurements of the chromoelectric
fields generated by a static quark-antiquark pair immersed in an uniform chromomagnetic field. Indeed, at $T=0$ and $\sqrt{gH} > \sqrt{gH_c}$
we lost  confinement due to the absence of a mass gap and the instauration of long-range color correlations.  However, since
the quantum vacuum is characterized by polarized  chromomagnetic domains we expect that the formation of a squeezed flux tube with
a non-zero string tension when the  static-source joining line is perpendicular to the external field, $\sqrt{\sigma_T} \simeq \sqrt{\sigma}$.
On the other hand, for a static quark-antiquark pair in the direction parallel to the external field the formation of the flux tube should be
strongly suppressed leading to $\sqrt{\sigma_L} \simeq 0$. Unfortunately, in the literature there are no lattice data that could support the above 
scenario. However, quite recently such an investigation has been presented in Ref.~\cite{DElia:2021} for full QCD with (2+1)-flavour at
the physical point immersed in extremely strong magnetic fields. Actually, the influence of external magnetic background fields on QCD
has attracted growing interest in recent years (see, for instance, Refs.~\cite{Kharzeev:2013,Miransky:2015,Andersen:2016,Hattori:2023}
and references therein). \\
Quantum chromodynamics with background magnetic fields can be studied directly by means of non-perturbative lattice simulations.
Continuum extrapolated results employing improved staggered quarks with physical masses have been used to map out the
QCD phase diagram up to field strength $eB = 9 \,  \text{GeV}^2$. It turned out that the applied magnetic field increases the light quark
condensate in the confined phase. This enhancement has been called magnetic catalysis. On the contrary, in the transition region
from the confined phase to the deconfined one the light quark condensate seems to decrease leading to the so-called inverse magnetic
catalysis. It is important to stress that the chiral and deconfinement critical temperatures alway coincide as happened in absence
of the external magnetic fields. Moreover, as a result of the non-monotonous dependence of the light quark condensate on the
 magnetic field strength and temperature, $T_c(eB)$ is reduced by the magnetic field. \\
 It is widely believed that the mechanisms behind the magnetic catalysis are quite transparent and these can be understood in terms
 of the dimensional reduction of the gauge system and the high degeneracy of the lowest Landau levels. Unfortunately, theoretical
 predictions are in reasonable agreement with the lattice data only for not too strong magnetic fields, $eB \lesssim 0.3 \; \text{GeV}^2$.
 On the contrary, we feel that there is a more mundane explanation of the magnetic catalysis at least in the confined phase.
 Indeed, one must bring in mind that in the QCD confined phase the physical states are colorless hadrons. Considering that
 an external magnetic field is coupled to the gauge system through the electromagnetic vector current, we may employ the
 old vector meson dominance idea to argue that the electromagnetic current is mainly coupled to vector mesons. Moreover,
 long time ago in Ref.~\cite{Cea:1986} it was proved the  duality relation between quark-antiquark bound states and asymptotically
 free quarks by using relativistic wave functions and couplings to both vector and axial current. Within these approximations we were
 able to evaluate as an external constant magnetic field contributes to the vacuum energy density and, thereby, the zero-temperature
 renormalized chiral condensate as usually measured in numerical lattice simulations:
\begin{equation}
\label{6.67}
\Sigma_{ud}(eB,T) \; = \;  \frac{m_{ud}}{m_{\pi}^4} \; \left \{< \psi \bar{\psi} >(eB,T) \;  - \;  < \psi \bar{\psi} >(0,0)  \right \} \; .
\end{equation}
Interestingly enough, we found that our theoretical estimate was in rather good agreement with the lattice 
data~\cite{Bali:2012a,Endrodi:2015,DElia:2021} for the average chiral condensate $\frac{1}{2}~[\Sigma_{u}(eB,0) ~+~\Sigma_{d}(eB,0)]$
in the whole range of applied magnetic fields $ eB \, \le \, 9 \, \text{GeV}^2$. Moreover, we also checked that our theoretical results
were able to track quite well the behaviour of  $\frac{1}{2}~[\Sigma_{u}(eB,0) ~-~\Sigma_{d}(eB,0)]$ versus $eB$ as reported in 
Ref.~\cite{Bali:2012a}. We have, also, evaluated the thermal corrections and found satisfying agreement 
with the continuum extrapolated lattice results~~\cite{Bali:2012a} in the confined phase. 
On the other hand, it should be evident that to understand the inverse magnetic catalysis
it is necessary to unravel the nature of the deconfined QCD vacuum in presence of external magnetic fields. To do this, it is mandatory to try to 
explain the behaviour of the critical deconfinement temperature as a function of the strength of the external magnetic field.
\begin{figure}
\begin{center}
\includegraphics[width=0.7\textwidth,clip]{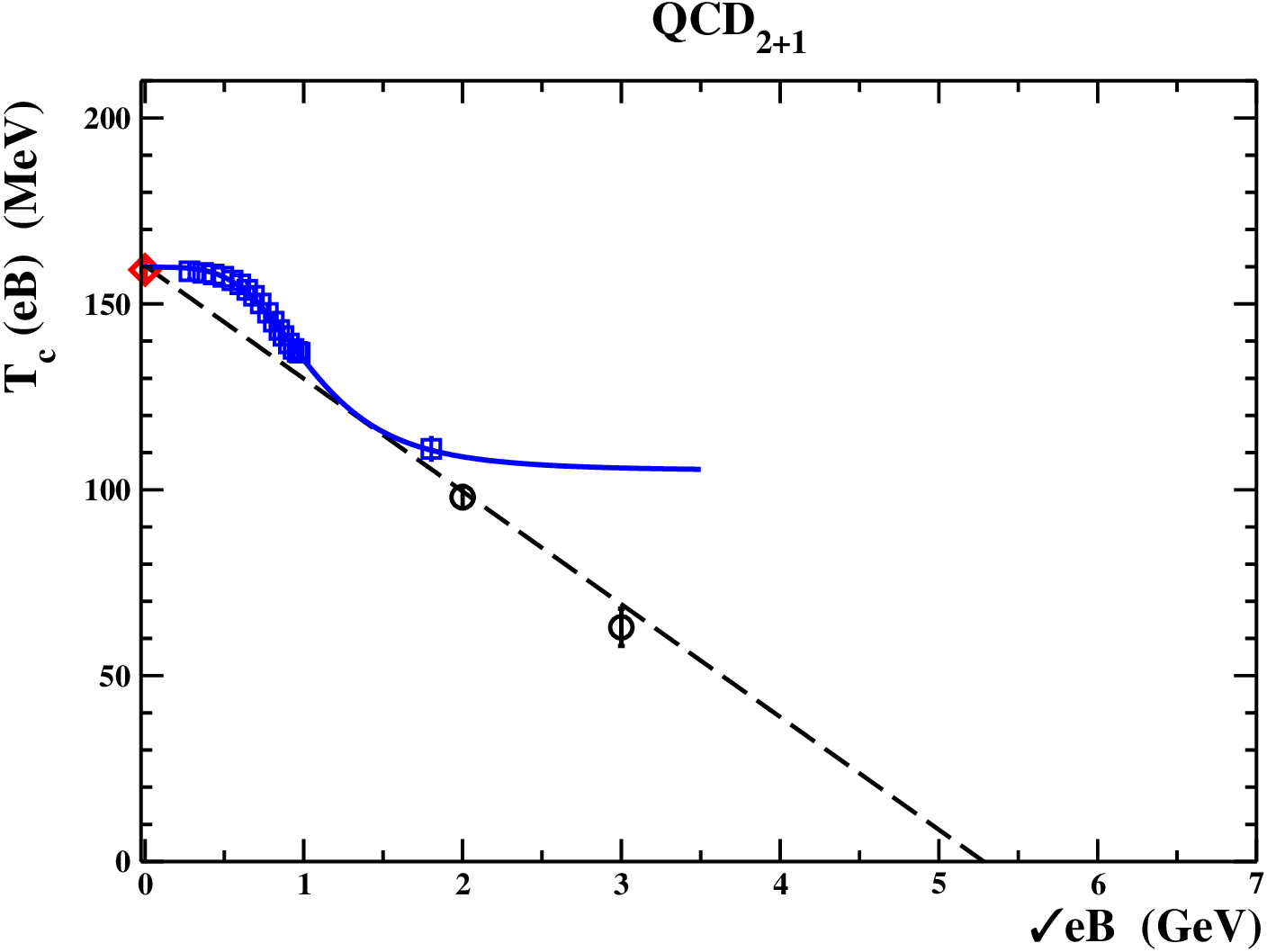}
\end{center}
\caption{\label{Fig7} 
The lattice data  for the deconfinement temperature versus the strength of the external magnetic field $\sqrt{eB}$. The continuous line
is our fit of the data to Eq.~(\ref{6.68}) for $eB \, \le \, 3.25 \, \text{GeV}^2$, the dashed line  is the fit to  Eq.~(\ref{6.71})  for 
$eB \, \ge \, 3.25 \, \text{GeV}^2$. }
\end{figure}
To this end, in Fig.~\ref{Fig7} we display the deconfinement temperature  versus $\sqrt{eB}$. The lattice data have been taken from Fig.~9,
left upper panel, in Ref.~\cite{Bali:2012b} for $eB \, \lesssim 1 \; \text{GeV}^2$, from Ref.~\cite{Endrodi:2015} at 
 $eB \, = \,  3.25 \; \text{GeV}^2$ and from Ref.~\cite{DElia:2022} at  $eB \, = \, 4, 9  \; \text{GeV}^2$. Looking at Fig.~\ref{Fig7} we see
 that the behaviour of the deconfinement temperature clearly indicates the presence of two different regimes. For not to strong magnetic
 fields $T_c(eB)$ smoothly  decreases by increasing the strength of the applied magnetic field and it seems to saturate in the limit
 $eB \, \rightarrow  \infty$. We agree with the discussion presented in Ref.~\cite{Endrodi:2015} where this effect is ascribed to the
 explicit breaking of the rotational symmetry by the external magnetic field and the expected dimensional reduction in the quark sector.
 Indeed, from the structure of the gluon propagator in very strong magnetic fields (see Ref.~\cite{Miransky:2015} and references therein)
 one argues that the limiting effective gauge theory corresponds to an anisotropic pure gauge theory due to the enhancement of the
 chromo-dielectric constant. This qualitatively explains the small reduction of the deconfinement temperature and the saturation to
an asymptotic value since for very strong magnetic fields the fermions are expected to be frozen into the lowest Landau levels.
Accordingly, in Ref.~\cite{Endrodi:2015} the lattice data were fitted to the phenomenological law:
\begin{equation}
\label{6.68}
T_c(eB)  \; = \;  T_c(0) \, \frac{1 \; + \; a_1 \, (eB)^2}{1 \; + \; a_2 \, (eB)^2} \;  \; .
\end{equation}
In fact, for $eB \, \le \, 3.25 \; \text{GeV}^2$ the data are very well described by Eq.~(\ref{6.68}) with~\cite{Endrodi:2015}:
\begin{equation}
\label{6.69}
  T_c(0) \; = \; 160(2) \; \text{MeV} \; \; , \; \;   a_1 \;  = \; 0.54(2) \; \; , \; \;  a_2 \;  = \; 0.82(2) \; \; .
\end{equation}
As a consistency check, we have fitted the data displayed in Fig.~\ref{Fig7} to Eq.~(\ref{6.68}) and found:
\begin{equation}
\label{6.70}
  T_c(0) \; = \; 159.9(1.5) \; \text{MeV} \; \; , \; \;   a_1 \;  = \; 0.55(15) \; \; , \; \;  a_2 \;  = \; 0.83(19) \; \; .
\end{equation}
It is reassuring to see that our fit is perfectly consistent with Eq.~(\ref{6.69}) albeit with larger errors. \\
However,  more recent studies presented in Ref.~\cite{DElia:2022} found that the critical deconfinement temperatures
suffer a further decrease in extremely strong magnetic fields $eB \, = \,  4, \, 9 \; \text{GeV}^2$ (see Fig.~\ref{Fig7}).
This kind of behaviour cannot be explained by the screening effects for  we already argued that the quarks are frozen into the lowest
Landau levels. However, we may offer a plausible explanation within our picture of the QCD vacuum as a disordered chromomagnetic condensate.
We rely on the one-loop calculations of the effective potential in QCD in external Abelian chromomagnetic and magnetic fields~\cite{Ozaki:2014}.
In fact, the chromomagnetic and magnetic fields are coupled to each other through the quark loops such that the vacuum energy is minimized 
if the external fields lie in the same direction. This means that external magnetic fields tend to polarize the chromomagnetic domains of
our QCD vacuum. As a consequence, according to our previous discussion for the SU(3) pure gauge theory, we are led to expect
the presence of  the Meissner effect where the critical deconfinement temperature should decrease according to:
\begin{equation}
\label{6.71}
T_c(eB)  \; = \;  T_c(0) \,\left [ 1 \; - \;  \frac{\sqrt{eB}}{\sqrt{eB_c}}  \right ] \;  \; .
\end{equation}
Actually, we find the the lattice data are consistent with Eq.~(\ref{6.71}) for  $eB \,  \gtrsim \, 3 \; \text{GeV}^2$ (see the dashed line in
Fig.~\ref{Fig7}):
\begin{equation}
\label{6.72}
  T_c(0) \; = \; 160.2(3.7) \; \text{MeV} \; \; , \; \;  \sqrt{eB_c} \;  = \; 5.28(31) \; \text{GeV}  \; \; .
\end{equation}
It is interesting to note that Eq. ~(\ref{6.71})  agrees with Eq.~(\ref{1.1}). However, before addressing to  any conclusions it should be desirable
 to investigate the behaviour of the critical temperature versus  applied chromomagnetic fields in QCD with dynamical quarks at the
 physical point. \\
A further indirect confirmation of our picture comes from the fact that at $eB \, = \, 9 \; \text{GeV}^2$ the deconfinement transition
turns into a first-order transition as in the pure gauge theory~\cite{DElia:2022}. We may, now, gain further insight on the nature
of the quantum vacuum by looking at the structure of the flux-tube chromoelectric fields. Remarkably, the authors of Ref.~\cite{DElia:2021}
investigated the transverse profile of the longitudinal chromoelectric field generated by static quark-antiquark pair directed along
the applied external magnetic field (longitudinal flux tube) as well as perpendicular to the magnetic field (transverse flux tube).
\begin{figure}
\begin{center}
\includegraphics[width=0.32\textwidth,clip]{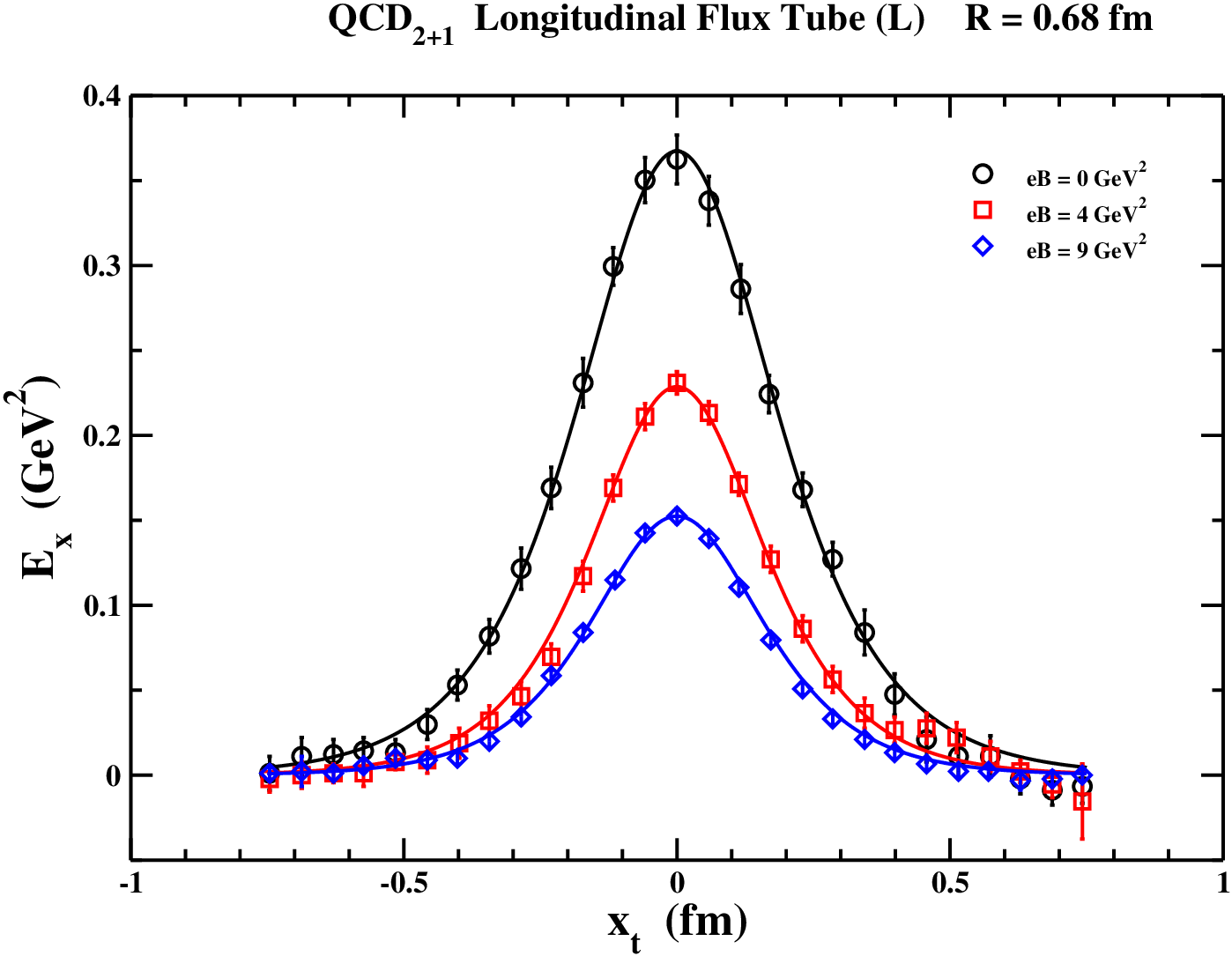}
\includegraphics[width=0.32\textwidth,clip]{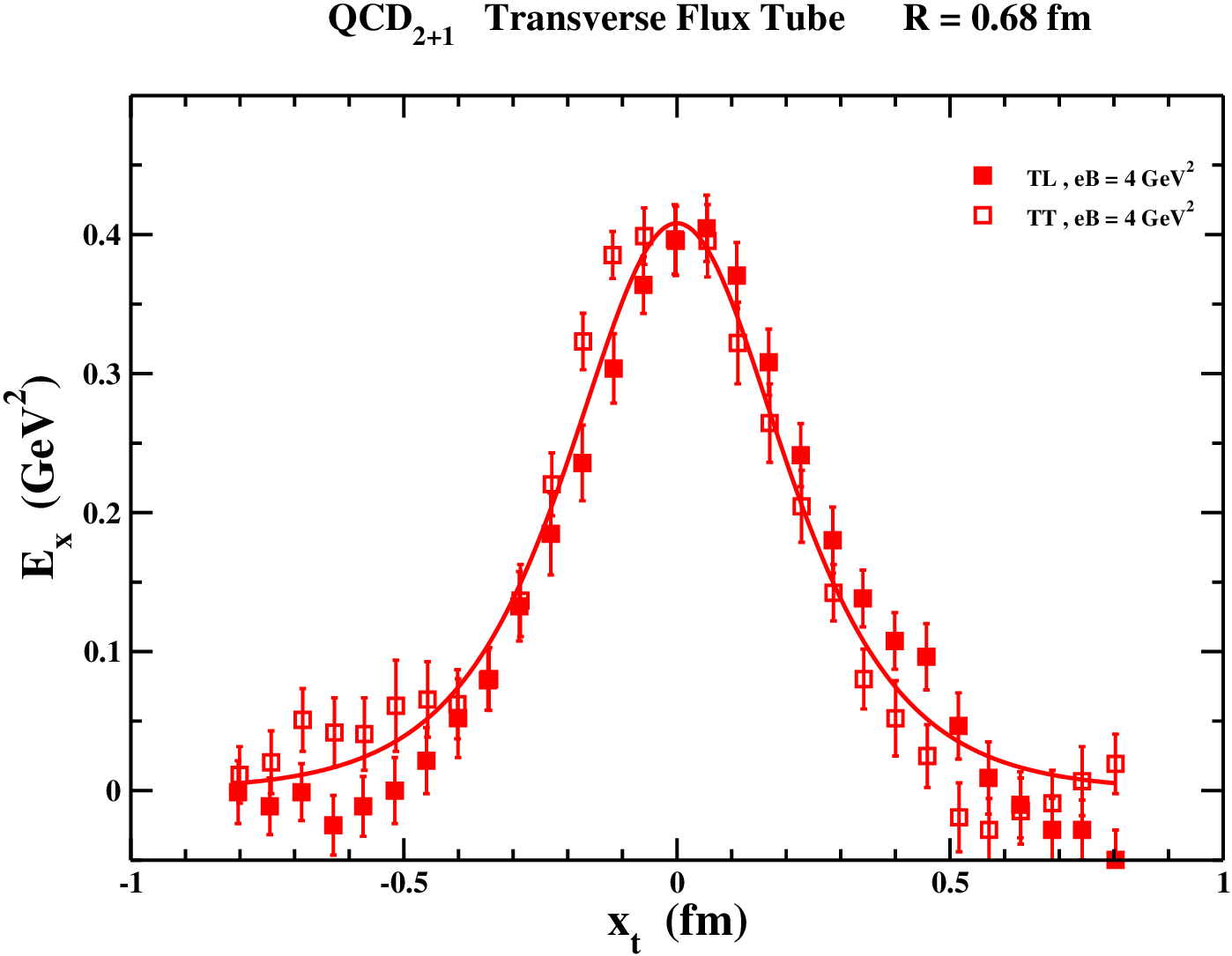} 
\includegraphics[width=0.32\textwidth,clip]{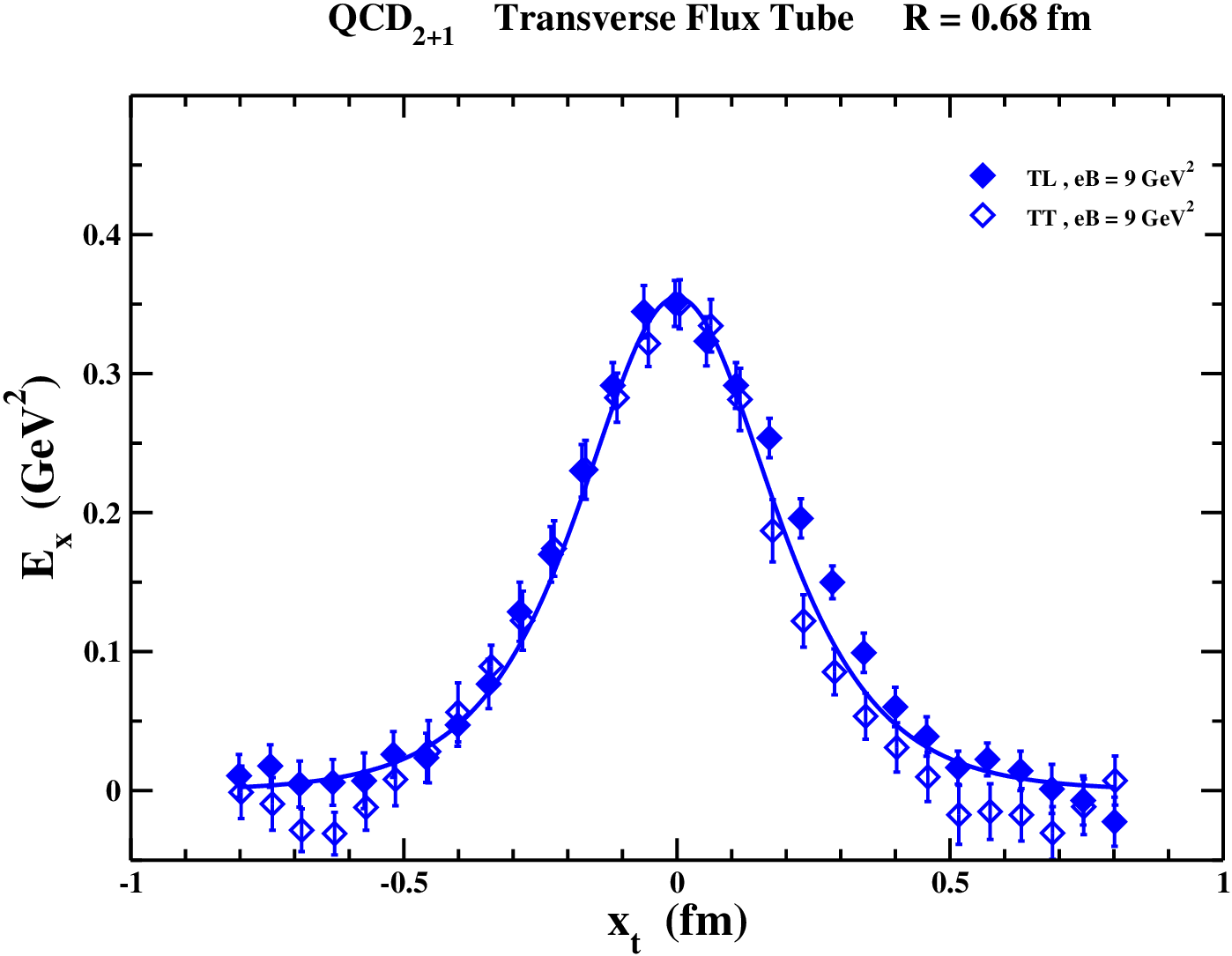} 
\end{center}
\caption{\label{Fig8} 
(Color online) Transverse profile of the flux-tube chromoelectric field for $eB \, = \, 0, \; 4, \, 9 \; \text{GeV}^2$; left panel, 
longitudinal flux tubes, middle and right panels,  transverse flux tubes. The data have been taken from Figs.~10, 11 and 14
in Ref.~\cite{DElia:2021} corresponding to a lattice size $a \, = \, 0.0572$ fm. We use the same nomenclature as in
Ref.~\cite{DElia:2021}. }
\end{figure}
In Fig.~\ref{Fig8} we display the profile of the chromoelectric field for longitudinal and transverse flux tubes for three different
values of the external magnetic field. Note that for $eB = 0$ the lattice data have been already displayed in Fig.~\ref{Fig5}, middle
panel. From the fit to our Eq.~(\ref{6.43}) we obtained:
\begin{equation}
\label{6.73}
 \text{eB} \; = \; 0 : \; v_{\phi}  \; \simeq \;  0.108 \; \; , \; \; \sqrt{gH_0} \; \simeq \; 1.35 \; \text{GeV} \; \; , \; \; \sqrt{\sigma} \; \simeq \; 314 \; \text{MeV}
\end{equation}
where for the string tension we are using Eqs.~(\ref{6.29}) and (\ref{6.30}).  Likewise, we get for the longitudinal flux tube (L):
\begin{equation}
\label{6.74}
 \text{eB} \; = \; 4 \; \text{GeV}^2: \;  v_{\phi}  \; \simeq \;  0.051 \; \; , \; \; \sqrt{gH_0} \; \simeq \; 1.55 \; \text{GeV} \; \; , \; \; \sqrt{\sigma} \; \simeq \; 
 170 \; \text{MeV} \; ,
\end{equation}
\begin{equation}
\label{6.75}
 \text{eB} \; = \; 9 \; \text{GeV}^2: \; v_{\phi}  \; \simeq \;  0.034 \; \; , \; \; \sqrt{gH_0} \; \simeq \; 1.55 \; \text{GeV} \; \; , \; \; \sqrt{\sigma} \; \simeq \; 
 113 \; \text{MeV} \; ,
\end{equation}
while for the transverse flux tubes (TT, TL):
\begin{equation}
\label{6.76}
 \text{eB} \; = \; 4 \; \text{GeV}^2: \;  v_{\phi}  \; \simeq \;  0.14 \; \; , \; \; \sqrt{gH_0} \; \simeq \; 1.25 \; \text{GeV} \; \; , \; \; \sqrt{\sigma} \; \simeq \; 370
  \; \text{MeV} \; ,
\end{equation}
\begin{equation}
\label{6.77}
 \text{eB} \; = \; 9 \; \text{GeV}^2: \; v_{\phi}  \; \simeq \;  0.10  \; \; , \; \; \sqrt{gH_0} \; \simeq \; 1.45 \; \text{GeV} \; \; , \; \; \sqrt{\sigma} \; \simeq \; 293
  \; \text{MeV} \; .
\end{equation}
The above results show quite clearly that the external magnetic field introduces strong anisotropies depending on the strength and orientation
of the background field. Interestingly enough, we find that the transverse string tension does not vary appreciably,
 $ \sqrt{\sigma_T}  \,  \simeq \,  \sqrt{\sigma}$. On the contrary, the longitudinal string tension   $ \sqrt{\sigma_L} $ decreases rapidly with
 the increase of the strength of the magnetic field consistently with the decrease of the deconfinement critical temperature, Eq.~(\ref{6.71}).
 More importantly, the decrease of the longitudinal string tension can be entirely ascribed to the drastic reduction of the azimuthal
 velocity $v_{\phi}$, while the fitted chromomagnetic condensate $\sqrt{gH_0}$ seems to do not vary appreciably. This means that
 for longitudinal flux tubes only the chromomagnetic current responsible for the squeezing of the chromoelectric fields via
 the Ampere law gets strongly suppressed. We see, thus, that the physical picture emerging from the behaviour of static
 quark-antiquark flux tube is in satisfying agreement with our scenario where the gauge system is driven in the deconfinement phase
 by means of polarization by the external background field of the vacuum chromomagnetic domains.
 Obviously, it remains an open question to see if and how our proposal to look at the thermal gauge system in the deconfined
 phase as a strongly correlated quantum liquid would eventually led to the inverse magnetic catalysis.
\section{\normalsize{Dynamical quarks and chiral symmetry breaking }}
\label{S7}
So far our analysis has been limited to the pure gauge sector of the theory. Now we would like to take into account the dynamical 
fermion fields. Let us consider a massive quark field. In this case, in the temporal gauge we must add to the gauge field Hamiltonian
the following Dirac Hamiltonian:
\begin{equation}
\label{7.1}
{\cal H}_D  =  \int d  \vec{x} \; \psi^{\dagger}(\vec{x}) 
 \left \{
  \vec{\alpha} \cdot [ \, -i \vec{\nabla} \; + \;  g \, \vec{A}^a(\vec{x})  \,  \frac{\lambda^a}{2}  \,]
 + \beta m 
 \right \} \psi(\vec{x})  \; .
\end{equation}
In Eq.~(\ref{7.1}) we are using the Bjorken-Drell~\cite{Bjorken:1964} convention for the Dirac matrices. We are interested in evaluating
the contributions of  dynamical quarks to the vacuum energy in the lowest order approximation.  This approximation, after 
separating the gauge fields $\vec{A}^a(\vec{x})$ into quantum fluctuations over the background field 
$\vec{\bar{A}}^a(\vec{x}) \, + \, \vec{U}^a(\vec{x})$, amounts to  neglecting in the Dirac Hamiltonian the contribution due to
the fluctuations. Moreover, since the background fields  $\vec{U}^a(\vec{x})$ generated by the dynamical condensation driven by
the unstable modes mainly affect the low-lying modes, in evaluating the vacuum energy we may further simplify the Dirac
Hamiltonian as:
\begin{equation}
\label{7.2}
{\cal H}^{(0)}_D  =  \int d  \vec{x} \; \psi^{\dagger}(\vec{x}) 
 \left \{
  \vec{\alpha} \cdot [ \, -i \vec{\nabla} \; + \;  g \, \vec{\bar{A}}^a(\vec{x})  \,  \frac{\lambda^a}{2}  \,]
 + \beta m 
 \right \} \psi(\vec{x})  \; .
\end{equation}
The role of the $\vec{U}^a(\vec{x})$ background field will be discussed more fully later on. Obviously, the presence of the
fermion fields modifies the  Gauss constraints. Within the above approximations we get for the Gauss law:
\begin{equation}
\label{7.3}
 D^{ab}_i(\vec{x}) \;  \frac{\delta}{\delta \eta^b_i(\vec{x}) } \; {\cal S}[A,\psi^\dagger,\psi] \; =  \;
 - \; i \, g \; \psi^{\dagger}(\vec{x}) \,  \frac{\lambda^a}{2} \, \psi(\vec{x}) 
 \; {\cal S}[A,\psi^\dagger,\psi] \; .
\end{equation}
Following our previous variational studies in U(1)~\cite{Cea:1985,Cea:1986b} and SU(2)~\cite{Cea:1997e} gauge theories in three
space-time dimensions, we may solve Eq.~(\ref{7.3}) by writing:
\begin{equation}
\label{7.4}
 {\cal S}[\eta,\psi^\dagger,\psi] \, = \, {\cal G}[\eta] \, \, 
 {\cal F}[\psi^\dagger,\psi] \; \exp \{ \Gamma_{\psi}[\eta,\psi^\dagger,\psi] \}
\end{equation}
with
\begin{equation}
\label{7.5}
 D^{ab}_i(\vec{x}) \; \frac{\delta {\cal G}[\eta]}{\delta \eta^b_i(\vec{x})} \, = 0  \;  ,
\end{equation}
namely the wavefunctional  ${\cal G}[\eta]$ depends on the transverse gauge fluctuations. It is, now, easy to see that:
\begin{equation}
\label{7.6}
\Gamma_{\psi}[\eta,\psi^\dagger,\psi]  \, = \,  - \; i \, g \;  \int d \, \vec{x} \;  d \, \vec{y} \; \psi^{\dagger}(\vec{x}) \,  \frac{\lambda^a}{2} \, \psi(\vec{x})
\;   [D^{-1}]^{ab}_j(\vec{x},\vec{y}) \;  \eta^b_j(\vec{y})
\end{equation}
solves the Gauss law  Eq.~(\ref{7.3}). As showed in  Ref.~\cite{Cea:1985}, it is quite easy to check that in evaluating the vacuum energy the functional
$\Gamma_{\psi}[\eta,\psi^\dagger,\psi]$ merely adds the Coulombic Hamiltonian:
\begin{equation}
\label{7.7}
{\cal H}_C \,  = \, \frac{g^2}{2} \,   \int d \, \vec{x} \;  d \, \vec{y} \; 
 \psi^{\dagger}(\vec{x}) \,  \frac{\lambda^a}{2} \, \psi(\vec{x}) \; 
 [D^{-2}]^{ab}(\vec{x},\vec{y}) \; 
 \psi^{\dagger}(\vec{y}) \,  \frac{\lambda^b}{2} \, \psi(\vec{y})
\end{equation}
that in the one-loop approximation can be neglected. To determine the fermion wavefunctional   ${\cal F}[\psi^\dagger,\psi] $, following
Refs.~\cite{Cea:1985,Cea:1986b,Cea:1997e} we shall employ the holomorphic representation~\cite{Berezin:1966,Faddeev:1976}.
Firstly,  for completeness, we show how this representation works for free massive Dirac fermions. The fermion field $\psi(\vec{x})$
can be expanded as:
\begin{equation}
\label{7.8}
\psi(\vec{x}) \; = \; \sum_{\vec{p},s} \; \left [  \psi^{(+)}_{\vec{p},s}(\vec{x}) \;  a^*_{\vec{p},s} \; + \; 
 \psi^{(-)}_{\vec{p},s}(\vec{x}) \;  b^*_{\vec{p},s}  \right ] 
\end{equation}
where  $ \psi^{(\pm)}_{\vec{p},s}(\vec{x})$ are the positive and negative-energy solutions of the Dirac equation:
\begin{equation}
\label{7.9}
 \left \{ - \, i \, \vec{\alpha} \cdot   \vec{\nabla}   \; + \  \beta \, m 
 \right \} \psi(\vec{x}) \; = \;  E \;  \psi(\vec{x}) \; .
\end{equation}
In the holomorphic representation    $a_{\vec{p},s},   a^*_{\vec{p},s},    b_{\vec{p},s},  b^*_{\vec{p},s}$  are independent Grassmann variables
that act in the space of states  $  {\cal F}[ a^*,b^*]$ polynomials in  $a^*_{\vec{p},s},  b^*_{\vec{p},s}$  with the following scalar product:
\begin{equation}
\label{7.10}
 < {\cal F}^*_1,  {\cal F}_2 >  = \int \prod_{\vec{p},s} d a^*_{\vec{p},s} \,  d a_{\vec{p},s} \, \exp \{ -  a^*_{\vec{p},s} \,  a_{\vec{p},s} \}
d \, b^*_{\vec{p},s} \,  d \, b_{\vec{p},s} \, \exp \{ -  b^*_{\vec{p},s} \,  b_{\vec{p},s} \}
{\cal F}^*_1[ a^*,b^*] \;  {\cal F}_2[ a^*,b^*] \; .
\end{equation}
In this representation the states, in general, can be written as:
\begin{equation}
\label{7.11}
{\cal F}[ a^*,b^*] \;  = \;  \prod_{\vec{p},s}  \left [  \alpha_{\vec{p},s}  + \beta_{\vec{p},s} \,  a^*_{\vec{p},s}  \right ] \, 
 \left [  \gamma_{\vec{p},s}  + \delta_{\vec{p},s}  \, b^*_{\vec{p},s}  \right ]  \; .
\end{equation}
It is easy to check that: 
\begin{equation}
\label{7.12}
 < {\cal F}^*,  {\cal F} >  \; = \; \prod_{\vec{p},s} \;  \left [  |\alpha_{\vec{p},s}|^2  +| \beta_{\vec{p},s}|^2   \right ] \, 
 \left [  |\gamma_{\vec{p},s}|^2  + |\delta_{\vec{p},s}|^2  \right ]  \; ,
\end{equation}
so that from  $< {\cal F}^*,  {\cal F} >  \, = 1$ we obtain:
\begin{equation}
\label{7.13}
 \alpha_{\vec{p},s} \, = \,  \sin  \theta_{\vec{p},s} \; , \;   \beta_{\vec{p},s}  \,  = \,  \cos  \theta_{\vec{p},s} \; , \; 
\gamma_{\vec{p},s}   \,  = \,  \sin  \phi_{\vec{p},s}  \; , \;    \delta_{\vec{p},s}    \,  = \, \cos  \phi_{\vec{p},s} \; .
\end{equation}
It also straightforward to see that the ground state is given by:
\begin{equation}
\label{7.14}
{\cal F}[ a^*] \;  = \;  \prod_{\vec{p},s} \;    a^*_{\vec{p},s}  \; ,
\end{equation}
corresponding to $ \theta_{\vec{p},s} = 0$,    $\phi_{\vec{p},s} = \frac{\pi}{2}$, with  vacuum energy:
\begin{equation}
\label{7.15}
E_0  \;  = \; - \, \frac{1}{2} \,  \sum_{\vec{p},s} \; \sqrt{\vec{p}^2 \; + \; m^2} \; = \; - \, V \, \int \; \frac{d \, \vec{p}}{(2 \pi)^3}
\; \sqrt{\vec{p}^2 \; + \; m^2} \; .
\end{equation}
Let us consider the Hamiltonian Eq.~(\ref{7.2}). To determine the ground state wavefunctional it is enough to solve the following Dirac
equation:
\begin{equation}
\label{7.16}
 \left \{
  \vec{\alpha} \cdot [ \, -i \vec{\nabla} \; + \;  g \, \vec{\bar{A}}^a(\vec{x})  \,  \frac{\lambda^a}{2}  \,]
 + \beta m 
 \right \} \psi(\vec{x})  \; = \; E \; \psi(\vec{x})  \; .
\end{equation}
One finds that the calculation of the spectrum reduces to solving  the Dirac equation for an electron in an uniform magnetic field
by replacing $eH$ with $ \pm \frac{gH}{2}$.  Accordingly, we find the following spectrum:
\begin{eqnarray}
\label{7.17}
\nonumber
 N_1 \; = \; (\vec{p},s) \; \; \; \;  \; \;  \; \; \; \;   E_{N_1} \; = \; \pm \; \sqrt{\vec{p}^2 \, + \, m^2}  \hspace{8.7 cm}
 \\
  N_2  =  (p_3, p_2, n, \alpha) \; \;  E_{N_2}  =  \pm \; \sqrt{p_3^2  + m^2  + \frac{gH_0}{2} \, (2n  + 1)  +  \alpha  } \;  \;  \;  \; 
  n \, = 0, 1, 2 ...  , \alpha  =  \pm \, 1 \;  \;  \; \; \; \; \; \; \; 
  \\
 \nonumber
   N_3  =  (p'_3, p'_2, n', \alpha') \; \;  E_{N_3}  =  \pm \; \sqrt{{p'_3}^2  + m^2  + \frac{gH_0}{2} \, (2n'  + 1)  +  \alpha' }   \; \;
  n' \, = 0, 1, 2 ...  , \alpha'  =  \pm \, 1 \;  \;  \; \; \; 
 \end{eqnarray}
Therefore, the vacuum wavefunctional turns out to be:
\begin{equation}
\label{7.18}
{\cal F}_0[ \psi^{\dagger}, \psi] \;  = \;  \prod_{N_1,N_2,N_3}  \;    a^*_{N_1}  \,  a^*_{N_2}  \,   a^*_{N_3}     \;
\end{equation}
with energy:
\begin{equation}
\label{7.19}
E^{quark}(gH_0)  =   - \, V \, \left \{\sqrt{\vec{p}^2 +  m^2}  +   \frac{gH_0}{4 \pi^2}  \int^{+ \infty}_{-\infty} d p_3 \sum_{n=0}^{\infty}   \sum_{\alpha=\pm1}
 \sqrt{p_3^2  + m^2  + \frac{gH_0}{2} \, (2n  + 1)  +  \alpha  }     \right \} .
\end{equation}
 Proceeding as we did in Sec.~\ref{S3} (see also Ref.~\cite{Cea:1985}, Appendix A), we may evaluate the sum over $n$ and the integration over
 $p_3$. After subtracting the energy of the perturbative vacuum we get:
\begin{equation}
\label{7.20}
\Delta E^{quark}(gH_0) \;  =  \;  V  \; \frac{gH_0}{8 \pi^2} \, \int_0^{\infty} \frac{d s}{s^2 } \; \exp [ - m^2  s] \; 
 \left \{\ \coth (\frac{gH_0 s}{2}) \; - \; \frac{2}{gH_0 s}    \right \}  \; .
\end{equation}
From Eq.~(\ref{7.20}) we deduce that the main contribution comes from quarks with $m \ll \sqrt{gH}$. In this case we find:
\begin{equation}
\label{7.21}
\Delta E^{quark}(gH_0) \;  \simeq  \;  V  \; N_f \; \frac{(gH_0)^2}{48 \pi^2} \,
 \left [ \ln  ( \frac{\Lambda^2}{gH_0})  \; +  \;  const    \right ]  \; ,
\end{equation}
where $N_f$ is the number of almost massless flavours. As a consequence, we may still look at the vacuum wavefunctional as composed
by almost independent chromomagnetic domains with energy ($N_c = 3$):
\begin{equation}
\label{7.22}
\Delta E_D(gH_0) \;  \simeq  \;  V_D  \; ( N_c  + 2  N_f ) \; \frac{(gH_0)^2}{48 \pi^2} \,
 \ln  ( \frac{\Lambda_H}{\sqrt{gH_0}})    \; .
\end{equation}
Remarkably, Eq.~(\ref{7.22}) applies also to the SU(2) gauge theory ($N_c = 2$).  We are led, thus, to conclude that, within our approximations,
the main effect of dynamical quarks is a slight increase of the vacuum energy in qualitative agreement with the discussion in the previous Section.
However, from one hand it is well ascertained that confinement leads to the chiral symmetry breaking~\cite{Casher:1979b}, on the other hand
our proposal for the QCD vacuum wavefunctional does not give rise to a non-zero chiral condensate for massless quarks. Indeed, the
Banks-Casher relation~\cite{Banks:1980}
\begin{equation}
\label{7.23}
< \bar{\psi} \, \psi > \;  =  \;  - \; \pi \; \rho(0) 
\end{equation}
relates the chiral condensate for massless quarks to the density of zero modes of the Dirac operator. So that,  for massless quarks we need
at least one zero mode localized in our chromomagnetic domains. However, for a given chromomagnetic domain the fermion
wavefunctional given by Eq.~(\ref{7.18}) is built from modes corresponding to fermions in an uniform Abelian chromomagnetic field
localized in the given domain. As a consequence, even for massless quarks the low-lying modes display a mass gap of order $\frac{1}{L_D}$
and, thereby, cannot give rise to a non-zero chiral condensate. At this point it is mandatory to take into account
the background field $U^a_i(\vec{x})$ induced by the condensation of the tachyonic modes that, as noticed before, could modify the fermion
low-lying spectrum. In particular, we are interested in localized non-trivial solutions of the massless Dirac equation:
\begin{equation}
\label{7.24}
  \vec{\alpha} \cdot  \left \{ \, -i \vec{\nabla} \; + \;  g \,[ \vec{\bar{A}}^a(\vec{x})  \; + \;  \vec{U}^a(\vec{x}) ] \, \frac{\lambda^a}{2}  \right \} \,  \psi(\vec{x}) 
   \; = \;  0  \; .
\end{equation}
In the first part of the present paper we showed that there are three different kind of unstable modes, namely the u-modes, v-modes and w-modes.
To gain insight into what is going on,  for the moment we simplify the problem by focusing on eventual zero modes due to the u-modes. In this case,
we try to solve Eq.~(\ref{7.24}) by writing:
\begin{equation}
\label{7.25}
\psi(\vec{x}) \; = \;  \left(
\begin{array}{c} \psi^{(1)}(\vec{x}) \\ \psi^{(2)}(\vec{x})  \\ 0 \end{array} \right) 
\end{equation}
where  $\psi^{(i)}(\vec{x})$ are Dirac spinors. We further simplify the problem by assuming that:
\begin{eqnarray}
\label{7.26}
\nonumber
\psi^{(1)}(\vec{x}) \; = \; h(\vec{x}) \; \left( \begin{array}{c} \chi^{(2)} \\ \,  \chi^{(2)} \end{array} \right)  \; , \\
\psi^{(2)}(\vec{x})  \; = \;  h(\vec{x}) \; \left( \begin{array}{c} \chi^{(1)} \\ \,  \chi^{(1)} \end{array} \right) \; , 
 \\
 \nonumber
\chi^{(1)} \; =   \;   \left( \begin{array}{c} 1 \\ \,  0  \end{array} \right)      \; \;  , \; \; 
\chi^{(2)} \; =   \;   \left( \begin{array}{c} 0 \\ \,  1  \end{array} \right)  
\; \; \; .
 \end{eqnarray}
Inserting Eq.~(\ref{7.26}) into Eq.~(\ref{7.24}) we obtain:
\begin{equation}
\label{7.27}
 - i \,  \vec{\alpha} \cdot   \vec{\nabla} \,   \psi^{(1)}(\vec{x}) \; + \; \frac{gH_0}{2} \, x_1 \,  \alpha_2  \, \psi^{(1)}(\vec{x})  \; + \; 
 g \, f^u(x_3) \, g^u_-(x_1,x_2) \,  \psi^{(1)}(\vec{x})    \; = \;  0  \; ,
\end{equation}
\begin{equation}
\label{7.28}
 - i \,  \vec{\alpha} \cdot   \vec{\nabla} \,   \psi^{(2)}(\vec{x}) \; - \; \frac{gH_0}{2} \, x_1 \,  \alpha_2  \, \psi^{(2)}(\vec{x})  \; + \; 
 g \, f^u(x_3) \, g^u_+(x_1,x_2) \,  \psi^{(2)}(\vec{x})    \; = \;  0  \; .
\end{equation}
Inspection of Eqs.~(\ref{7.27}) and (\ref{7.28}) shows that:
\begin{equation}
\label{7.29}
h(\vec{x}) \; = \; \varphi(x_1,x_2) \, \theta(x_3) \; ,
\end{equation}
where the functions $\varphi(x_1,x_2)$ and  $\theta(x_3)$ must satisfy the following equations:
\begin{equation}
\label{7.30}
 ( \partial_1 \; - \; i \, \partial_2 \; + \frac{gH_0}{2} \, x_1 ) \, \varphi(x_1,x_2)  \;  = \; 0
\end{equation}
\begin{equation}
\label{7.31}
 [ i \,  \partial_3  \; + \; g \, f^u(x_3) \, g^u_+(x_1,x_2) \,]  \theta(x_3) \; = \; 0
\end{equation}
with the constraint that $\theta(x_3)$  must be a real function. Solving Eq.~(\ref{7.30}) we obtain:
\begin{equation}
\label{7.32}
 \varphi(x_1,x_2)  \;  = \;  \exp ( i \, p_2 \, x_2) \; \exp \left [ - \frac{gH_0}{4} \, ( x_1 \, + \, \frac{p_2}{\frac{gH_0}{2}})^2 \right ] \; ,
\end{equation}
i.e.  $\varphi(x_1,x_2)$ belongs to the lowest Landau level wavefunctions. It is evident from Eq.~(\ref{7.32})  that the zero-mode
wavefunction is not localized in the chromomagnetic domain.  However, due to the energy degeneracy we may employ a suitable
linear combination of the  lowest Landau level wavefunctions such that the resulting wavefunction gets localized in the given
chromomagnetic domain. The best way to illustrate this point is to solve the Dirac equation in the symmetric gauge
$\vec{A}_{sym} = ( - \frac{1}{2} x_2 H_0, \frac{1}{2} x_1 H_0) $ that is related to the vector potential in the Landau gauge
$\vec{A}_{Landau} = ( -0, x_1 H_0) $ by a gauge transformation:
\begin{equation}
\label{7.33}
 \vec{A}_{sym}(x_1, x_2)  \;  = \; \vec{A}_{Landau}(x_1, x_2) \; - \; \vec{\nabla} \Phi(x_1, x_2)  \; , \; \Phi(x_1, x_2) \; = \; \frac{1}{2} H_0 \, x_1 x_2 \; .
\end{equation}
In our case the relevant gauge transformation is implemented by the SU(3) matrix:
\begin{equation}
\label{7.34}
 \Lambda(\vec{x} ) \;  = \;  \exp \{ i \, g \,  \Phi(\vec{x}) \, \frac{\lambda_3}{2} \}   \; .
\end{equation}
In the symmetric gauge the eigenstates of the Dirac Hamiltonian are eigenstates of the angular momentum labelled by the integer $m$.
In particular, the zero-mode wave functions are given by:
\begin{equation}
\label{7.35}
 \varPsi_{0,m}(x_1,x_2)  \;  = \;  \frac{1}{\sqrt{ 2^{m + 1} \pi \, m! \, a_0^2}}   \; \left ( \frac{x_1 \, + \, i \, x_2}{a_0} \right )^m \; \exp \left [ -  \frac{( x_1^2 \, + \, x_2^2)}{4 \, a_0}  \right ] \; ,
\end{equation}
where $a_0 = \sqrt{\frac{2}{g H_0}}$.  Using the normalized wavefunctions:
\begin{equation}
\label{7.36}
 \varphi_{0, p_2}(x_1,x_2)  \;  = \;  \left( \frac{g H_0}{2 \pi} \right )^{\frac{1}{4}} \;  \frac{1}{\sqrt{2 \pi}} \;
  \exp ( i \, p_2 \, x_2) \; \exp \left [ - \frac{gH_0}{4} \, ( x_1 \, + \, \frac{p_2}{\frac{gH_0}{2}})^2 \right ] \; ,
\end{equation}
one can show that~\cite{Cea:1986b}:
\begin{equation}
\label{7.37}
\exp \left ( i \, \frac{x_1 x_2}{2 \, a_0^2} \right ) \; \varPsi_{0,m}(x_1,x_2)  \;  = \;  \int d \, p_2  \; d_m(p_2) \; \varphi_{0, p_2}(x_1,x_2)  \;  
\end{equation}
with:
\begin{equation}
\label{7.38}
d_m(p_2) \; =  \;   \left( \frac{a_0^2}{ \pi} \right )^{\frac{1}{4}} \;  \frac{1}{\sqrt{ 2^{m }  \, m! }} \; \exp ( - a_0^2 p_2^2) \; 
\; H_m(a_0 p_2)  \;  ,
\end{equation}
$H_m(z)$ being Hermite polynomials. Note that the phase factor in Eq.~(\ref{7.37}) is due to the gauge invariance. Since:
\begin{equation}
\label{7.39}
 < r  > \; = \;  \sqrt{ < x_1^2 \; + \; x_2^2 >} \; = \;   \frac{1}{\sqrt{ g H_0 }}   \;  \sqrt{ 2 ( 2 m + 1)}  \;  \;  ,
\end{equation}
we see that the condition $< r > \lesssim \frac{L_D}{2}$ requires $m = 0$.  As a consequence the zero-mode wavefunction is given by:
\begin{equation}
\label{7.40}
\varPsi_{z-m}(x_1,x_2)  \;  = \;   \varPsi_{0, 0}(x_1,x_2) \; = \; 
\sqrt{\frac{g H_0}{ 4 \pi}}   \;  \exp \left [ - \frac{g H_0}{8} \, ( x_1^2 \, + \, x_2^2) \right ] \; .
\end{equation}
Let us consider now Eq.~(\ref{7.31}). By using the results in Section~\ref{S3} we can rewrite this equation as:
\begin{equation}
\label{7.41}
\left [ \partial_3  \; - \; i \; \sqrt{g H_0} \, \exp \{ i \delta^u(x_1, x_2) \}  \; \tanh \{ \frac{\sqrt{g H_0}}{2} \,  x_3\} \right ]  \theta(x_3) \; = \; 0
\end{equation}
The formal solution of this last equation is:
\begin{equation}
\label{7.42}
 \theta(x_3) \; = \;  \left [  \text{sech}  \{ \frac{\sqrt{g H_0}}{2} \,  x_3\} 
 \right ]^{- i \sqrt{2} \exp \{ i \delta^u(x_1, x_2) \} } \; \; .
\end{equation}
As discussed at length in Sect.~\ref{S4}, inside the chromomagnetic domain the phase $\delta^u(x_1, x_2)$ is rapidly varying so that
 $\exp \{ i \delta^u(x_1, x_2) \} $ averages to zero. Therefore, we are led to conclude that
\begin{equation}
\label{7.43}
 \theta(x_3) \; \simeq  \;  1 \; \; \; , \;  \; \;  | x_3 | \, < \; \frac{L_D}{2} \; \; .
\end{equation}
On the other hand, in the transition layer between two adjacent domains the phase $\delta^u(x_1, x_2)$ must vary smoothly going from
one domain to the adjacent domain. Since: 
\begin{equation}
\label{7.44}
|  \theta(x_3) | \; =  \;   \left [  \text{sech}  \{ \frac{\sqrt{g H_0}}{2} \,  x_3\} 
 \right ]^{ \sqrt{2} \sin \{ \delta^u(x_1, x_2) \} } \; \; ,
\end{equation}
we see that for $\sin \{ \delta^u(x_1, x_2) \}  < 0$ the zero-mode wavefunction is exponentially suppressed. In this case we can write:
\begin{equation}
\label{7.45}
h_{z-m}(x_1, x_2, x_3 )  \; \simeq  \;  \frac{1}{\sqrt{L_D}}  \;
\sqrt{\frac{g H_0}{ 4 \pi}}   \;  \exp \left [ - \frac{g H_0}{8} \, ( x_1^2 \, + \, x_2^2) \right ] \; \; , \; \; 
| x_3 | \; \lesssim \; \frac{L_D}{2} \; \; .
\end{equation}
We have also checked that even including the effects due to the induced background fields $\vec{v}^a(\vec{x})$ and $\vec{w}^a(\vec{x})$
do not modify the zero-mode wavefunction given by Eq.~(\ref{7.45}). We may, thus, conclude that an arbitrary chromomagnetic 
domain could accommodate two zero modes as given by Eqs.~(\ref{7.26}) and (\ref{7.45}). So that, according to the Banks-Casher relation
we have:
\begin{equation}
\label{7.46}
< \bar{\psi} \, \psi >  \; \simeq  \;  - \; \frac{2 \pi}{L_D^3} \; \; .
\end{equation}
Note that:
\begin{equation}
\label{7.47}
| < \bar{\psi} \, \psi > |^{\frac{1}{3}}  \; \sim  \;  \frac{1}{L_D} \; \sim \; 10^2 \; \text{MeV} \; 
\end{equation}
that is of the correct order of magnitude with respect to several lattice determinations (see, eg, Ref.,~\cite{Faber:2017} and
references therein). Moreover, according to the general discussion in Ref.~\cite{Casher:1979}, the rate of change of chirality
leads to an effective quark mass $m_{eff} \sim \frac{1}{L_D} \sim 10^2$~MeV that compares reasonable well to  phenomenology. \\
We would like to conclude the present Section by discussing the intriguingly results presented in Ref.~\cite{Iritani:2015} where
it was shown that the magnitude of the chiral condensate is reduced inside the color flux tube generated by a static quark-antiquark
pair. This result is interpreted as an evidence of partial restoration of the chiral symmetry inside hadrons. However, our explanation on the
formation of the chiral condensate leads to different conclusions. Indeed, in the previous Section we have seen that the squeezing
of the color fields into a narrow flux tube is due to chromomagnetic currents almost uniform along the flux tube that, in turns, originate
from the polarization of the chromomagnetic domains. As a consequence, along the flux tube one must image to have a larger coherent 
chromomagnetic domain with an effective volume that increases according to $V_D \, \frac{R}{L_D}$, where $R$ is the distance
between the static color sources. It follows, then, that the chiral condensate is reduced by approximatively a factor $ \frac{L_D}{R}$
for large enough $R$. This result, not only explains naturally the suppression of the chiral condensate inside the flux tube, but also
seems to be in qualitative agreement with Fig.~3 in Ref.~\cite{Iritani:2015} where it is displayed the ratio of the chiral condensate at
the center of the flux tube versus the separation distance $R$.
\section{\normalsize{Summary and concluding remarks }}
\label{S8}
Quantum chromodynamics is widely accepted as the fundamental  theory of strong interactions. However, the confinement of quarks
and gluons into hadrons, namely color singlet states, has evaded a satisfying understanding. Actually, the absence of asymptotic colored 
particle-states leads to the time-honoured confinement problem in QCD. The purpose of this paper was to gain insight on color
and quark confinement starting from first principles. The first part of this paper was devoted to extend to the SU(3) gauge theory our
previous work in the pure gauge SU(2) theory in presence of an Abelian chromomagnetic field. We enlightened  for the first time
the presence in SU(3)  of three different kinds of unstable  modes. We set up a perturbative-variational scheme by employing gauge-invariant
wavefunctionals that allowed to minimize the ground-state energy and, at the same time, to stabilize the instabilities that affected
the low-order evaluation of the vacuum energy. As in the SU(2) theory, the condensation of the tachyonic modes by quantum
fluctuations lead to the dynamical generation of three different background fields with a peculiar kink structure typical of
the (1+1)-dimensional charged scalar fields subject to dynamical symmetry breaking. The resulting stabilized ground-state wavefunctional
turned out to be not energetically favoured with respect to the perturbative quantum vacuum. However, we have shown that the stabilized 
ground-state wavefunctional looked like a collection of independent domains each characterized by almost the same chromomagnetic
condensate pointing in arbitrary space and color directions. Separating these chromomagnetic domains will be Bloch walls that
carry an energy per unit area leading to a further increase in the vacuum energy. Nevertheless, we argued that the number of gauge field
configurations that realize the vacuum wavefunctional were large enough to span a set of finite volume in the functional space
of physical states. After introducing the configurational entropy, we found that there was a order-disorder quantum phase transition
of the Berezinskii-Kosterlitz-Thouless type at an energy scale of order $\frac{1}{L_D}$ separating the perturbative vacuum from
our disordered chromomagnetic condensate quantum vacuum. It is important, at this point, to stress that our proposal for the 
ground-state wavefunctional should account for the QCD vacuum at large distances. On the other hand, at very small distances
our wavefunctional should smoothly be replaced by the perturbative ground state realizing in this way the Bjorken's femptouniverse. \\
In the second part of the paper, we addressed the problem to see if our vacuum wavefunctional could be a viable proposal
for the QCD quantum vacuum. Indeed, we showed that the disordered  chromomagnetic condensate vacuum displayed a non-zero
gluon condensate of the right order of magnitude. In addition, we argued that our proposal for the QCD vacuum had both a finite
energy gap and absence of color correlations for distances larger than $L_D$ leading to the confinement of colors. We have, also,
advanced a vivid picture on the formation of the flux-tube chromoelectric fields for a static quark-antiquark pair. The squeezing
of the chromoelectric fields into a narrow tube were caused by  Lorentz forces generated by the chromomagnetic currents originating
from the background fields induced by the dynamical condensation by quantum fluctuations of the tachyonic mode.
This allowed us to determine the color structure and the transverse profile of the flux-tube chromoelectric fields that was
compared favorably to several non-perturbative numerical simulations in lattice QCD.
Our proposal for the quantum QCD vacuum allowed us to understand the origin of the color Meissner effect observed in lattice
simulations of the pure gauge SU(3) theory as well as in QCD with two massive degenerate staggered quarks. Moreover, our
proposal allowed to reach a physical explanation on the evidence of the Meissner effect from lattice simulations
of QCD with (2+1)-flavours at the physical point in very strong external magnetic fields. This led us to the prevision of a new
quantum phase transition in the QCD phase diagram where at the critical magnetic field the gauge system undergoes a deconfinement
transition  to a quantum liquid made of quarks and gluons with strong chromomagnetic correlations. Finally, we showed
that the inclusion of dynamical quarks did not modify substantially the chromomagnetic domain structure of our vacuum 
wavefunctional. Remarkably, we suggested that for massless quarks the background fields dynamically generated by the
condensation of the unstable modes could allow  the formation of fermion zero modes localized on the chromomagnetic domains that,
in turns, led to the spontaneous  breaking of the chiral symmetry with an ensuing  chiral condensate of the correct
order of magnitude. \\
To conclude, the purpose of this paper has been to understand confinement from first principles. As a matter of fact, we are
aware that the analysis we followed, albeit quite involved, has been useful leading to a proof of concept that confinement
can be understood. The comparison of our approach to hadron phenomenology and non-perturbative simulations
of QCD on the lattice seems to suggest that we are on the right track, even though there are still many things to do and
questions to answer. Nevertheless, with the present paper we hope to stimulate further studies to reach a complete
quantitative understanding of confinement in quantum chromodynamics.

\begin{thebibliography}{199}
%
\bibitem{Joos:1979}
H. Joos, Introduction to quark  confinement in QCD,
Acta Phys. Austr. Suppl. {\bf XXI} (1979) 407
%
\bibitem{Mandelstam:1980}
S. Mandelstam, General Introduction to Confinement, 
Phys. Rep. C {\bf 67} (1980) 109
%
\bibitem{Bander:1981} 
M. Bander, Theories of Quark Confinement, 
Phys. Rep.  {\bf 75} (1981) 206
%
\bibitem{Zachariasen:1986}
F. Zachariasen, Classical Picture of Confinement,
 Mod. Phys. Lett. {\bf A01} (1986) 255
%
\bibitem{Haymaker:1999}
R.~W. Haymaker,  Confinement Studies in lattice QCD,
Phys. Rep.  {\bf 315}  (1999)   153 
%
%
\bibitem{Greensite:2003}
J.~Greensite, 
Prog. Part. Nucl. Phys. {\bf 51}  (2003) 1
%
\bibitem{Kogut:2004}
J. B. Kogut and M. A. Stephanov,
The Phases of Quantum Chromodynamics: From Confinement to Extreme Environments,
Cambridge University Press, Cambridge UK (2004)
%
%
\bibitem{Ripka:2004}
G.~Ripka,  Dual superconductor models of color confinement, 
Lect. Notes  Phys. {\bf 639} (2004)  Springer-Verlag - Berlin - Heidelberg
%
\bibitem{Shifman:2009}
M. Shifman and M. Unsal,
Confinement in Yang-Mills: Elements of a Big Picture, 
Nucl. Phys. Proc. Suppl. {\bf 186} (2009) 235
%
%
\bibitem{Greensite:2011}
J.~Greensite,  An Introduction to the Confinement Problem,
Lect. Notes Phys. {\bf 821}  (2011) 
Second Edition, Springer Nature Switzerland AG 2011, 2020  
%
%
\bibitem{tHooft:1976}
G.~'t~Hooft, The confinement phenomenon in quantum field theory,  in 
 Proc. Eur. Phys. Soc. Conf. in  High Energy Physics, 
 A. Zichichi, G. Leone and European Physical Society, Compositori Editors (1976) pag. 1225 
%
\bibitem{tHooft:1980}
G.~'t~Hooft, Confinement and topology in non-Abelian gauge theories,
Acta Phys. Austr. Suppl. {\bf 22} (1980) 1531
%
\bibitem{tHooft:1982}
G.~'t~Hooft, The Topological Mechanism for Permanent Quark Confinement in a Non-Abelian Gauge Theory,
 Phys.  Scripta {\bf 25} (1982) 133
%
\bibitem{Mandelstam:1976}
S.~Mandelstam,  Vortices and quark confinement in non-Abelian gauge theories,
Phys. Rep.  C {\bf 23} (1976) 245 
%
%
\bibitem{Baker:1991}
M. Baker, J. S. Ball and  F. Zachariasen,
Dual QCD: a Review,
Phys. Rep.  {\bf 209} (1991) 73 
%
\bibitem{Kondo:2015}
K.-I. Kondo,  S. Kato, A. Shibata  and T. Shinohara,
Quark confinement: Dual superconductor picture based on a non-Abelian Stokes theorem and
reformulations of Yang-Mills theory,
Phys. Rep. {\bf 579} (2015) 1 
%
\bibitem{Gribov:1978}
V.~N.~Gribov,  Quantization of non-Abelian Gauge Theories,
Nucl. Phys. {\bf B139} (1978)  1 
%
\bibitem{Gribov:2001}
V.~N.~Gribov,  The Gribov Theory of Quark Confinement,
Editor J. Nyiri, World Scientific Publishing Co. Pte. Ltd., Singapore (2001)
%
\bibitem{Zwanziger:1989}
D.~Zwanziger, 
Local and renormalizable action from the gribov horizon,
Nucl. Phys. {\bf B323} (1989) 513
%
%
\bibitem{Vandersickel:2012}
N. Vandersickel and D.~Zwanziger, 
The Gribov problem and QCD Dynamics,
Phys. Rep. {\bf 520} (2012) 175
%
\bibitem{Greensite:2004a}
J.~Greensite, S.~Olejnik and D.~Zwanziger, 
Coulomb energy, remnant symmetry, and the phases of non-Abelian gauge theories,
Phys.\ Rev.\  {\bf D 69} (2004) 074506 
%
\bibitem{Gattnar:2004}
J.~Gattnar, K.~Langfeld and H.~Reinhardt,
Signals of Confinement in Green Functions of SU(2) Yang-Mills Theory,
 Phys.\ Rev.\ Lett.\  {\bf 93} (2004)  061601 
%
\bibitem{Greensite:2005}
J.~Greensite, S.~Olejnik and D.~Zwanziger, 
Center Vortices and the Gribov Horizon,
JHEP {\bf 0505} (2005)  070 
%
\bibitem{Feuchter:2004}
C.~Feuchter and H.~Reinhardt,
Variational solution of the Yang-Mills Schr\"odinger equation in Coulomb gauge,
 Phys.\ Rev.\ D {\bf 70} (2004) 105021
%
\bibitem{Reinhardt:2005}
H.~Reinhardt and C.~Feuchter, 
Yang-Mills wave functional in Coulomb gauge,
Phys.\ Rev.\ D {\bf 71} (2005) 105002 
%
%
\bibitem{Reinhardt:2012}
H.~Reinhardt, D. R. Campagnari, J. Heffner and M. Pak,
Variational approach to Yang-Mills theory with non-Gaussian wave functionals, 
Prog. Part. Nucl. Phys.  {\bf 67} (2012) 173 1
%
%
\bibitem{Feynman:1981}
R.~P.~Feynman, 
The qualitative behavior of Yang-Mills theory in 2 + 1 dimensions,
Nucl. Phys. {\bf B188} (1981)  479 
%
\bibitem{Cea:1997a}
P.~Cea, L.~Cosmai and A. D. Polosa, 
The Lattice Schr\"odinger Functional and the Background Field Effective Action,
Phys. Lett.  {\bf B392} (1997) 177 
%
\bibitem{Cea:1997b}
P.~Cea, L.~Cosmai and A. D. Polosa, 
Finite Size Analysis of the U(1) Background Field Effective Action, 
Phys. Lett.  {\bf B397} (1997)  229 
%
%
\bibitem{Cea:1997c}
P.~Cea and  L.~Cosmai, 
Lattice background effective action: A Proposal,
Nucl.Phys. B Proc. Suppl. {\bf 53}  (1997) 574
%
%
\bibitem{Cea:2003}
P.~Cea and L.~Cosmai, 
Abelian chromomagnetic fields and confinement,
JHEP {\bf 0302} (2003)  031
%
\bibitem{Cea:2005}
P.~Cea and L.~Cosmai, 
Color dynamics in external fields,
JHEP {\bf 0508} (2005)  079 
%
\bibitem{Cea:2006}
P.~Cea and L.~Cosmai, 
Deconfinement phase transition in external fields,
PoS {\bf LAT2005}  (2006)  289 
%
\bibitem{Cea:2007}
  P.~Cea, L.~Cosmai, M.~D'Elia, 
 QCD dynamics in a constant chromomagnetic field, 
 JHEP {\bf 0712} (2007) 097 
%
%
\bibitem{Savvidy:1977}
G. K. Savvidy, 
Infrared Instability of the Vacuum State of Gauge Theories and Asymptotic Freedom,  
Phys. Lett.  {\bf B71} (1977)  133 
%
%
\bibitem{Matinyan:1978}
S. G. Matinyan and G. K. Savvidy, 
Vacuum Polarization Induced by Intense Gauge Field,
Nucl. Phys. {\bf B134} (1978)  539 
%
%
\bibitem{Pagels:1978}
H. Pagels and E. Tomboulis, 
Vacuum of the quantum Yang-Mills theory and magnetostatics,
Nucl. Phys. {\bf B143} (1978)  485 
%
%
\bibitem{Savvidy:2020}
G. K. Savvidy, 
From Heisenberg-Euler Lagrangian to the discovery of Chromomagnetic Gluon Condensation,
Eur. Phys. J. C {\bf 80} (2020)  165 
%
\bibitem{Nielsen:1978a}
N.~K.~Nielsen and P.~Olesen,
Unstable Yang-Mills Field Mode,
 Nucl. Phys. {\bf B144} (1978) 376 
%
%
\bibitem{Nielsen:1978b}
N.~K.~Nielsen and P.~Olesen,
Electric vortex lines from the Yang-Mills theory,  
Phys. Lett.  {\bf B79} (1978)  304
%
%
\bibitem{Leutwyler:1981}
H. Leutwyler,
Constant Gauge Fields and Their Quantum Fluctuations,
 Nucl. Phys. {\bf B179} (1981) 129 
%
%
\bibitem{Ambjorn:1979}
J. Ambjorn, N. K. Nielsen and P. Olesen,
A hidden Higgs lagrangian in QCD, 
 Nucl. Phys. {\bf B152} (1979) 75 
%
\bibitem{Nielsen:1979a}
N.~K.~Nielsen and M. Ninomiya,
A bound on bag constant and Nielsen-Olesen unstable mode in QCD,
 Nucl. Phys. {\bf B156} (1979) 1 
%
\bibitem{Nielsen:1979b}
H. B.~Nielsen and P.~Olesen,
A quantum Liquid Model for the QCD Vacuum: Gauge and Rotational Invariance of Domained and Quantized Homogeneous
Color Fields,
 Nucl. Phys. {\bf B160} (1979) 380 
%
%
\bibitem{Ambjorn:1980a}
J. Ambjorn  and P. Olesen,
On the formation of a random color magnetic quantum liquid in QCD,
 Nucl. Phys. {\bf B170} (1980) 60
%
%
\bibitem{Ambjorn:1980b}
J. Ambjorn  and P. Olesen,
A Color Magnetic Vortex Condensate in QCD,
 Nucl. Phys. {\bf B170} (1980) 265 
%
\bibitem{Nielsen:1980}
 H. B. Nielsen,  Confinement with Special Emphasis on the Copenhagen Vacuum, in Particle Physics,
eds. I. Andric, I. Dadic and N. Zovko, North-Holland Publishing Company, Amsterdam -New-York - Oxford (1981), pag. 67
%
%
\bibitem{Olesen:1981}
 P. Olesen,
On the QCD Vacuum,
Physica Scripta {\bf 23} (1981) 1000 
%
%
\bibitem{Cea:1987}
P. Cea,
Stability Analysis of the Nielsen-Olesen Unstable Modes,
 Phys. Lett. {\bf B193} (1987)  268 
%
\bibitem{Cea:1988}
P. Cea, 
SU(2) gauge theory in a constant chromomagnetic background field,
Phys. Rev. {\bf D37} (1988) 1637
%
%
\bibitem{Cea:1991}
P. Cea and L. Cosmai,
 Constant background fields and unstable modes on the lattice, 
Phys. Lett.  {\bf B264} (1991)  415
%
%
\bibitem{Cea:1993}
P. Cea and L. Cosmai,
 Unstable modes in three-dimensional SU(2) gauge theory,  
Phys.\ Rev.\ D {\bf 48} (1993) 3364 
%
%
\bibitem{Cea:1997d}
P.~Cea and  L.~Cosmai, 
Exploring the Unstable Modes Dynamics by the Lattice Schr\"odinger
Functional,
Nucl.Phys. B Proc. Suppl. {\bf 53}  (1997) 578
%
%
\bibitem{Cea:1998}
P.~Cea and L.~Cosmai, 
Unstable Modes and Confinement in the Lattice Schr\"odinger Functional Approach,
Mod.\ Phys.\ Lett.\ A {\bf 13} (1998)  861
%
\bibitem{Cea:1999b}
 P.~Cea and L.~Cosmai, 
Probing the nonperturbative dynamics of the SU(2) vacuum, 
 Phys.\ Rev.\ D {\bf 60} (1999)  094506 
%
%
\bibitem{Jackiw:1984}
R. Jackiw, Topological investigations of quantized gauge theories,
in  Relativity, Groups and Topology~II, Proceedings of Les~Houches Summer School, edited by B. S. De~Witt
and R. Stora, North-Holland, Amsterdam  (1984), pag. 221
%
\bibitem{Fetter:1971}
A. L. Fetter and J. D. Walecka, Quantum Theory of Many-Particle Systems, 
McGraw-Hill, New York (1971)
%
\bibitem{Lowdin:1950}
P.  L\"owdin, On the NonOrthogonality Problem Connected with the Use of Atomic Wave
Functions in the Theory of Molecules and Crystals,
J. Chem. Phys. {\bf 18} (1950) 365
%
\bibitem{Coleman:1976}
S. Coleman,
Classical Lumps and Their Quantum Descendants,
New Phenomena in Subnuclear Physics - Part A,
Edited by A. Zichichi, Plenum Press - New York and London (1976)  pag. 297
%
\bibitem{Rajaraman:1982}
R. Rajaraman, Solitons and Instantons. An Introduction to Solitons and Instantons in Quantum Field Theory, 
North-Holland,  Amsterdam, New York, Oxford, Tokyo (1982)
%
\bibitem{Manton:2004}
N. Manton and P. Sutcliffe, Topological Solitons,
Cambridge University Press, Cambridge, New York (2004)
%
\bibitem{Weinberg:2012}
E. J. Weinberg, Classical Solutions in Quantum Field Theory. 
Solitons and Instantons in High Energy Physics,  Cambridge University Press, 
Cambridge, New York (2012)
%
\bibitem{Feynman:1988}
R. P. Feynman, Difficulties in Applying the Variational Principle to Quantum Field Theories,
in Variational Calculations in Quantum Field Theory, edited by L. Polley and D. E. L. Pottinger,
World Scientific, Singapore - New Jersey - Hong Kong (1988) pag. 28
%
\bibitem{Bjorken:1982}
J.D. Bjorken, Elements of Quantum Chromodynamics, in 
W.B. Atwood, J.D. Bjorken,  S.J. Brodsky and R. Stroynowski,
Lectures on Lepton Nucleon Scattering and Quantum
Chromodynamics, Springer Science+Business Media, LLC (1982) pag. 423
%
%
\bibitem{Kittel:1949}
C. Kittel, Physical Theory of Ferromagnetic Domains,
Rev. Mod. Phys. { \bf 21} (1949) 541
%
%
\bibitem{Weiss:1907}
P. Weiss,
L'hypothese du champ moleculaire et la propriet\'e ferromagnetique,
J. de Phys. {\bf 6} (1907) 661
%
\bibitem{Landau:1935}
L. Landau and E. Lifshitz, On the theory of the dispersion of magnetic permeability in ferromagnetic
bodies, Phys. Zeitsch. Sow.  {\bf 8} (1935) 153
%
\bibitem{Bloch:1932}
F. Bloch, Zur Theorie des Austauschproblems und der Remanenzerscheinung der Ferromagnetika,
Zeit. f. Phys. {\bf 74} (1932) 295
%
\bibitem{Coleman:1985}
S. Coleman, The Uses of Instantons, in S. Coleman, Aspects of symmetry, selected Erice Lectures, Cambridge
University Press, Cambridge (1985) pag. 265
%
\bibitem{Teper:1979}
M. J. Teper, Instantons, $\theta$ Vacua, Confinement .... A pedagogical Introduction,
Lectures given at Rutherford Laboratory and the University of Oregon, RL-80-004 (1979)
%
\bibitem{Berezinskii:1971}
V. L. Berezinskii, Destruction of long-range order in one-dimensional and two-dimensional systems
having a continuous symmetry group I. Classical systems,
Sov. Phys. JETP  {\bf 32} (1971) 493
%
\bibitem{Berezinskii:1972}
V. L. Berezinskii, Destruction of long-range order in one-dimensional and two-dimensional systems
having a continuous symmetry group I. Quantum systems,
Sov. Phys. JETP   {\bf 34} (1972) 610
%
\bibitem{Kosterlitz:1972}
J. M. Kosterlitz  and D. J. Thouless,  
Long range order and metastability in two-dimensional solids and superfluids,
J. Phys. C: Solid State Phys. {\bf 5} (1972) L124
%
\bibitem{Kosterlitz:1973}
J. M. Kosterlitz and D. J. Thouless, 
Ordering, metastability and phase transitions in two-dimensional systems,
J. Phys. C: Solid State Phys. {\bf 6} (1973) 1181
%
\bibitem{Kosterlitz:1974}
 J. M. Kosterlitz, 
 The critical properties of two-dimensional xy-model,
 J. Phys. C: Solid State Phys. {\bf 7} (1974) 1046
%
\bibitem{Kosterlitz:2016}
J. M. Kosterlitz,
Kosterlitz-Thouless physics: a review of key issues,
Rep. Prog. Phys. {\bf 79}  (2016) 026001
%
%
\bibitem{Elitzur:1975}
S. Elitzur,  Impossibility of Spontaneously Breaking Local Symmetries,
 Phys. Rev.  D {\bf  12} (1975) 3978
%
%
\bibitem{Narison:2018}
S. Narison,
QCD parameter correlations from heavy quarkonia,
Int. J. Phys. A {\bf 33} (2018) 1850045
%
\bibitem{Bali:2014}
G. S. Bali, C. Bauer and A. Pineda,
Model Independent Determination of the Gluon Condensate in Four Dimensional SU(3)
Gauge Theory,
Phys. Rev. Lett. {\bf 113} (2014) 092001
%
\bibitem{Necco:2002}
S. Necco and R. Sommer,
The $N_f=0$ heavy quark potential from short to intermediate distances,
Nucl. Phys.  {\bf B622} (2002) 328
%
\bibitem{Jackson:1999}
J. D. Jackson,
Classical Electrodynamics, John Wiley \& Sons Inc., New York (1999)
%
\bibitem{Felsager:1983}
B. Felsager, Geometry, Particles and Fields,
Odense University Press, Copenhagen (1983)
%
%
\bibitem{Cea:2016}
P. Cea, L. Cosmai , F. Cuteri and A. Papa,
Flux tubes at finite temperature,
JHEP {\bf 06} (2016) 033
%
\bibitem{Cea:2017}
P. Cea, L. Cosmai , F. Cuteri and A. Papa,
Flux tubes in the QCD vacuum,
Phys. Rev. D {\bf 95} (2017) 114511
%
\bibitem{Baker:2019}
M. Baker, P. Cea, V. Chelnokov, L. Cosmai , F. Cuteri and A. Papa,
Isolating the confining color field in the SU(3) flux tube,
Eur. Phys. J. C {\bf 79} (2019) 478
%
\bibitem{Baker:2020}
M. Baker, P. Cea, V. Chelnokov, L. Cosmai , F. Cuteri and A. Papa,
The confining color field in  SU(3) gauge theory,
Eur. Phys. J. C {\bf 80} (2020) 514
%
\bibitem{Baker:2022}
M. Baker, V. Chelnokov, L. Cosmai , F. Cuteri and A. Papa,
Unveiling confinement in pure gauge SU(3): flux tubes, fields, and magnetic currents, 
Eur. Phys. J. C {\bf 82} (2022) 937
%
\bibitem{Clem:1975}
J. R. Clem,
Simple model for the vortex core in a type II superconductor,
J. Low Temp. Phys. {\bf 18} (1975) 427
%
\bibitem{Bornyakov:2004}
V. G. Bornyakov, et al.,  Dynamics of monopoles and flux tubes in two-flavor
dynamical QCD, Phys. Rev. D {\bf 70} (2004)  074511
%
\bibitem{Yanagihara:2019}
R. Yanagihara, et al., Distribution of stress tensor around static quark-anti-quark
from Yang-Mills gradient flow,
Phys. Lett.  {\bf B789} (2019) 210
%
\bibitem{Schindler:2022}
A. Schindler, Gradient Flow: Perturbative and Non-Perturbative Renormalization,
EPJ Web Conf. {\bf 274} (2022) 01005
%
\bibitem{Suzuki:2013}
H. Suzuki, Energy-momentum tensor from Yang-Mills gradient flow,
Prog. Theor. Exp. Phys. {\bf 2013} (2013) 083B03
%
\bibitem{DElia:2021}
M. D'Elia,  L. Maio, F. Sanfilippo and A. Stanzione,
Confining and chiral properties of QCD in extremely strong magnetic fields,
Phys. Rev. D {\bf 104} (2021) 114512
%
\bibitem{Casher:1979}
A. Casher, H. Neuberger and S. Nussinov,
Chromoelectric-flux-tube model of particle production,
Phys. Rev. D {\bf 20}  (1979) 179
%
\bibitem{Schwinger:1951}
J. Schwinger, On Gauge Invariance and Vacuum Polarization,
Phys. Rev. {\bf 82} (1951) 664
%
\bibitem{Brezin:1970}
E. Brezin and C. Itzykson,
Pair Production in vacuum by an Alternating Field,
Phys. Rev. D {\bf  2} (1970) 1191
%
\bibitem{Wong:1994}
C.-Y. Wong, Introduction to High-Energy Heavy-Ion Collisions,
Word Scientific Publishing Co. Pte. Ltd., Singapore, New Jersey, London, Hong kong (1994)
%
\bibitem{Teper:1998}
M. Teper, Glueball masses and other properties of SU(N) gauge theories in D=3+1: a review
of lattice results for theorists,
arXiv:hep-th/9812187 (1998)
%
\bibitem{Kharzeev:2013}
D. Kharzeev, K. Landsteiner, A. Schmit and H.-U. Yee Editors,
Strongly Interacting Matter in Magnetic Fields,
Lect. Notes  Phys. {\bf 871} (2013) Springer-Verlag, Berlin, Heidelberg
%
\bibitem{Miransky:2015}
V. A. Miransky and I. A. Shovkovy,
Quantum field theory in a magnetic field: From quantum chromodynamics to graphene and Dirac semimetals,
Phys. Rep. {\bf 576} (2015) 1
%
\bibitem{Andersen:2016}
J. O. Andersen, W. R. Naylor and A. Tranberg,
Phase diagram of QCD in a magnetic field: A review,
Rev. Mod. Phys. {\bf 88} (2016) 025001
%
\bibitem{Hattori:2023}
K. Hattori, K. Itakura and S. Ozaki,
Strong-Field Physics in QED and QCD: From Fundamentals to Applications,
Prog. Part. Nucl. Phys. {\bf 133} (2023) 104068
%
\bibitem{Cea:1986}
P. Cea and G.Nardulli,
Bound states and asymptotically free quarks,
Phys. Rev. D {\bf 34} (1986) 1863
%
\bibitem{Bali:2012a}
G. S. Bali, et al., QCD quark condensate in external magnetic fields,
Phys. Rev. D {\bf 86} (2012) 071502
%
%
\bibitem{Bali:2012b}
G. S. Bali, et al., The QCD phase diagram for external magnetic fields,
JHEP {\bf 02} (2012) 044
%
\bibitem{Endrodi:2015}
G. Endr\"odi,
Critical point in the QCD phase diagram for extremely strong background magnetic field,
JHEP {\bf 07} (2015) 173
%
%
\bibitem{DElia:2022}
M. D'Elia, L. Maio, F. Sanfilippo and A. Stanzione,
Phase diagram of QCD in a magnetic background,
Phys. Rev. D {\bf 105} (2022)  034511
%
\bibitem{Ozaki:2014}
S. Ozaki, QCD effective potential with strong U(1)$_{em}$ magnetic fields,
Phys. Rev. D {\bf 89} (2014)  054022
%
\bibitem{Bjorken:1964}
J. D. Bjorken and S. D. Drell, Relativistic Quantum Mechanics,
McGraw-Hill Book Company, New York (1964)
%
%
\bibitem{Cea:1985}
P. Cea, 
Variational approach to (2+1)-dimensional QED,
Phys. Rev. D {\bf 32} (1985) 2785
%
%
\bibitem{Cea:1986b}
P. Cea, 
Variational approach to (2+1)-dimensional QED with topological mass term,
Phys. Rev.  D {\bf 34} (1986) 3229
%
%
\bibitem{Cea:1997e}
P. Cea, 
Vacuum stability for Dirac fermions in three dimensions,
Phys. Rev. D {\bf 55} (1997) 7985
%
\bibitem{Berezin:1966}
F. A. Berezin, The Method of Second Quantization,
Academic Press, New York  and London (1966)
%
\bibitem{Faddeev:1976}
L. D. Faddeev, Introduction to Functional Methods, in
R. Balian and J. Zinn-Justin, Methods in Field Theory,
North-Holland Publishing Company, Amsterdam, New York, London (1976) pag. 1
%
\bibitem{Casher:1979b}
A. Casher, 
Chiral Symmetry Breaking in Quark Confining Theories,
Phys. Lett. {\bf B83} (1979) 395
%
\bibitem{Banks:1980}
 T. Banks and A. Casher,
Chiral Symmetry Breaking in Confining Theories,
Nucl. Phys. {\bf B169} (1980) 103
%
\bibitem{Faber:2017}
M. Faber and R. H\"ollwieser,
Chiral symmetry breaking on the lattice,
Prog. Part. Nucl. Phys. {\bf 97} (2017) 312
%
\bibitem{Iritani:2015}
T. Iritani, G. Cossu and S. Hashimoto,
Partial restoration of chiral symmetry in the color flux tube,
Phys. Rev. D {\bf 91} (2015) 094501
%
%
\end{thebibliography}
\end{document}